\documentclass[11pt]{article}
\usepackage{url,bm,setspace}
\usepackage{amsmath,amssymb,amsthm,amsfonts,natbib,bm,setspace,graphics,float,graphicx,url,caption,multicol,longtable,lscape,verbatim,enumerate,color}

\usepackage[ruled,lined,boxed]{algorithm2e}

\usepackage[letterpaper, margin=1in]{geometry}
\date{}

\newtheorem{prop}{Proposition}

\newcommand{\bbeta}{{\bm \beta}}

\newcommand{\tR}{\mathbb{R}}

\newcommand{\bJ}{\mathbf{J}}

\newcommand{\cN}{\mathcal{N}}

\newcommand{\bS}{\mathbf{S}}

\newcommand{\bv}{\bm{v}}

\newcommand{\bd}{\bm{d}}

\newcommand{\cV}{\mathcal{V}}

\newcommand{\bSigma}{\bm{\Sigma}}
\newcommand{\bsigma}{\bm{\sigma}}

\newcommand{\bLambda}{\bm{\Lambda}}

\newcommand{\bA}{\mathbf{A}}

\newcommand{\ba}{\bm{a}}

\newcommand{\bU}{\mathbf{U}}

\newcommand{\rank}{\mathrm{rank}}
\newcommand{\diag}{\mathrm{diag}}
\newcommand{\tr}{\mathrm{tr}}
\newcommand{\bX}{\mathbf{X}}

\newcommand{\hatbW}{\hat{\bW}}
\newcommand{\bY}{\mathbf{Y}}

\newcommand{\bW}{\mathbf{W}}

\newcommand{\bF}{\mathbf{F}}

\newcommand{\bR}{\mathbf{R}}

\newcommand{\bK}{\mathbf{K}}
\newcommand{\bk}{\bm{k}}
\newcommand{\hatbR}{\hat{\bR}}
 
\newcommand{\bx}{\bm{x}}

\newcommand{\bz}{\bm{z}}

\newcommand{\hatbSigma}{\hat{\bm{\Sigma}}}

\newcommand{\bZ}{\mathbf{Z}}
\newcommand{\hatbZ}{\hat{\bZ}}

\newcommand{\bone}{\bm{1}}

\def\E{{\mathbb E}\,}
\def\Cov{{\rm Cov}\,}

\newcommand{\tn}[1]{\textnormal{#1}}
\newcommand{\iv}{{1\hspace{-2.5pt}\tn{l}}}

\newcommand{\cA}{\mathcal{A}}

\def\E{{\rm E}\,}

\def\Cov{{\rm Cov}\,}

\newcommand{\cJ}{\mathcal{J}}

\begin{document}

\title{ ACE of Space: Estimating Genetic Components of High-Dimensional Imaging Data}


\author{BENJAMIN B. RISK$^{1}$\footnote{To whom correspondence should be addressed: benjamin.risk@emory.edu}, HONGTU ZHU$^{2}$\\[4pt]
\smaller{$^{1}$Department of Biostatistics \& Bioinformatics, Emory University, Atlanta, GA, USA}\\
\smaller{$^{2}$Department of Biostatistics, University of North Carolina at Chapel Hill, Chapel Hill, NC, USA}}

\markboth%
{B. B. Risk and H. Zhu}
{ACE of Space}

\maketitle


\begin{abstract}
{It is of great interest to quantify the contributions of genetic variation to brain structure and function, which are usually measured by high-dimensional imaging data (e.g., magnetic resonance imaging). In addition to the variance, the covariance patterns in the genetic effects of a functional phenotype are of biological importance, and covariance patterns have been linked to psychiatric disorders.  The aim of this paper is to develop a scalable method to estimate heritability and the non-stationary covariance components in high-dimensional imaging data from twin studies. Our motivating example is from the Human Connectome Project (HCP). Several major big-data challenges arise from estimating the genetic and environmental covariance functions of functional phenotypes extracted from imaging data, such as cortical thickness with 60,000 vertices. Notably, truncating to positive eigenvalues and their eigenfunctions from unconstrained estimators can result in large bias. This motivated our development of a novel estimator ensuring positive semidefiniteness.  Simulation studies demonstrate large improvements over existing approaches, both with respect to heritability estimates and covariance estimation.  We applied the proposed method to cortical thickness data from the HCP. Our analysis suggests fine-scale differences in covariance patterns, identifying locations in which genetic control is correlated with large areas of the brain and locations where it is highly localized.}\\

\end{abstract}

\doublespacing

\section{Introduction}

It is of great interest to quantify  the contribution of genetic effects to brain structure and function, but 
scientific understanding of such effects is in its infancy (\citealt{chen2013genetic}). 
Measuring the relative size of genetic and environmental effects on brain traits may 
  provide insight into the etiology of neurological and neurodegenerative disorders (\citealt{kendler2001twin}). 
One method to  estimate genetic variation in brain phenotypes is to use twin studies.  A major goal of the young-adult Human Connectome Project (HCP) is to map the heritability and genetic underpinnings of brain traits (\citealt{van2013wu}). This dataset includes imaging data on approximately 1200 subjects with over 200 twin pairs.  

The traditional heritability model uses  monozygotic and dizygotic twins to decompose  genetic and environmental components of univariate phenotypes, and it provides guidance for molecular genetic studies \citep{van2012continuing}. 
Brain structure and function, however, are often  measured by high-dimensional imaging data and commonly represented as functional phenotypes.    For instance, various shape analysis methods have been developed to characterize brain cortical and subcortical structures in humans.
Functional  phenotypes extracted from shape analysis may be effective for the  identification of causal genes and  a mechanistic understanding of the pathophysiological processes of neurological disorders  \citep{zhao2016annual}. 
 
 As an illustration, we consider the cortical thickness dataset obtained from the HCP, where each cortical thickness functional phenotype is measured at approximately  $60,\!000$ locations on the cortical surface. The cerebral cortex is the outer layer of the brain and consists of a highly folded sheet of gray matter varying in thickness from approximately two to four millimeters. 
 Cortical thickness is important to cognition and intelligence, and cortical thinning may be associated with dementia \citep{dickerson2009cortical}.
 
 Correlations in cortical thickness between subregions measured across a population are biologically important, but the extent to which these correlations result from genetic influences is poorly understood  \citep{evans2013networks}. These correlations, also called cortical networks, may reflect coordinated developmental pathways and recapitulate certain white-matter tracts and functional networks \citep{Alexander-Bloch2013}. Cortical correlations have been associated with psychiatric and neurological disorders including depression and Alzheimer's \citep{wang2016disorganized,he2008structural}. Cortical correlations are typically estimated between different regions using a parcellation, but additional insight may be gained by examining higher resolution spatial covariance functions. In this paper, we will  use the HCP  data set to measure the heritability and  various covariance structures (e.g., genetic) of cortical thickness in the healthy human brain. In particular, we focus on the additive genetic covariance function, $\Sigma_a(v,v')$, described in Section \ref{sec:ACE}. The clinical meaning of $\Sigma_a(v,v')$ is the covariance in the genetic component of cortical thickness between two locations. For example, a positive value indicates an individual with a thicker cortex in location $v$ tends to have a thicker cortex in location $v'$ due to genetic factors. 
 
The aim of this paper is to develop a scalable method that improves estimates of heritability and three non-stationary covariance functions -- genetic, shared (i.e., common) environmental, and unique environmental -- of high-dimensional functional phenotypes. \cite{luoetal} proposed a heritability model for twin functional data based on Fisher's Additive genetic, Common environmental, and unique Environmental (ACE) model (see Section \ref{sec:ACE}). However, high-dimensional functional data with thousands or more grid points results in several major big-data challenges.  The first big-data challenge is the dimensionality of the covariance matrices, each of which consists of   $V(V+1)/2 \gg n$ unknown parameters, where $V$ is the number of grid points for a given functional phenotype and $n$ is the number of twin pairs.   It is computationally intractable to use joint maximum likelihood estimation (MLE) to estimate these $V \times V$ covariance matrices. Alternatively, one may resort to a pairwise analysis by  estimating the genetic and environmental correlations of each grid pair, for example using bivariate MLE separately for all possible grid pairs. However, the computational costs are still very high with no software available to perform such a large-scale implementation, and there are additional statistical issues described below.

The second big-data challenge is to develop a method in which the covariance functions are positive semidefinite (PSD), where estimators lacking this property may be less accurate.  Pairwise approaches do not ensure the PSD properties of the three non-stationary covariance functions. 
In contrast, joint estimation approaches can result in PSD estimates, and as we will show, large improvements in accuracy. However, joint approaches become computationally more difficult as $V$ increases. Methods to estimate covariance matrices and functions in the high dimensional, low sample size setting have been recently proposed \citep{xiao2016fast} but are not  applicable to multiple covariance functions from twin functional data. Our major contributions are as follows: 
\begin{itemize}
 \item We propose novel estimators of non-stationary genetic and environmental effects in twin studies with high dimension, low sample size data. 
 \vspace{-0.3cm}
  \item We propose estimates of measurement-error corrected heritability, where measurement error can be estimated based on the smoothness of the underlying biological processes.
 \vspace{-0.3cm}
  
 \item We  automate smoothing using kernels that incorporate local information based on geodesic distance and use generalized cross validation   (GCV) to select  the bandwidth, which in our application leads to a higher effective resolution.
 \vspace{-0.3cm}

 \item We estimate the covariance patterns in genetic effects in cortical thickness in the Human Connectome Project, which provides detailed insight into cortical networks. 

  \end{itemize}
    
 \vspace{-0.3cm}

The remainder of this paper is organized as follows. In Section 2, we present Fisher's model for heritability and a recent functional extension. We then present our estimators and algorithm. In Section 3, we conduct a simulation study. In Section 4, we analyze the HCP cortical thickness data and conclude with a discussion in Section 5. 

\section{Methods}

\subsection{The ACE of space model}\label{sec:ACE}

Fisher's ACDE model proposes additive genetic (A), dominant genetic (D), common environmental (C), and unique environmental (E) components of variation in a phenotype. Additive genetic effects can be estimated by assuming the correlation between genetic traits is 1 for monozygotic (MZ) twins and 0.5 for dizygotic (DZ) twins, which is based on genetic theory. The correlation for dominant effects is 1 for MZ and 0.25 for DZ, but dominant and additive effects are not simultaneously identifiable in the basic twin study design. The ACE model controls for effects due to the shared environment and is most appropriate for polygenic phenotypes. Heritability estimates from the ACE model are called narrow-sense heritability, denoted by $h^2$, which contrasts with broad-sense heritability, $H^2$, which includes dominant effects. 

Let $i\in \{1,\dots,n\}$, where $i$ indexes the family and $n$ the total number of families. Let $n_1$ denote the number of MZ pairs and $\cN_1$ the set of families with MZ pairs; $n_2$ and $\cN_2$ denote the number and set, respectively, of DZ pairs; and $n_3$ and $\cN_3$ denote singletons with no relatives in the dataset. Let $m_i = 2$ if the $i$th family contains twins and  $m_i = 1$ if the $i$th family comprises a singleton, and let $j$ index the individual in the $i$th family. For clarity, we here assume that all observations $y_{ij}$ belong to one of these three classes, but inclusion of non-twin siblings is discussed in Section \ref{sec:CovarianceFunctions}. We define the standard ACE model using the mixed model formulation \citep{rabe2008biometrical} as follows: 
\begin{align}\label{eq:ACEmodel}
 y_{ij} = X_{ij}^T \bbeta + \sqrt{0.5} \iv(i \in \cN_2) a_{ij} 
+ \left\{ \iv(i \in \{\cN_1 \cup \cN_3\}) + \sqrt{0.5} \iv(i \in \cN_2) \right\}a_i + c_i + e_{ij},
\end{align}
where $\iv(\cdot)$ is an indicator function;  $X_{ij} \in \tR^p$ are fixed covariates, which are typically effects we want to control for that are not of primary interest; $\bbeta \in \tR^p$ is a vector of coefficients; $a_{ij} \overset{iid}{\sim} N(0,\sigma_a^2)$ and $a_i \overset{iid}\sim N(0,\sigma_a^2)$ are the additive genetic effects; $c_i \overset{iid}\sim N(0,\sigma_c^2)$ is the common environmental effect; $e_{ij}\overset{iid}\sim N(0,\sigma_e^2)$ is the random effect for the total unique variance, which is equal to the sum of the unique environmental effects plus measurement error; and $a_{ij}, a_i, c_i$, and $e_{ij}$ are mutually independent. In the next section, we will decompose $e_{ij}$ into the unique environmental effect and measurement error. The ACE model can also be formulated as a structural equation model.
The formulation in \eqref{eq:ACEmodel} assumes that there are no dominant effects (no non-additive genetic effects), gene-gene interactions (epistasis), gene-environment interactions, and no assortative mating.  The standard approach for heritability estimates in neuroimaging is to pre-smooth the data and then estimate a separate model at each location, where the amount of smoothing is fixed a priori.

Next, we apply the functional model in \cite{luoetal} to a spatial domain, which we call the ACE of space. 
Let $\cV$ denote a spatial domain and $v$ an arbitrary location. In our application, $\cV$ is the cortical surface, in which the subjects are aligned in a common domain consisting of two 2D manifolds (one for each hemisphere) embedded in 3D space, and data (e.g., cortical thickness) is measured at approximately 60,000 locations (vertices). We use the spherical representation of each manifold in which each cerebral ``hemisphere'' is represented by its own sphere, with null values in the non-cortical areas corresponding to the connection between the two hemispheres (Figure S.8 in the Supplementary Materials). This is the common template in the processed data in which cortical thickness at a given vertex represents the same location across subjects relative to the aligned cortical folding patterns of the sulci and gyri.  The location of each vertex is denoted by a triple $(\theta_v,\phi_v,(L/R)_v)$, where the last coordinate denotes the cerebral hemisphere. The locations with data are denoted as $\cV_0 = \{(\theta_1,\phi_1,(L/R)_1),\dots,(\theta_V,\phi_V,(L/R)_V)\}$. For conciseness, we hereafter denote these locations with single indices $v \in \{1,\dots,V\}$ or $v \in \cV_0$. 
Modeling will incorporate the local spatial correlation using kernel regression with geodesic distance, as measured using the great circle distance, and the kernel is equal to zero when vertices are in different hemispheres. See Appendix C. More generally, long-distance correlations will be estimated from the data. See Section \ref{sec:CovarianceFunctions}. For display, the vertices are mapped to the Conte69 atlas, which is an average of cortical folding patterns from subjects in an independent dataset.

Let $e_{ij,G}(v)$ denote the unique environmental effect generated from a Gaussian process, and let $e_{ij,L}(v)$ denote the measurement error. Then, the functional ACE model is
\begin{align}\label{eq:fsem}
 y_{ij}(v) &= X_{ij}(v)^T \bbeta(v) + \sqrt{0.5} \iv(i \in \cN_2) a_{ij}(v) \\
 \nonumber &\;\;\;+ \left\{ \iv(i \in \{\cN_1 \cup \cN_3\}) + \sqrt{0.5} \iv(i \in \cN_2) \right\}a_i(v) + c_i(v) + e_{ij,G}(v)+e_{ij,L}(v).
\end{align}
It is assumed that $a_{ij}(v)$, $a_i(v)$, $c_i(v)$, $e_{ij,G}(v)$, and $e_{ij,L}(v)$ are mutually independent mean-zero Gaussian processes with covariance functions $\bSigma_a(v,\cdot)$, $\bSigma_a(v,\cdot)$, $\bSigma_c(v,\cdot)$, $\bSigma_{e,G}(v,\cdot)$, and $\bSigma_{e,L}(v,\cdot)$, respectively. We assume $\bSigma_{e,L}(v,v') = 0$ for $v \ne v'$. We have
\begin{align*}
 \Cov (y_{ij}(v),y_{ij}(v')) &= \bSigma_a(v,v') + \bSigma_c(v,v') + \bSigma_{e,G}(v,v') + \bSigma_{e,L}(v,v'), \\
 \Cov (y_{i1}(v),y_{i2}(v')) &= \left\{ \iv(i \in \{\cN_1\}+ 0.5 \iv(i \in \cN_2) \right\} \bSigma_a(v,v') + \bSigma_c(v,v').
\end{align*}
To simplify notation, we let $\bSigma_a(v,v) = \sigma_a^2(v)$, $\bSigma_c(v,v) = \sigma_c^2(v)$, $\bSigma_{e,G}(v,v) = \sigma_{e,G}^2(v)$, and $\bSigma_{e,L}(v,v) = \sigma_{e,L}^2(v)$. Then narrow-sense heritability is defined:
\begin{align}\label{eq:h2}
 h^2(v) = \sigma_a^2(v)/\left\{\sigma_a^2(v) + \sigma_c^2(v) + \sigma_{e,G}^2(v)\right\},
\end{align}
which is corrected for the measurement error due to  $\sigma_{e,L}^2(v)$. 

Pairwise covariance estimators were proposed in \cite{luoetal} based on kernel regression and applied to 150 locations on the corpus callosum for 129 twin pairs. We examine these estimators, called S-FSEM (Symmetric estimators from the Functional Structural Equation model), in simulations (Section \ref{sec:Simulations}). This approach can result in negative estimates of variance parameters, particularly for $V \gg n$. Here, we develop a method that jointly estimates the covariance function (under PSD constraints) for thousands of locations, which can improve both estimates of heritability and correlation patterns. 

\subsection{Covariance estimation}\label{sec:CovarianceFunctions}

When the estimate of a covariance function is not PSD, it is common to truncate to the positive eigenvalues and their associated eigenfunctions. 
This approach generally results in lower MISE than the original symmetric functions \citep{hall2008modelling}. 
Truncation was used to estimate a genetic covariance function in a pedigree model for cow growth in \cite{lei2015functional}. In the ACE of space model, we initially truncated to positive eigenvalues of the FSEM, called PSD-FSEM, but this resulted in large biases as examined in Section \ref{sec:Simulations}. This may be more problematic in our $V \gg n$ application. This motivated estimation of the ACE model under PSD constraints (PSD-ACE), which we summarize in six steps. 

\begin{itemize}
 \item \emph{Step 1.} Calculate point-wise MLEs of unknown parameters in  model \eqref{eq:ACEmodel}. 
 \vspace{-0.25cm}
 \item \emph{Step 2.} Smooth the parameter estimates using bandwidths selected using GCV and calculate the fixed effect residuals. 
 \vspace{-0.25cm} 
 \item \emph{Step 3.} Estimate the measurement error.
  \vspace{-0.25cm}
 \item \emph{Step 4.} Use the fixed-effect residuals and estimates of measurement error as input to an initial estimator of the covariance functions, which has a convenient closed-form solution, and estimate the bandwidths using GCV.
  \vspace{-0.25cm}
 \item \emph{Step 5.} Estimate the rank of the covariance functions from the number of positive eigenvalues (aided by scree plots) and truncate to the corresponding positive eigenvalues/vectors. 
  \vspace{-0.25cm}
 \item \emph{Step 6.} Estimate the covariance functions under PSD constraints initialized from Step 5. 
 \vspace{-0.25cm}
\end{itemize}
\underline{Step 1} is straightforward, e.g., \cite{rabe2008biometrical}. In our implementation, we numerically optimize the log of the variance parameters, which constrains their estimates to be positive. \underline{Step 2} aims to decrease the mean square error (MSE) of the point-wise MLEs since the MLEs in one location tend to be similar to those in adjacent locations. The smoothed MLEs (SMLEs) are calculated using a biweight (quartic) kernel with bandwidth $h$, denoted $k_h(v,v_0)$, based on the geodesic distance between locations $v$ and $v_0$. The biweight kernel is defined as 
\begin{align}\label{eq:biweight}
 k_h(v,v_0) = \frac{15}{16h}\left\{ 1 - (d(v,v_0)/h)^2 \right\}^2 \iv_{d(v,v_0)\le h}
\end{align}
and is used throughout.  We use the convention that $v$ is the focal vertex, which here we restrict to $v \in \cV_0$, and $v_0 \in \cV_0$ are locations that can contribute to the estimate at the focal vertex $v$.  A sparse $V \times V$ smoothing matrix is formed, $\widetilde{\bK}_h$, with normalized entries 
$(\widetilde{\bK}_h)_{v,v_0} = k_h(v,v_0)/w(v), 
$ 
with $w(v) = \sum_{v_0 \in \cV_0} k_h(v,v_0)$. Let $\hat{\bsigma}^2_{e\; MLE}$ be a vector of length $V$ of the MLE estimates of the unique environmental plus measurement error variance. Then define $\hat{\bsigma}^2_{e\;SMLE-\!h} = \tilde{\bK}_h \hat{\bsigma}_{e\;MLE}^2$. GCV is an approximation to leave-one-location out cross-validation and is calculated as
\begin{align}
GCV(h) = \frac{1}{V}  \sum_{v=1}^V \left\{ \frac{\hat{\sigma}_{e\;MLE}^2(v) - \hat{\sigma}^2_{e\;SMLE-\!h}(v)}{1 - \tr(\tilde{\bK}_h)/V} \right\}^2,
\end{align}
where $\tr$ is the trace operator. Then the GCV is calculated for a grid of values of $h$ and the value minimizing the GCV is chosen. Note that this approach allows a separate bandwidth to be chosen for each parameter, which contrasts with maximum weighted likelihood. This procedure is repeated for $\hat{\bbeta}_{MLE}$, and the residuals from the smoothed estimates are calculated. 

\underline{Step 3} uses the difference between the estimates of the total variance from the SMLEs and estimates of the variance due to genetic and environmental effects derived based on smoothness assumptions. 
 We use kernel regression with the biweight kernel to estimate the smooth sum of covariance functions in which diagonal elements are excluded as described in Appendix A.1, which has similarities to estimating the nugget effect in spatial statistics. \underline{Step 4} is related to the closed-form solutions presented in the FSEM in \cite{luoetal}, whose method is described in Appendix A.2. Our modification lends itself to GCV for bandwidth selection and is described in Appendix A.3. We call this modification the sandwich estimator in the spirit of \cite{xiao2016fast}, abbreviated as SW hereafter.  \underline{Step 5} chooses the rank based on the scree plot. When $V$ is greater than the number of subjects, such as our data application, the rank is clearly determined by the number of twins and subjects due to constraints on the maximum possible rank (e.g., Figure S.11 of the Appendix). Hence, the approach is arguably less ad-hoc in our application than its use in standard PCA.  
 
\underline{Step 6} greatly improves upon the initial estimates. Let $m_i$ denote the number of individuals in a family. We order the families such that $\cN_1 = \{1,\dots,n_1\}$, $\cN_2 = \{n_1+1,\dots,n_2\}$, and $\cN_3 = \{n_2+1,\dots,n\}$, and order the individuals such that the twins are indexed by $j=1$ and $2$. Let $N = \sum_{i=1}^n m_i$. For some observed location $v_0$, let $\widehat{R}_{i,j}(v_0)= y_{ij}(v_0)-X_{ij}(v_0)^T \widehat\bbeta(v_0)$, the fixed effect residuals obtained from Step 2, and $\hat{\sigma}_{e,L}^2(v_0)$ be the estimate of the measurement error obtained from Step 3. For $i  \in \{\cN_1 \cup \cN_2\}$, define
\begin{align}\label{eq:Uhatij}
\widehat{U}_{i,j}(v_0,v_0') = \widehat{R}_{i,j}(v_0)\widehat{R}_{i,j}(v_0') - \hat{\sigma}_{e,L}^2(v_0)\iv( v_0 = v_0') ,
\end{align}
\begin{align}
\widehat{U}_i(v_0,v_0') = \left(\widehat{R}_{i,1}(v_0)\widehat{R}_{i,2}(v_0') + \widehat{R}_{i,1}(v_0')\widehat{R}_{i,2}(v_0)\right) / 2.
\end{align}

We consider the discrete problem restricting the covariance functions $\bSigma_a(\cdot,\cdot)$ and $\bSigma_c(\cdot,\cdot)$ to observed locations $\cV_0$. Let $\bSigma_a$ denote the $V \times V$ matrix. Define $d_a = \rank(\bSigma_a)$, $d_c = \rank(\bSigma_c)$, and $d_{e,G} = \rank(\bSigma_{e,G})$. Consider the class of $V \times V$ (symmetric) PSD matrices $\cA$ of rank $d_a$. We can define this class as $\cA = \{\bA = \bZ_a \bZ_a^T: \bZ_a \in \tR^{V \times d_a}\}$. 
Let $\bz_v^a$ denote the $v$th row of $\bZ_a$. Similarly define $\bz_v^c$ and $\bz_v^{e,G}$ for $\bZ_c \in \tR^{V \times d_c}$ and $\bZ_{e,G} \in \tR^{V \times d_{e,G}}$. For focal locations $v,v' \in \cV_0$ and contributing locations $v_0,v_0' \in \cV_0$ (i.e., have non-zero weight when nearby), define the PSD-ACE objective function 
 \begin{align}\label{eq:objfun_psd}
  \cJ^{PSD} &= \underset{\bZ_a \in \tR^{V \times d_a},\bZ_c \in \tR^{V \times d_c},\bZ_{e,G} \in \tR^{V \times d_{e,G}}}{\min} \\  
  \nonumber &\frac{1}{N} \sum_{i=1}^n \sum_{j=1}^{m_i} \sum_{v,v'} \sum_{v_0, v_0'} \left\{ \widehat{U}_{ij} (v_0,v_0') - {(\bz_v^a)}^T {\bz_{v'}^a}- {(\bz_v^c)}^T {\bz_{v'}^c} - {(\bz_v^{e,G})}^T {\bz_{v'}^{e,G}} \right\}^2 k_h(v,v_0) k_h(v',v_0') \\
\nonumber &+ \frac{1}{n_1} \sum_{i=1}^{n_1} \sum_{v,v'} \sum_{v_0,v_0'}  \left\{ \widehat{U}_{i} (v_0,v_0') - {(\bz_v^a)}^T {\bz_{v'}^a}- {(\bz_v^c)}^T {\bz_{v'}^c}  \right\}^2 k_h(v,v_0) k_h(v',v_0') \\
\nonumber &+ \frac{1}{n_2} \sum_{i=n_1+1}^{n_1+n_2} \sum_{v,v'} \sum_{v_0, v_0' }  \left\{ \widehat{U}_{i} (v_0,v_0') - 0.5 {(\bz_v^a)}^T {\bz_{v'}^a}- {(\bz_v^c)}^T {\bz_{v'}^c} \right\}^2 k_h(v,v_0) k_h(v',v_0').
 \end{align}
We note that this decomposition is not identifiable because the minimum is not unique. However, we are not interested in the decompositions beyond their usefulness as a memory efficient representation of the covariance matrices. The key observation is that this re-parameterization allows the calculation of an analytic gradient that is scalable. Then we can optimize \eqref{eq:objfun_psd} by initializing from the closed-form solution in Step 5. Extensive simulations support the estimation steps outlined here. The steps and iterations of the gradient descent algorithm can be seen as progressively decreasing the MSE: truncating the eigenvalues/vectors decreases the MSE relative to the symmetric (closed form) estimator, and then the gradient steps result in additional improvements. This will be seen in the simulations. See also the discussion in Section 3.2. In our applications, we did not have issues with convergence, but an approach using diagonalization after each iteration could also be pursued. 

Also note this objective function uses the product of two bivariate functions as the kernel for covariance estimation, i.e., $k_h(v,v';v_0,v_0') = k_h(v,v_0)k_h(v',v_0')$, which results in computational simplifications that enable estimation for large datasets. We again use the biweight kernel, and note its finite support decreases computational expense.

The computational complexity of   (\ref{eq:objfun_psd}) is $O(V^4)$, which makes it impracticable for modestly sized $V$, much less $V=60,\!000$. However, we can derive a gradient descent algorithm in which there is a one time cost of $O(V^3)$ and updates are $O(V^2)$; the formulas for the analytic gradients appear in Appendix A.4. We initially use a modestly sized learning rate. In each iteration, the algorithm checks if the norm of the gradient increased relative to the previous iteration, and if so, halves the learning rate. Convergence is assessed relative to the initial size of the gradient; see Algorithm \ref{algorithm1}.
\begin{algorithm}
\smaller
\SetKwInOut{Input}{Inputs}
\SetKwInOut{Output}{Output}
\SetKwFor{For}{for}{}{endfor}
\SetKwFor{While}{while}{}{endwhile}
\caption{Covariance estimation}\label{algorithm1}
\Input{The $N \times V$ data matrix $\bY$ and design matrix $\bX$; tolerance $\epsilon$, e.g., 0.0001; learning rate $\lambda$, e.g., 0.1.}
\KwResult{$\hatbSigma^{PSD-\!ACE}_a$, $\hatbSigma^{PSD-\!ACE}_c$, and $\hatbSigma^{PSD-\!ACE}_{e,G}$.}
\begin{enumerate}
  \item Estimate measurement error, $\hat{\bsigma}_{e,L}$, and residuals, $\hatbR$, using SMLE with input $\bY$ and $\bX$. (Steps 1-3)
  \item Calculate $\hatbSigma_a^{S-\!SW}$, $\hatbSigma_c^{S-\!SW}$, and $\hatbSigma_{e,G}^{S-\!SW}$  in which the bandwidths are chosen using GCV. These bandwidths will be used in subsequent estimators. (Step 4)
  \item Choose the rank $d_a$ based on the scree plot for $\hatbSigma_a^{S-\!SW}$. Use the selected eigenvalue/eigenvector pairs to generate an initial value $\bZ_a^{(0)}$. Repeat this process for $\bZ_c^{(0)}$ and $\bZ_{e,G}^{(0)}$.  (Step 5)
  \item Calculate gradients $\nabla_a^{(0)}$, $\nabla_c^{(0)}$, and $\nabla_{e,G}^{(0)}$ using the initial values and calculate $\alpha_0 = \sqrt{||\nabla_a^{(0)}||_F^2 + ||\nabla_c^{(0)}||_F^2 + ||\nabla_{e,G}^{(0)}||_F^2}$, and let $n=0$. 
  \item While $\sqrt{||\nabla_a^{(n)}||_F^2 + ||\nabla_c^{(n)}||_F^2 + ||\nabla_{e,G}^{(n)}||_F^2} > \epsilon \alpha_0$, increment $n$ and calculate $\bZ_a^{(n)} = \bZ_a^{(n-1)} - \lambda \nabla_a^{(n-1)}$, and similarly for $\bZ_c^{(n)}$ and $\bZ_{e,G}^{(n)}$. 
  \item If $\sqrt{||\nabla_a^{(n)}||_F^2 + ||\nabla_c^{(n)}||_F^2 + ||\nabla_{e,G}^{(n)}||_F^2} > \sqrt{||\nabla_a^{(n-1)}||_F^2 + ||\nabla_c^{(n-1)}||_F^2 + ||\nabla_{e,G}^{(n-1)}||_F^2}$, then return to the previous step and decrease the learning rate, $\lambda^* = \lambda/2$. Re-calculate  $\bZ_a^{(n)} = \bZ_a^{(n-1)} - \lambda^* \nabla_a^{(n-1)}$, and similarly for $\bZ_c^{(n)}$ and $\bZ_{e,G}^{(n)}$. 
\end{enumerate}
\end{algorithm}

For clarity, we described \eqref{eq:objfun_psd} for twins and unrelated singletons; however, we can also use information from non-twin siblings. By looking at the expected values of the products of residuals in \eqref{eq:Uhatij}, it can be seen that the non-twin siblings can be treated in the same way as singletons, and thus are included in the first summand in \eqref{eq:objfun_psd}. In a similar manner, we can treat non-twin siblings as singletons in steps 3--5. Note we do not make the assumption that non-twin siblings have the same common environment as their twin siblings.

This procedure results in three covariance matrices, which are the covariance functions evaluated at $V \approx 60,\!000$ locations. However, it may be desirable to obtain their corresponding covariance functions, which can be used to evaluate the heritabilities and covariances at unobserved grid locations; see Appendix A.5. The approach can also be used to estimate the covariance functions from a sample of points from the spatial domain to decrease computational expense, particularly with respect to memory usage. We can partition the locations, estimate the covariance functions for each subset, interpolate to all locations, and subsequently combine the estimates. Code on the author's GitHub is available to implement this low-memory approach.   

\section{Simulations}\label{sec:Simulations}

\subsection{Simulation design}

We simulated functional spatial data for 100 MZ pairs, 100 DZ pairs, and 200 singletons with 1,002 grid points  on the unit sphere with the sizes of the variances motivated from the HCP cortical thickness data and the basis functions chosen to result in a realistic range of variances and covariances. We defined covariance functions using sixth-order even spherical harmonics, which comprise twenty-eight basis functions, $\bx_1(v),\dots,\bx_{28}(v)$. This order was chosen to result in a mixture of high and low frequency fluctuations. Higher frequencies capture quick-changes in correlation structure, which is motivated by the patterns observed in Section \ref{sec:HCP}. We then defined $\bSigma_a$ using five basis functions as $ 
 \bSigma_a(v,v') = \alpha_a \sum_{k = \{1,7,13,19,25\}} \bx_k(v) \bx_{k}(v'), 
$
where $\alpha_a$ was chosen such that $  \sum_{v} \bSigma_a(v,v)/{V} = 0.015$. The value of 0.015  is approximately equal to the average of the MLE of the genetic variance (in mm$^2$) in cortical thickness across all locations, as calculated in Section \ref{sec:HCP}. Similarly,  we define 
$
 \bSigma_c(v,v') = \alpha_c \sum_{k = \{2,8,14,20,26\}} \bx_k(v) \bx_k(v'), 
$
where $\alpha_c$ was chosen such that $ {V}^{-1} \sum_{v} \bSigma_c(v,v) = 0.010$. The value of 0.010 is approximately equal to the average common environmental variance in Section \ref{sec:HCP}. 
Next, $\bSigma_{e,G}(v,v')$ was defined with basis functions $k \in \{1, 3, 9, 15, 21, 27\}$ and scaled so that  the average variance was 0.12. This resulted in heritability that ranged from 0.016 to 0.498 with mean 0.126. Finally, $\bsigma_{e,L}(v)$ was defined from the diagonal of the matrix formed from the basis functions $k \in \{1, 4, 10, 16, 22, 28\}$ and scaled to have average equal to 0.03. Estimation included a design matrix with a column of ones and a continuous covariate, while their true coefficients were equal to zero.

Here, we define the quantities $\textrm{ISE}^{(t)} =  \sum_{v=1}^V \sum_{v'=1}^V \left\{ \hatbSigma_a^{(t)}(v,v') - \bSigma_a(v,v') \right\}^2/{V^2}$ and $\textrm{MISE}= \sum_t \textrm{ISE}^{(t)}/{T}$ as measures of error, where $t$ denotes the $t$th simulation. We also present normalized versions in Appendix B.
 
We compare three estimators of the covariance functions: (i) the symmetric FSEM proposed in \cite{luoetal}, S-FSEM, defined in Appendix A.2;  (ii) the PSD analogue of S-FSEM based on truncating to positive eigenvalues (PSD-FSEM); and (iii)  the PSD-ACE estimator. We also compared three additional estimators of the covariance functions:  (iv) the symmetric sandwich estimator (S-SW) defined in Appendix A.3 and used in initialization in Step 4 of the PSD-ACE; (v) the PSD-SW based on truncating to positive eigenvalues; and (vi) PSD-ACE oracle estimator (PSD-ACE-O), based on Algorithm 1 but using the true ranks. The S-SW results are very similar to S-FSEM, and the PSD-SW results are very similar to PSD-FSEM; additionally, PSD-ACE-O results are very similar to PSD-ACE (Tables S.1 and S.2 and Figures S.1-S.4 of the Appendix). We include the primary estimators S-FSEM, PSD-FSEM, and PSD-ACE in the main manuscript. Based on an inspection of the scree plots from a few hundred simulations, we chose eight, eight, and six eigenvalues for $\hatbSigma_a$, $\hatbSigma_c$, and $\hatbSigma_{e,G}$, respectively (where the true ranks were 5, 5, and 6) for the PSD-ACE.  
We also compared the variances and heritabilities with the point-wise MLE and the maximum weighted likelihood estimator (MWLE) with bandwidth selected using 5-fold CV as in \cite{luoetal} (defined in Appendix A.6).

\subsection{Simulation results}\label{sec:SimResults}

\begin{figure}[ht]
 \includegraphics[width=0.3\textwidth]{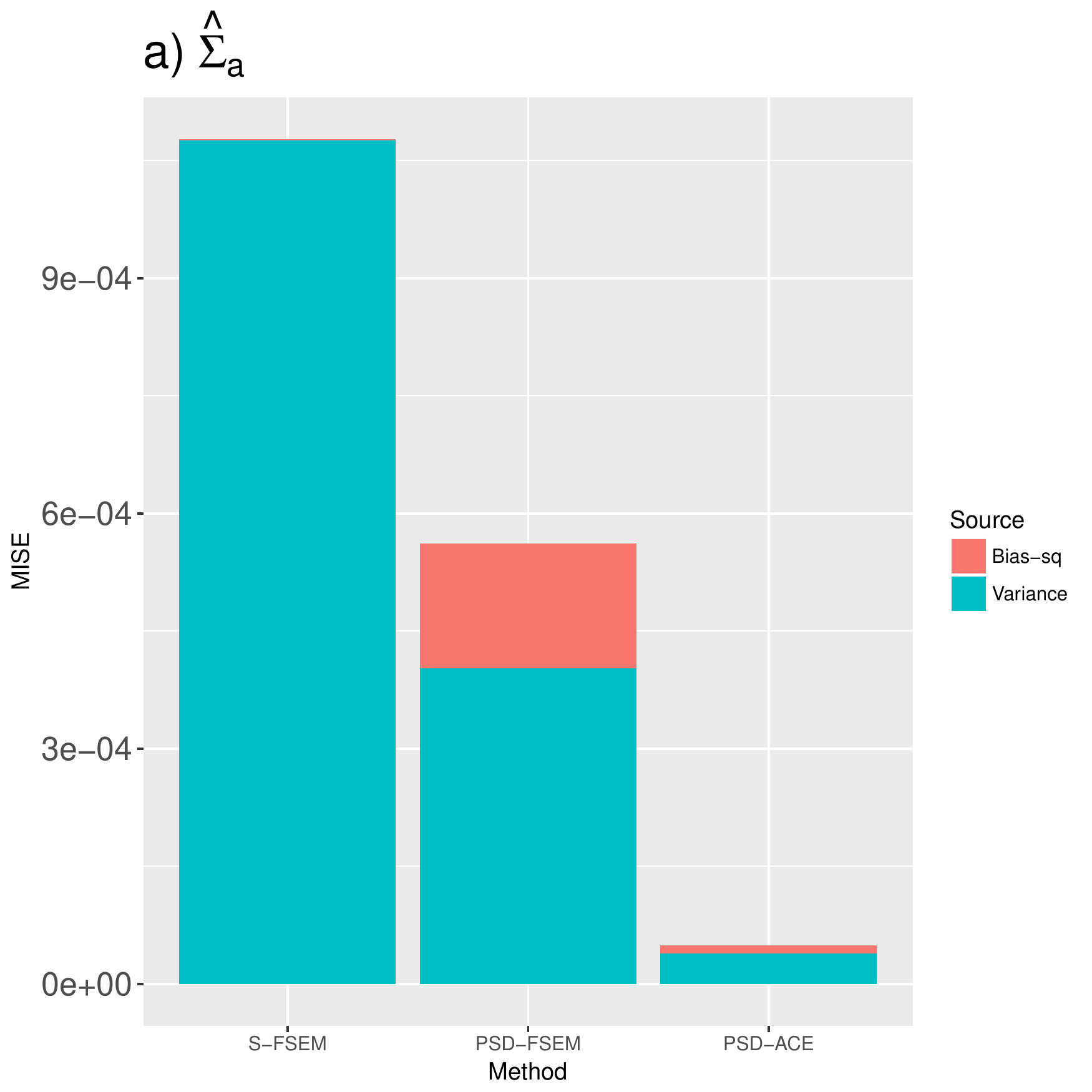}
 \includegraphics[width=0.3\textwidth]{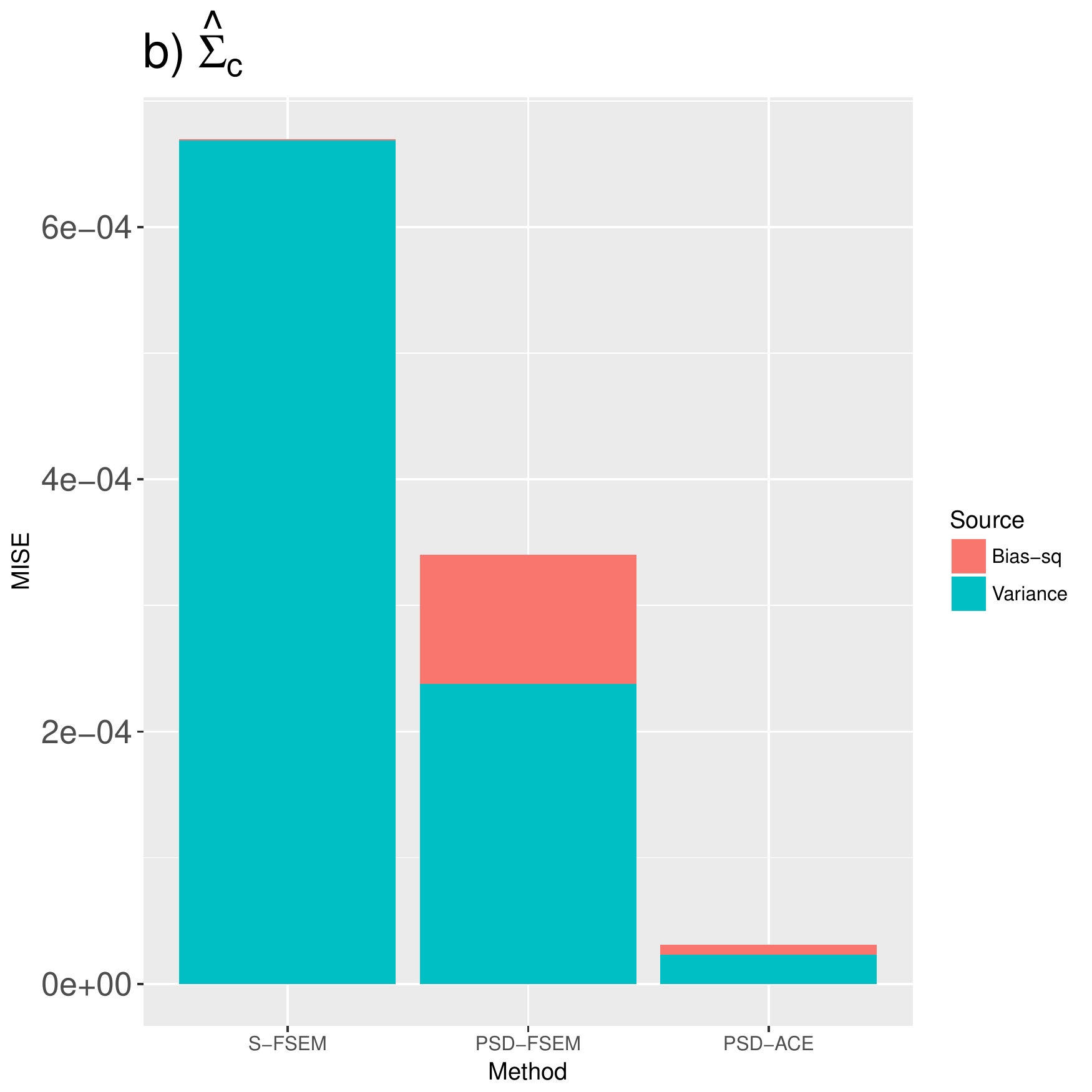} 
 \includegraphics[width=0.3\textwidth]{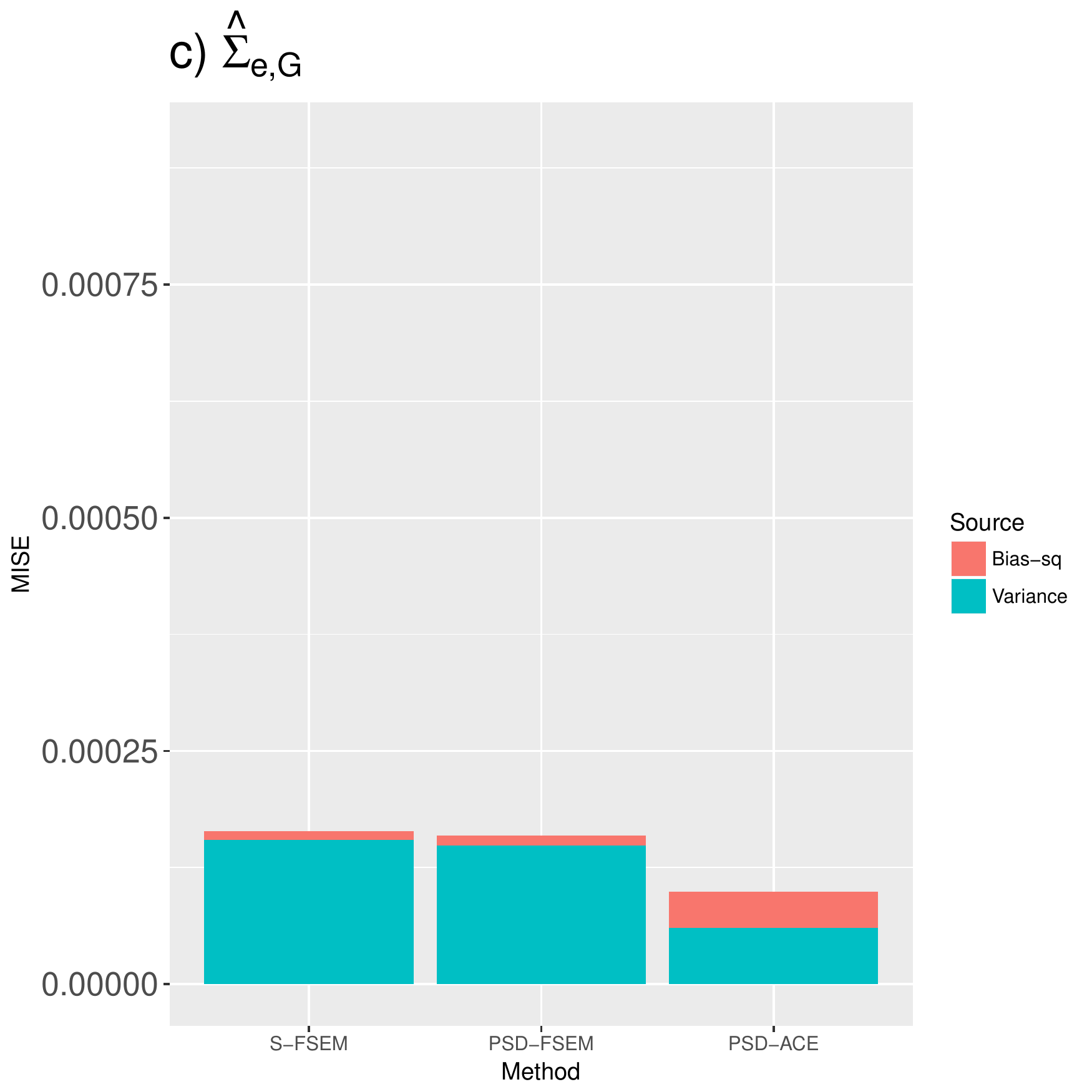} 
 \includegraphics[width=0.3\textwidth]{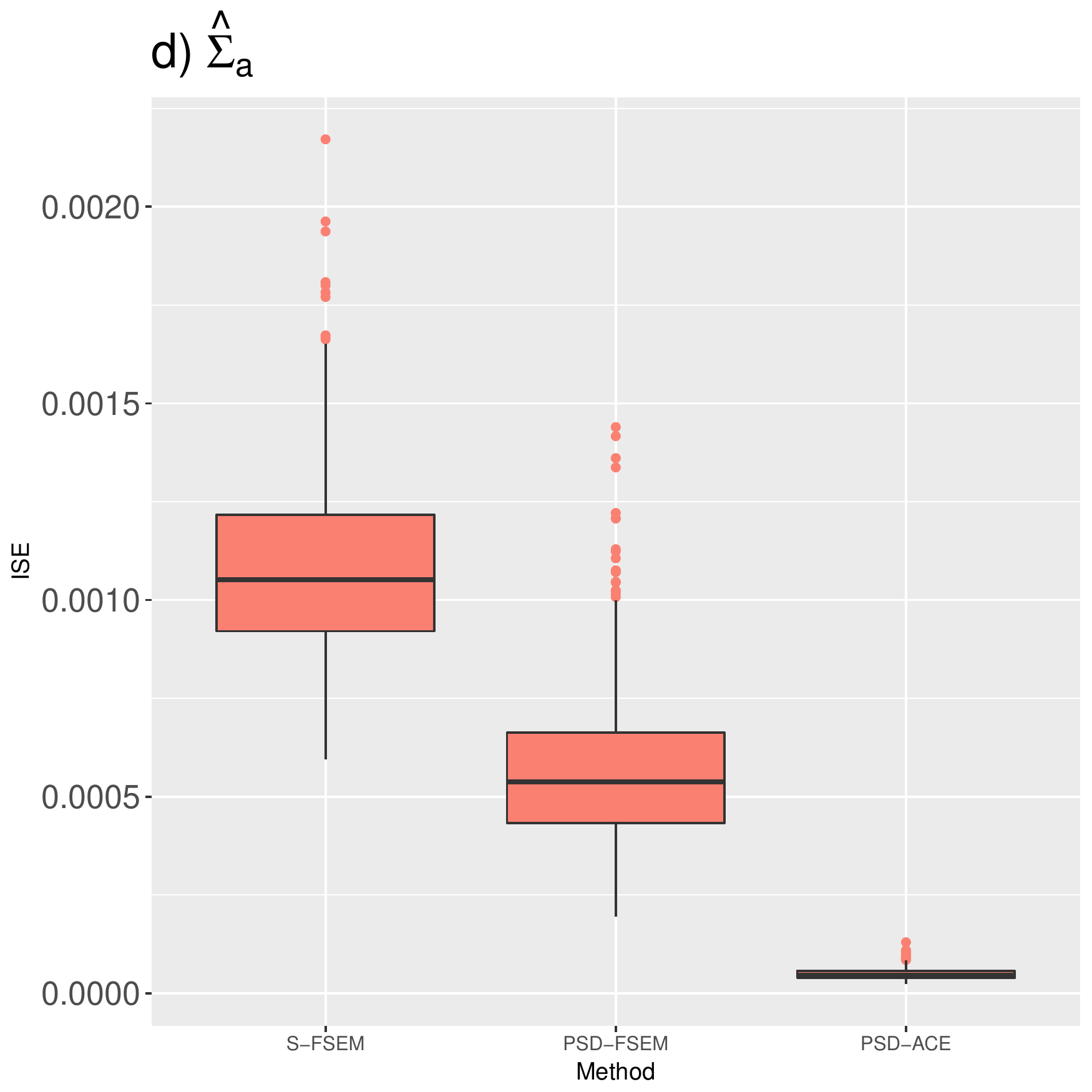}
 \includegraphics[width=0.3\textwidth]{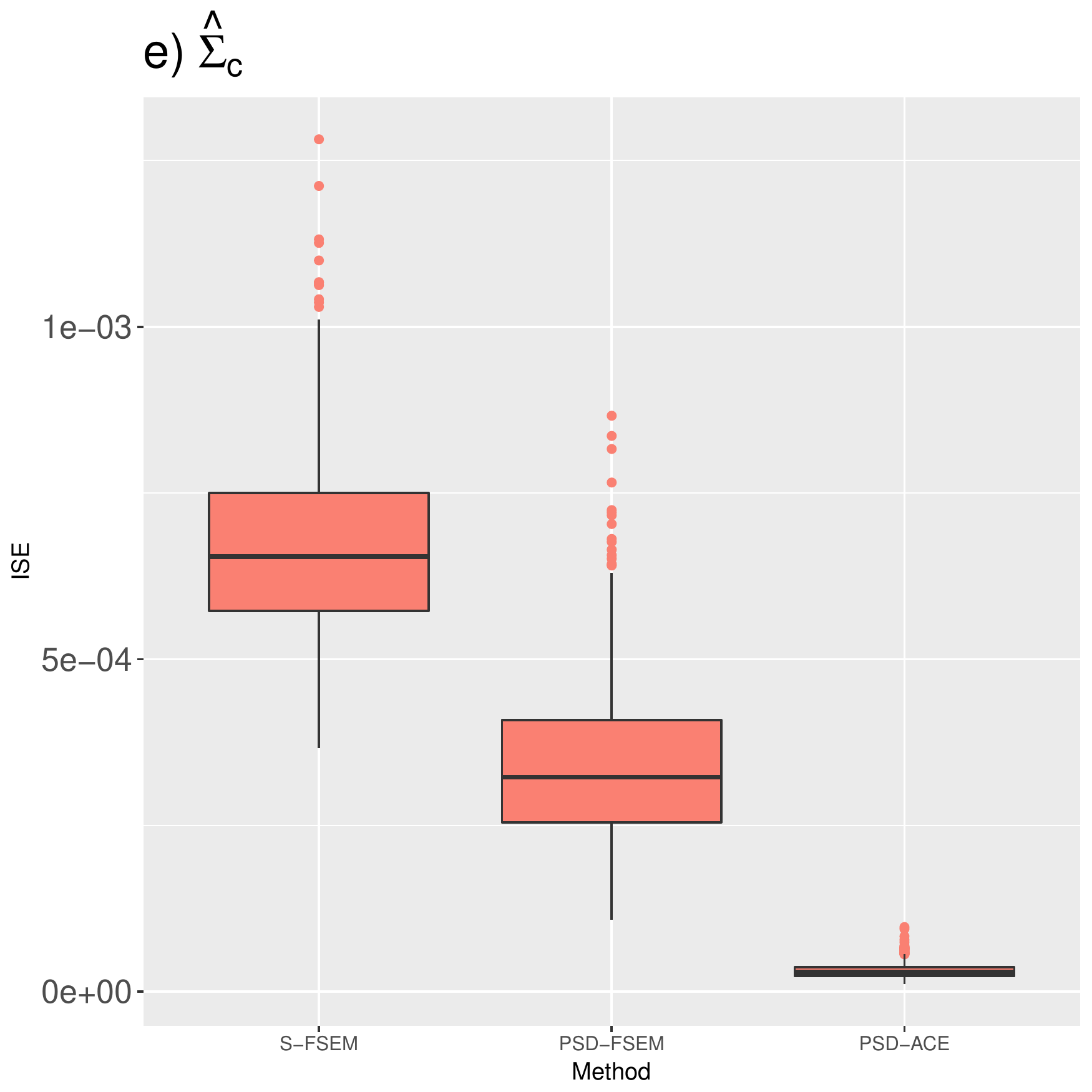}
 \includegraphics[width=0.3\textwidth]{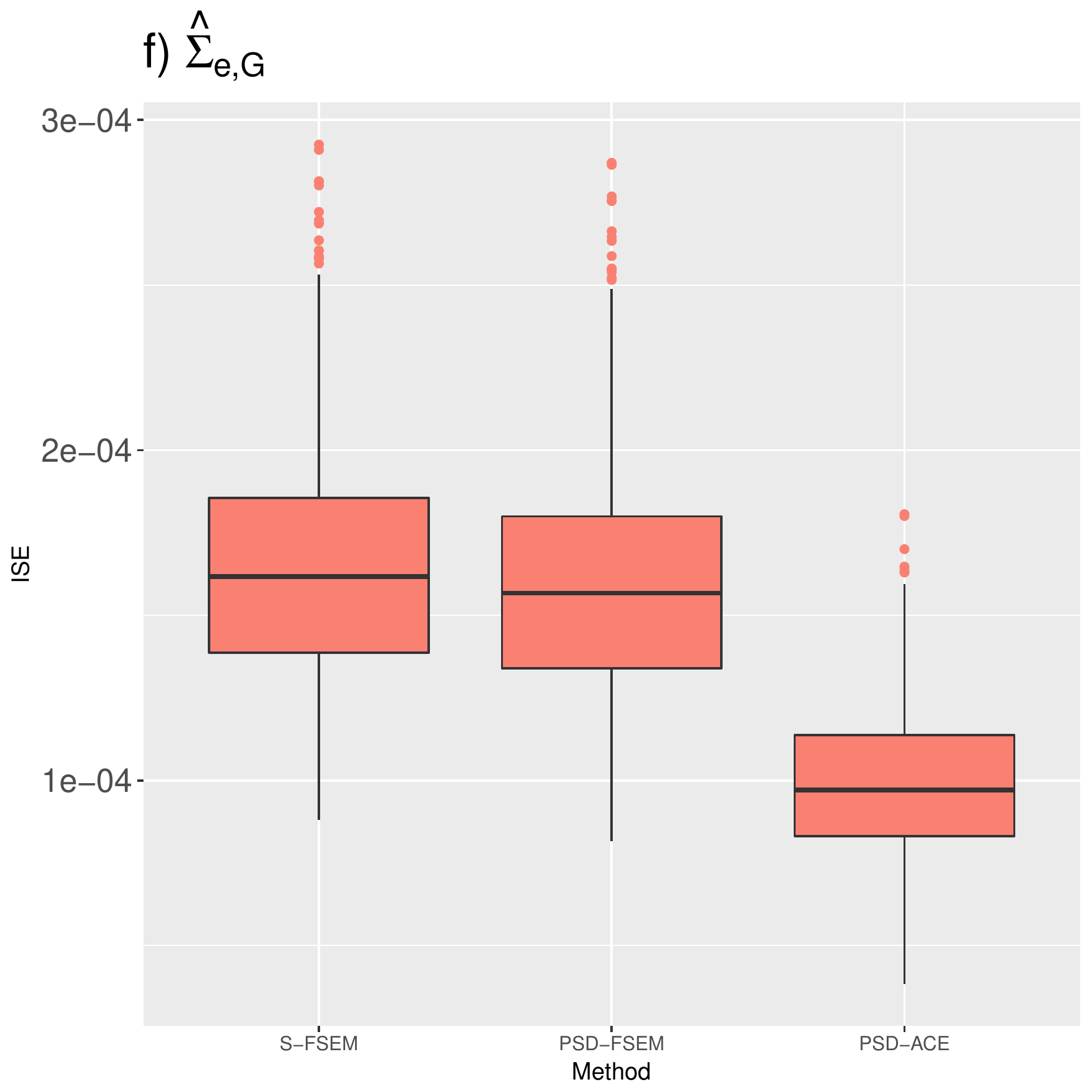}
\caption{MISE of covariance functions for the S-FSEM, PSD-FSEM (truncated eigenvalues of S-FSEM), and PSD-ACE. Panels d, e, and f depict boxplots of the ISE from 1,000 simulations. See Figure S.1 in the Appendix for a normalized version of this figure.}\label{fig:boxplotcovfun}
\end{figure}

Overall, the MISEs for $\hatbSigma_a$ and $\hatbSigma_c$ are much lower for PSD-ACE than the other methods (Figure \ref{fig:boxplotcovfun}, Figure S.1 of the Appendix). In all individual simulations, PSD-ACE has a lower ISE than S-FSEM and PSD-FSEM for all covariance matrices. For $\hatbSigma_a$ and $\hatbSigma_c$, S-FSEM has the lowest bias but largest variance, whereas PSD-FSEM has high bias but smaller variance, while PSD-ACE has some bias but less than PSD-FSEM and dramatically lower variance (Figure \ref{fig:boxplotcovfun}a and b). A different pattern emerges for $\hatbSigma_{e,G}$. The S-FSEM version of this estimator is PSD in most simulations. Consequently, the S-FSEM and PSD-FSEM versions are very similar. For $\hatbSigma_{e,G}$, PSD-ACE again has the best MISE driven by lower variance, but now has more bias relative to PSD-FSEM. 
To visualize the bias, we can examine plots of the covariance between a focal vertex, i.e., a seed, and the 1,002 vertices of the discretized spatial domain, which corresponds to a row of the covariance matrix of the $V$ locations. PSD-FSEM tends to inflate differences between locations, in particular having higher values near the seed, whereas the constrained estimates tend to shrink the differences towards zero (Figure \ref{fig:wsSigmaACov2}, Figure S.2 of the Appendix).

\begin{figure}
\centering \includegraphics[width=0.7\textwidth]{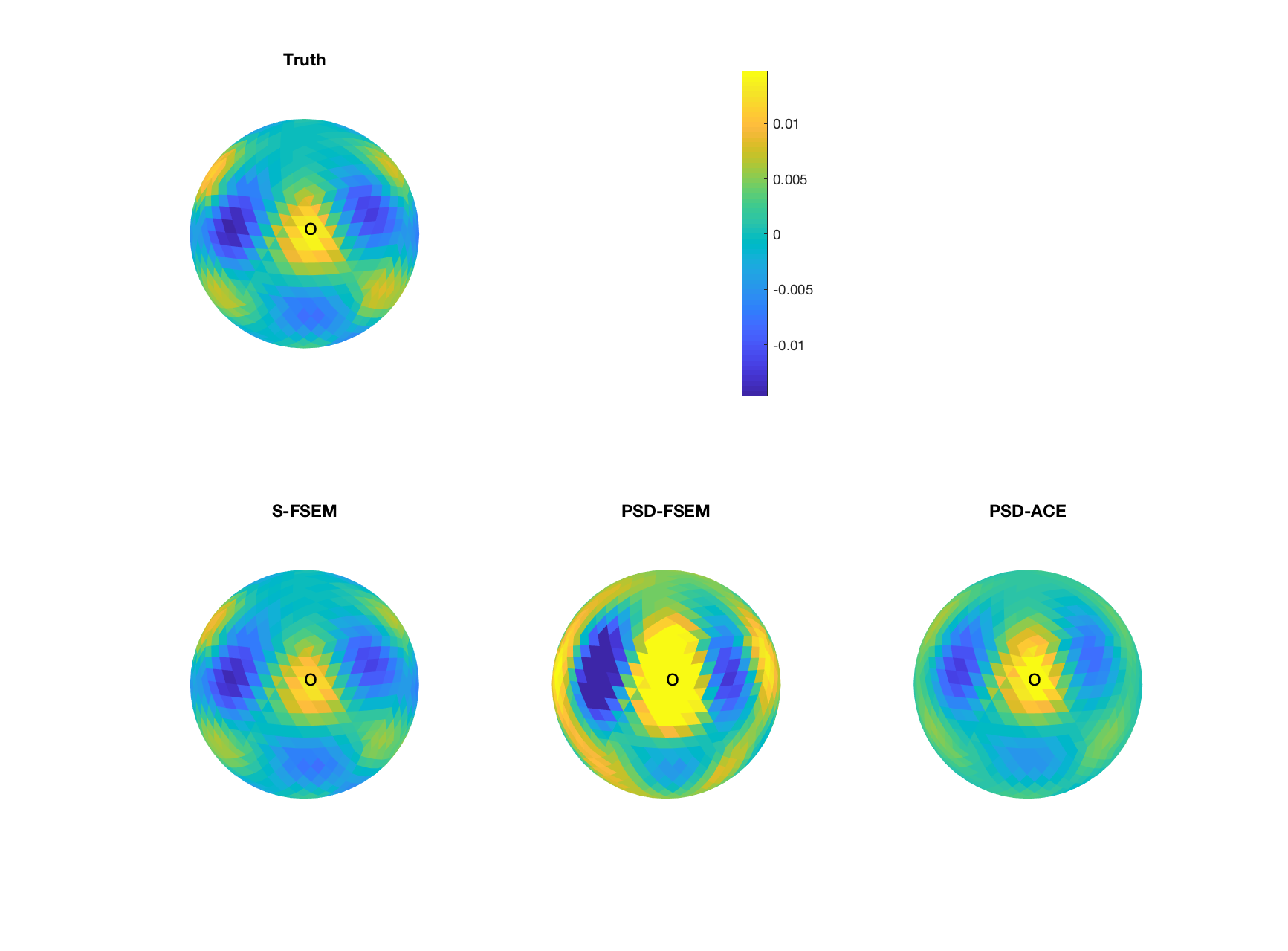}
\caption{Visualizing bias in covariance estimation. Plotted is the average across 1,000 simulations of $\hatbSigma_c^{(t)}(v,\cdot)$, where $t$ denotes the simulation, and the covariance between a randomly selected seed location ($v=888$) and 1,002 locations is evaluated.}\label{fig:wsSigmaACov2}
\end{figure}

When we restrict attention to the MISE of the variance functions (diagonals of the previous matrices), we again observe that PSD-ACE has the best MISE, and in particular outperforms the point-wise likelihood methods (Figure \ref{fig:varianceetc}, Figure S.3 of the Appendix). Many of the estimates of variance and heritability for S-FSEM are negative, contributing to large variances, which results in large MISEs. The average genetic variance across all locations is biased upwards for PSD-ACE (for $\hat{\bsigma}_a$, 
the average is 0.020 whereas the true average is 0.015), but this represents a dramatic improvement relative to PSD-FSEM (0.091), while MLE and MWLE are the least biased (0.015, 0.015) (see also Figures S.4-S.7 of the Appendix). Note that in Figure \ref{fig:varianceetc} c,  $\hat{\sigma}_{e,G}^2$ for MLE and MWLE depict estimates of measurement error plus unique environmental variance, $\sigma_{e,G}^2 + \sigma_{e,L}^2$, since measurement error is not identifiable in the point-wise approach. Consequently, MLE and MWLE have large bias. 

For heritability, PSD-ACE is less biased than MLE and MWLE due to the ability to disentangle measurement error and unique environmental variance. Variance and bias accumulate in the heritability estimates such that the relative improvements of PSD-ACE over other approaches are even greater (Figure \ref{fig:varianceetc} d and e, Table S.2 of the Appendix), and the unidentifiability of measurement error in the pointwise MLE and MWLE results in downwardly biased estimates of heritability (Figures S.6 and S.7 of the Appendix). 
 
\begin{figure}[h!]
\includegraphics[width=0.3\textwidth]{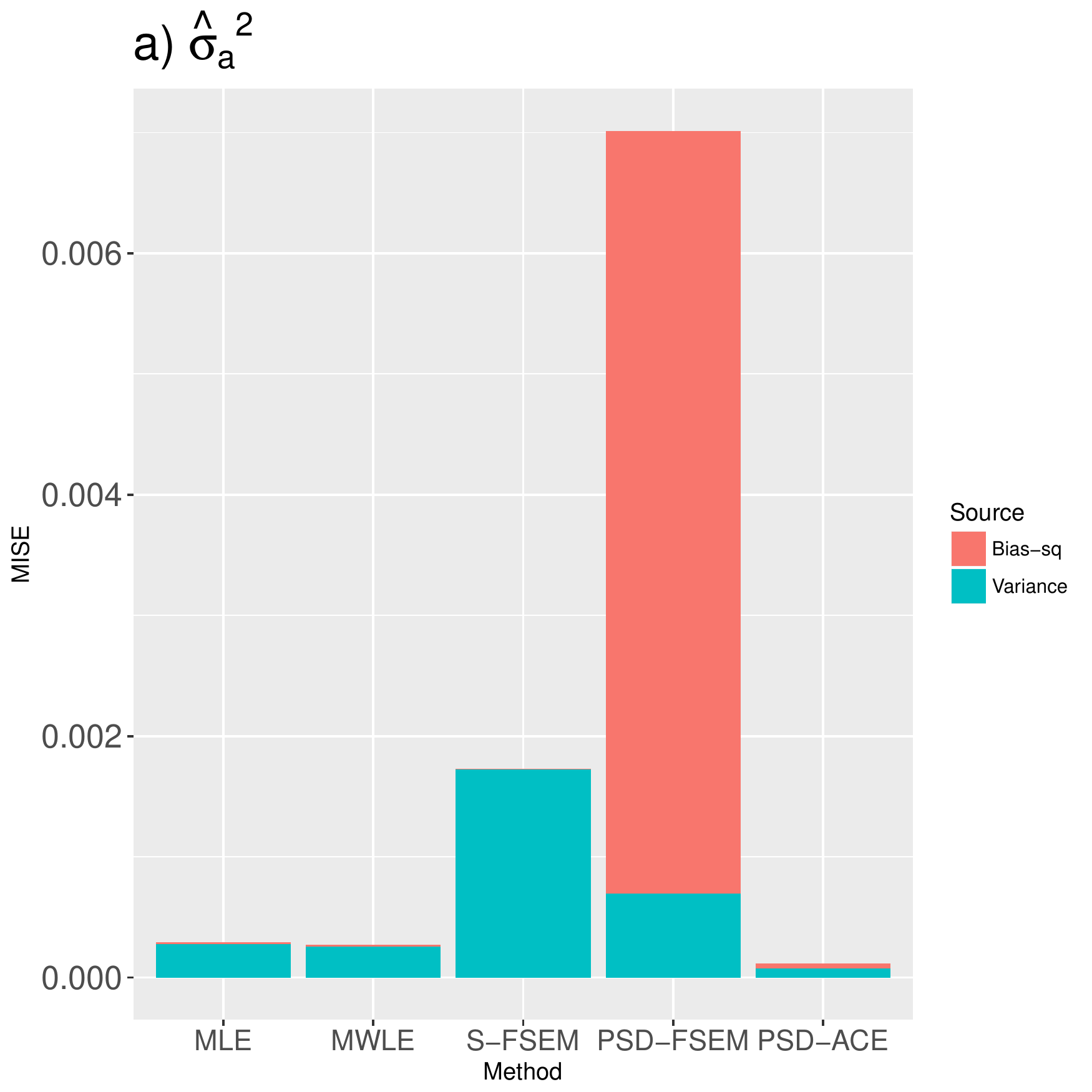}
\includegraphics[width=0.3\textwidth]{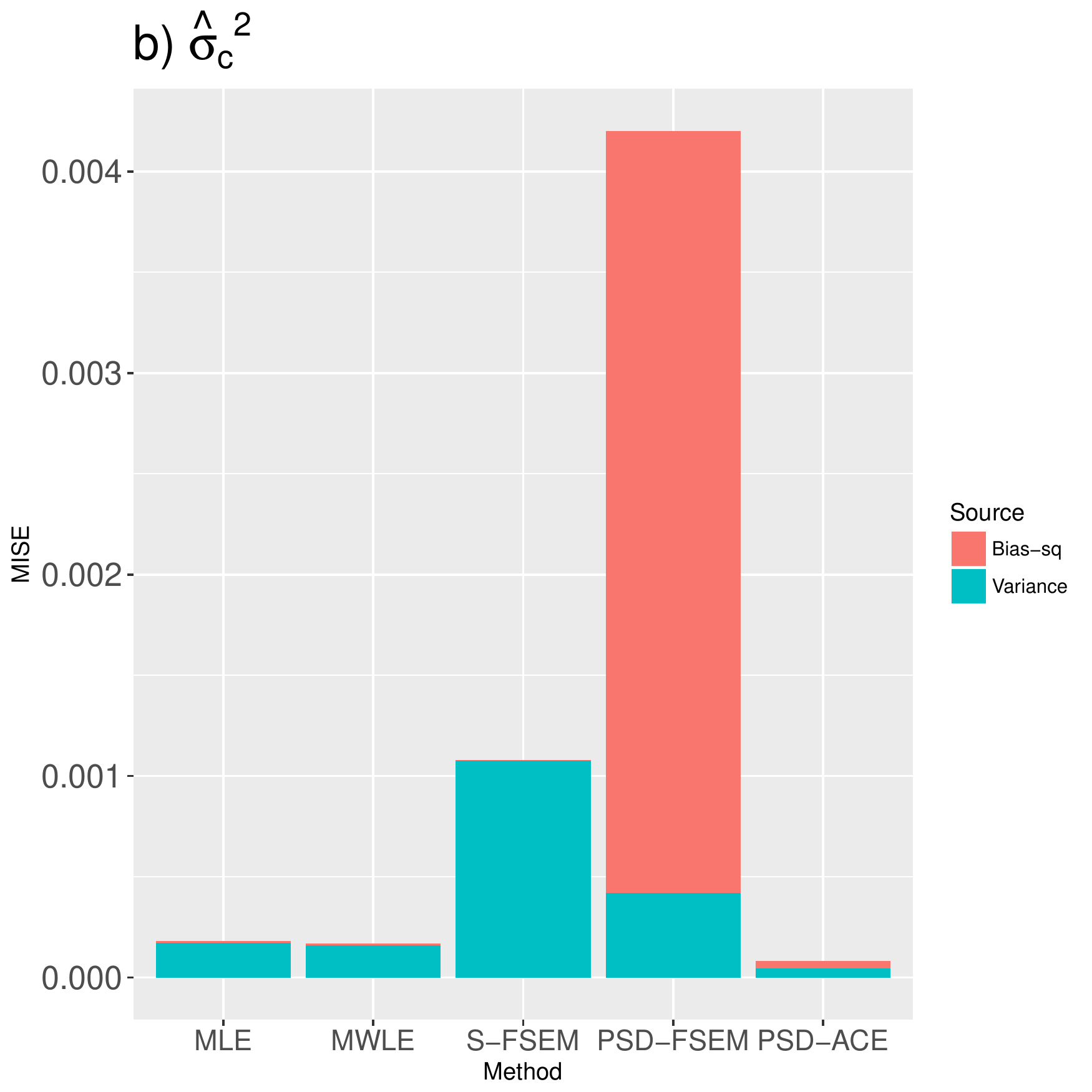}
\includegraphics[width=0.3\textwidth]{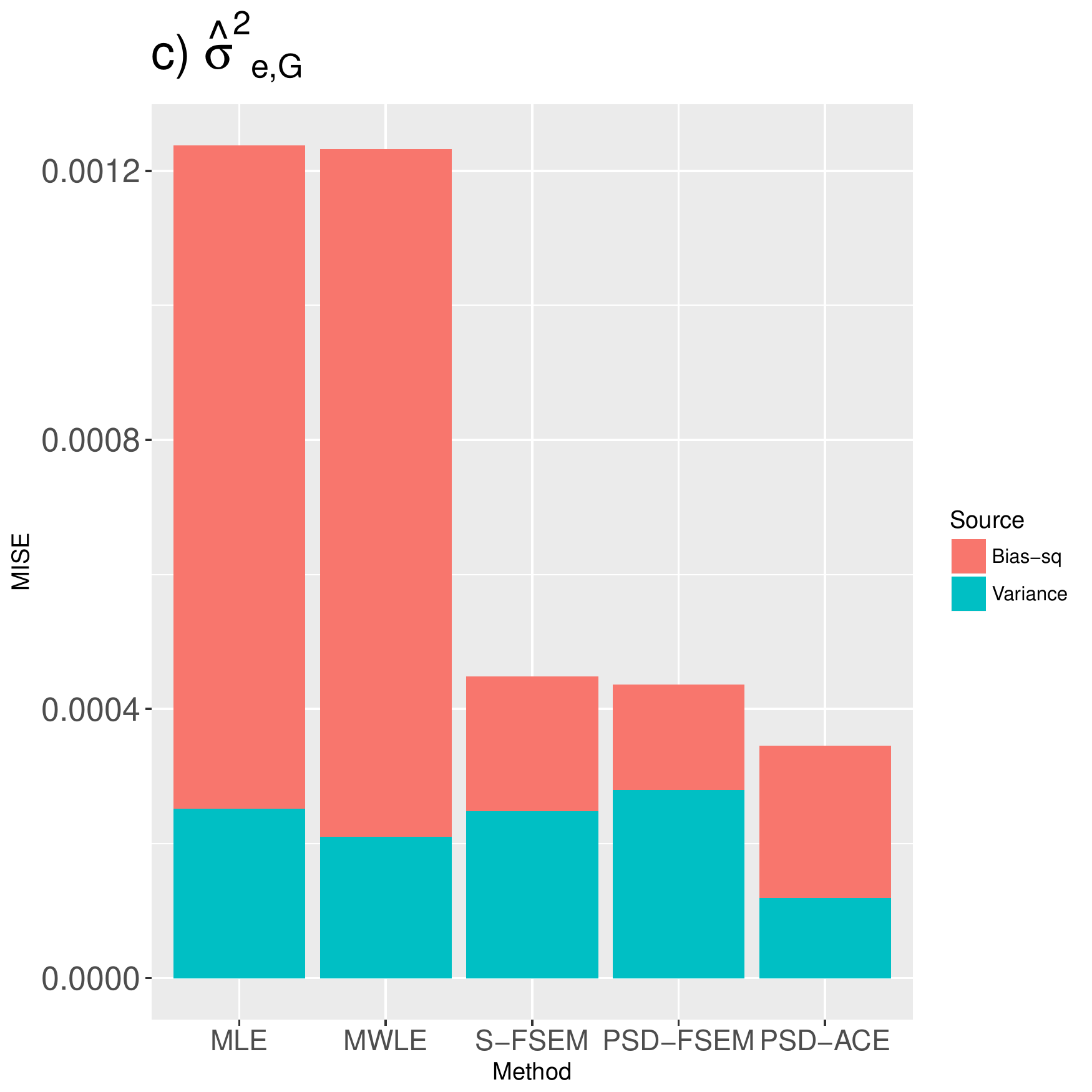}

\includegraphics[width=0.3\textwidth]{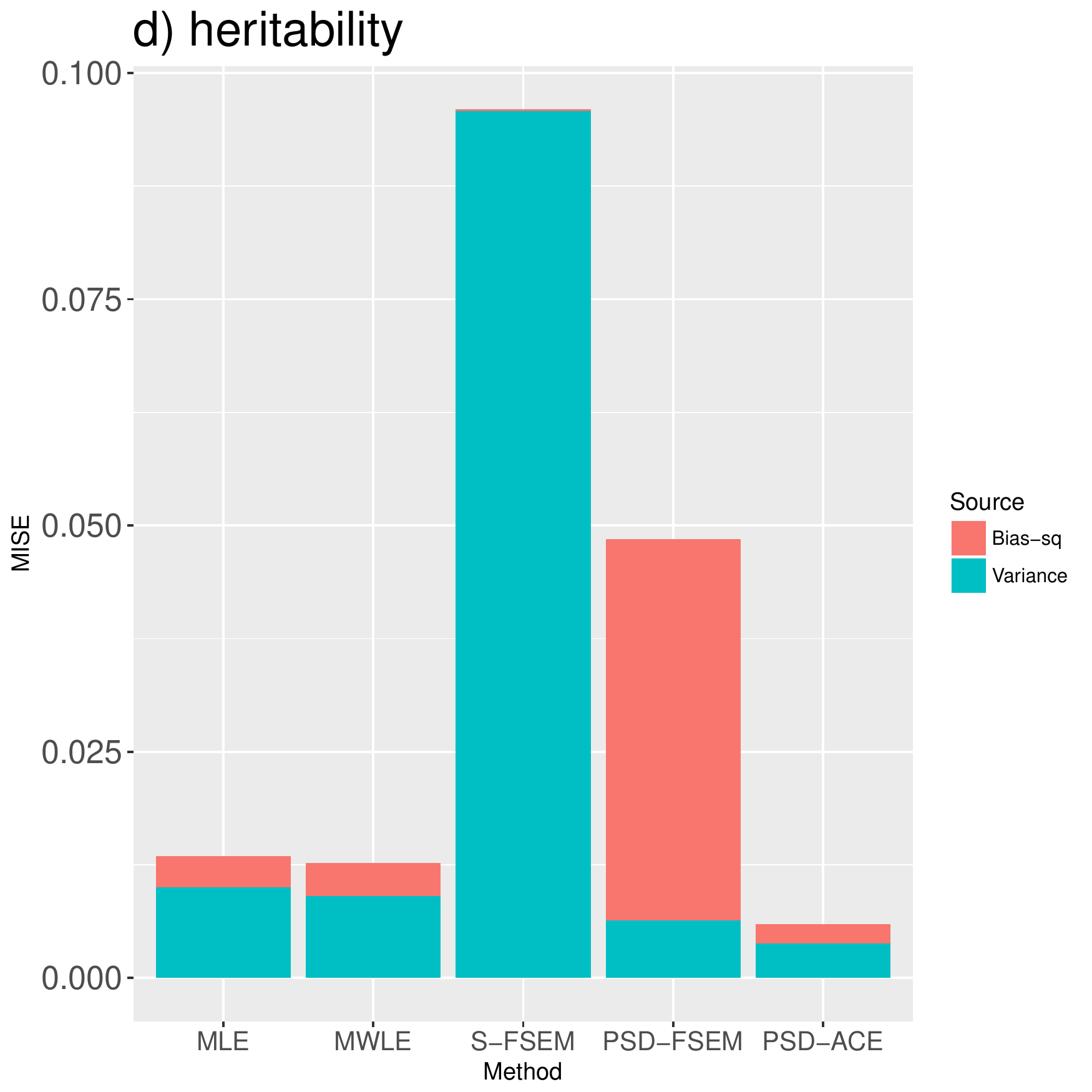}
\includegraphics[width=0.3\textwidth]{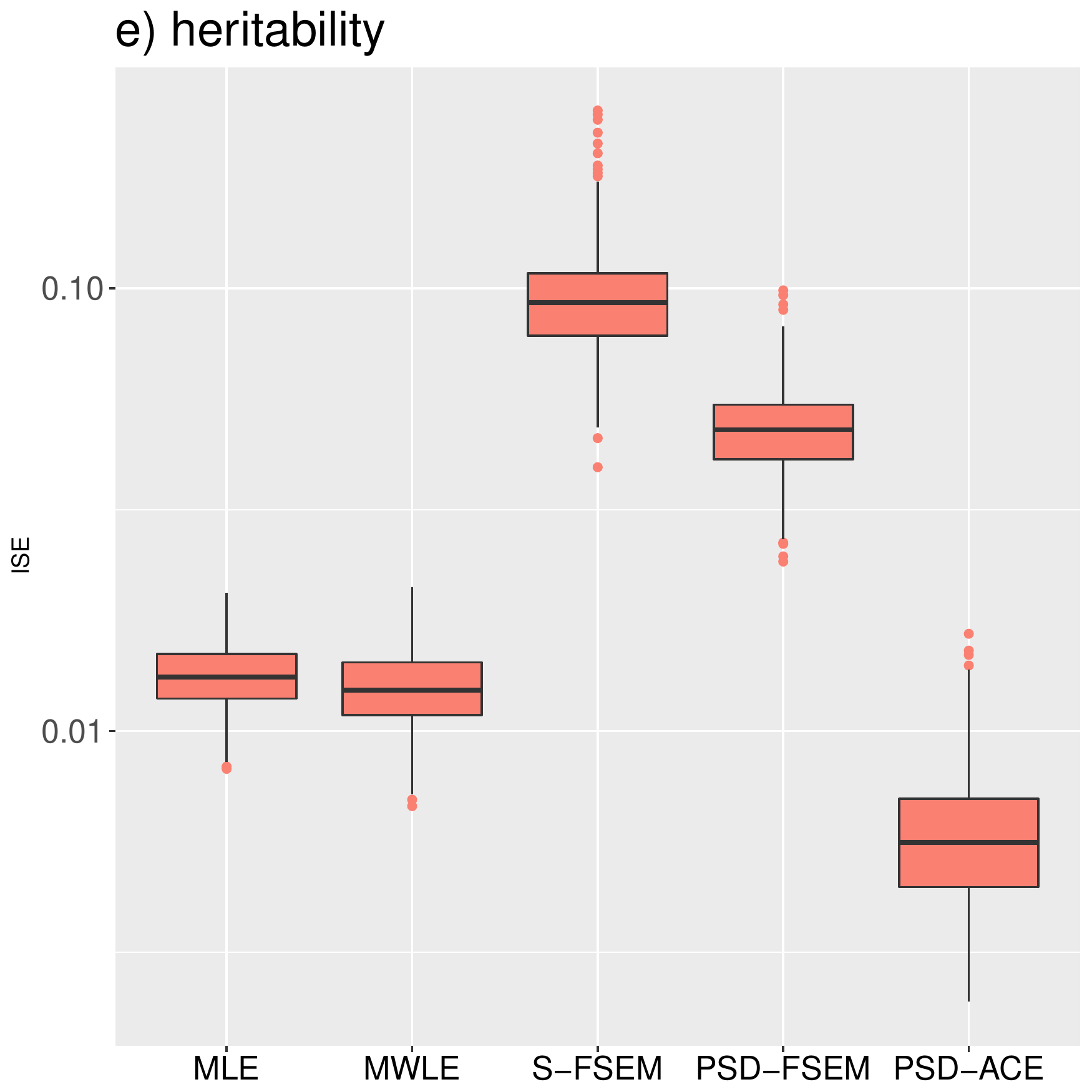}
\caption{MISE of $\hat{\sigma}_a^2(v)$, $\hat{\sigma}_c^2(v)$, $\hat{\sigma}_{e,G}^2$, and heritability across $V$=1,002 locations. (See Figures S.4-S.7 for a visualization of the bias across space.) In panel c), point-wise MLE and point-wise MWLE estimates of $\sigma_e^2(v) = \sigma_{e,G}^2(v)+\sigma_{e,L}^2(v)$ are presented because they do not separate $\sigma_{e,G}^2$, which leads to bias. Panel e) includes boxplots of the ISE from 1,000 simulations, where the y-axis is on the log10 scale due to large differences between methods. See Figure S.3 in the Appendix for a normalized version of this figure.}\label{fig:varianceetc}
\end{figure}

Note that the PSD-ACE improved upon the initial estimators in all simulations. Previous literature has noted that when imposing positive definite constraints with the parameterization $\bSigma = \bZ \bZ^T$, non-uniqueness can lead to issues with convergence when optima are close to each other \citep{pinheiro1996unconstrained}. Here, this does not appear to detrimentally affect parameter estimation, where we used an adaptive learning rate. We used a strict convergence criteria (ratio of the norm of the current gradient to the initial gradient less than 0.0001), and found that when the algorithm did not converge (in the sense that the size of the gradient failed to get smaller for vanishingly small learning rates), the estimate appeared to have adequately minimized the objective function (e.g., \ref{fig:boxplotcovfun}).   
In practice, we found that convergence can be improved by increasing the ranks in PSD-ACE, but this does not necessarily improve the estimate.

In summary, PSD-ACE has the lowest MISE albeit with more bias than the S-FSEM. Restricting the analysis to heritability, the bias in PSD-ACE was less than the MLE and MWLE, while also having the lowest overall MISE. Truncating covariance functions (PSD-FSEM) from symmetric estimates led to a large amount of bias in $\hat{\sigma}_a^2(v)$ and $\hat{\sigma}_c^2(v)$.

\section{Application to HCP}\label{sec:HCP}
We used the 32k (per hemisphere) preprocessed cortical thickness data from the 1200-subject HCP data release. We controlled for age, gender, and total intracranial volume (Appendix C of the Appendix). The HCP dataset contains cortical thickness for 1094 subjects, which includes twins, non-twin siblings, and unrelated individuals. In this sample, the age (mean$\pm$sd) was 28.8$\pm$3.7 years. There were 595 females versus 499 males with 75\% White, 15\% African-American, 6\% Asian/Native Hawaiian/Pacific Islander, and 4\% ``other''. Of these, 452 were genotyped, which revealed that 31 of 109 twins pairs that self-reported DZ twin status were in fact MZs. In contrast, 151 out of 151 genotyped twin pairs that self-reported MZ were in fact MZ. Thus, the set of subjects that self-reported MZ was 100\% accurate, while the set of subjects that self-reported DZ was only 72\% accurate. Consequently, we included all self-reported MZs but excluded self-reported DZs without genotype data. This resulted in 151 MZ and 78 DZ pairs. 

For S-FSEM and PSD-ACE, we included \emph{all} 1094 subjects, where non-twin siblings were treated as singletons as discussed in Section \ref{sec:CovarianceFunctions}. We excluded non-twin siblings from the MLEs and MWLEs, where the likelihoods assume independence or require the assumption that siblings can be treated in the same manner as DZ pairs (i.e., the common environmental variance effects, $c_{i}(v)$, are the same for siblings of different ages and twins). Similarly, we elected to include one randomly selected member of a family for the families with siblings and no twins, which resulted in 676 subjects in the MLE and MWLE analyses. 

We estimated heritabilities using the point-wise MLE, point-wise MWLE with 5-fold leave-one-family out CV, S-FSEM, and PSD-ACE. Overall, the selected bandwidths were notably small; for details, see Appendix C.4. An inspection of the scree plots of the eigenvalues clearly indicates the ranks of the covariance functions are determined by the number of twin pairs and individuals: for $\hatbSigma_a^{SW}$ and $\hatbSigma_c^{SW}$, the rank equals the number of twin families (229); for $\hatbSigma_{e,G}^{SW}$, the rank equals $N-n_1$ (943) (Figure S.10 of the Appendix). 

Step 6 is the most computationally intensive step. In the Appendix, we provide scripts with the option to partition the data to decrease memory overhead, where the number of partitions can be tuned to meet a user's memory limits. Here, we used the full data on a high memory server (required approximately 1.8 TB) and 24 CPUs. The PSD-ACE took approximately 26 hours to fit, while the S-FSEM took approximately 0.5 hours, and the PSD-FSEM took approximately 1 hour.

We assessed the sensitivity of the PSD-ACE to the selected ranks. We re-ran Step 6 with the ranks reduced by 10 for each covariance, and found negligible changes; see Appendix C.4.  This is expected because the PSD-ACE is initiated from the ordered positive eigenvalues/vectors of the symmetric estimates, such that excluding the smallest eigenvalues/vectors should have negligible impacts. We also assessed convergence of the PSD-ACE by running an additional 600 iterations and found negligible changes; see Appendix C.5. 

In general, heritability was higher near the central sulci and medial areas near the corpus callosum (Figure \ref{fig:h2_psdace}; also see the annotated PSD-ACE in Figure S.13 of the Appendix). The heritability was higher in PSD-ACE than MLE and MWLE, and there were many zeros in the MLE and MWLE estimates but not PSD-ACE. S-FSEM had some negative heritabilities due to negative estimates of $\sigma_a^2(v)$ and tended to have estimates that were higher than MLE and MWLE but lower than PSD-ACE. The mean $\pm$ sd across all vertices was $0.272 \pm 0.034$ for PSD-ACE and $0.164 \pm 0.305$ for S-FSEM, while MLE and MWLE were $0.088 \pm 0.093$ and $0.088 \pm 0.089$, respectively. Using $\hat{\bsigma}_{e,L}$ from Step 3 in the PSD-ACE estimation, we can also construct estimates of  measurement-error corrected heritability for MLE and MWLE: $0.108 \pm 0.115$ and $0.108 \pm 0.110$, respectively. Here, the measurement error only accounts for a small proportion of the differences.

\begin{figure}[ht]
\includegraphics[width=0.5\textwidth]{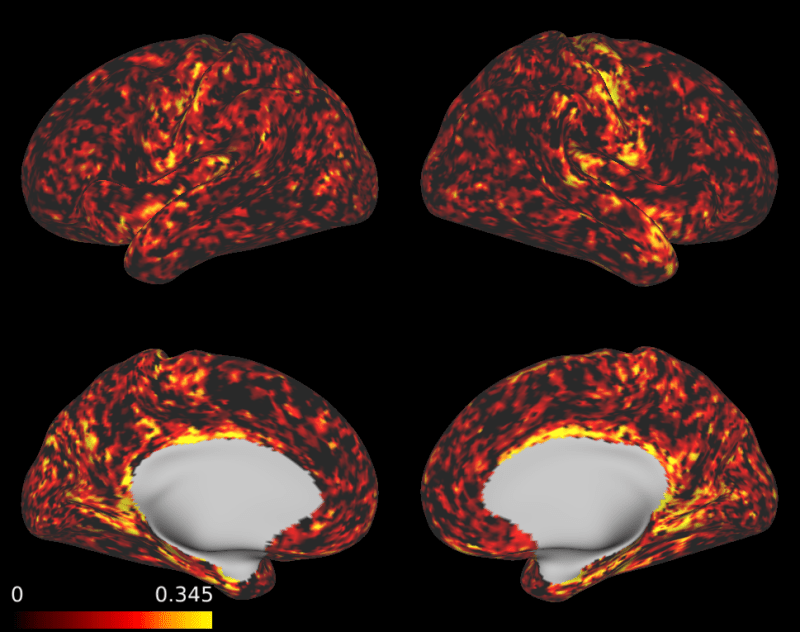}\includegraphics[width=0.5\textwidth]{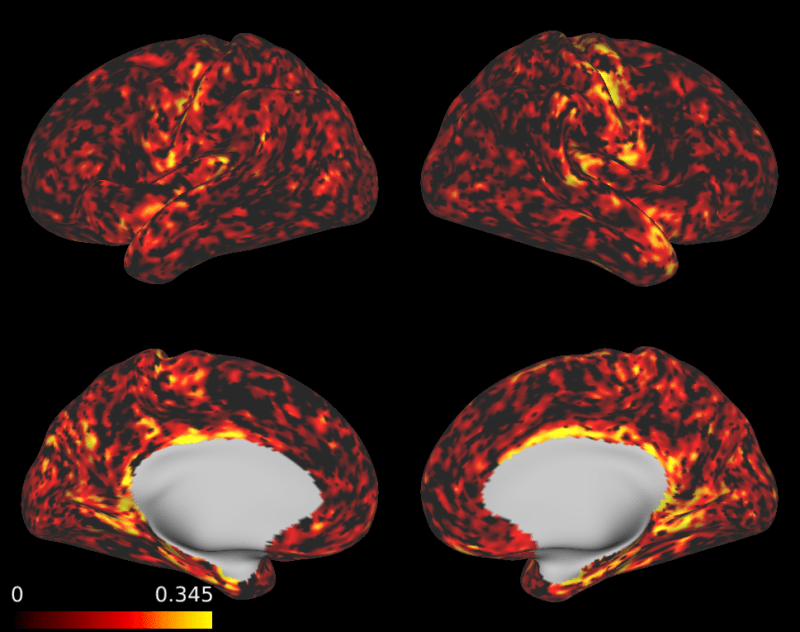}

\includegraphics[width=0.5\textwidth]{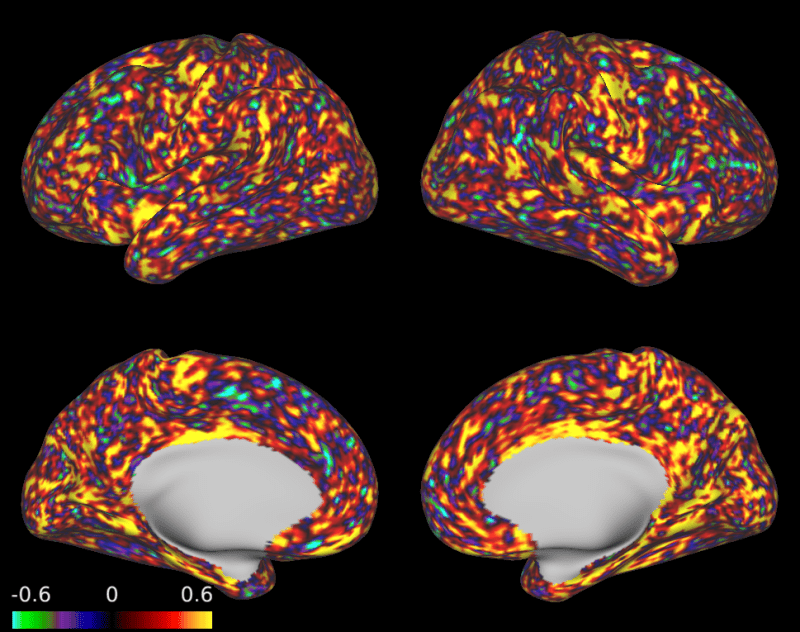}\includegraphics[width=0.5\textwidth]{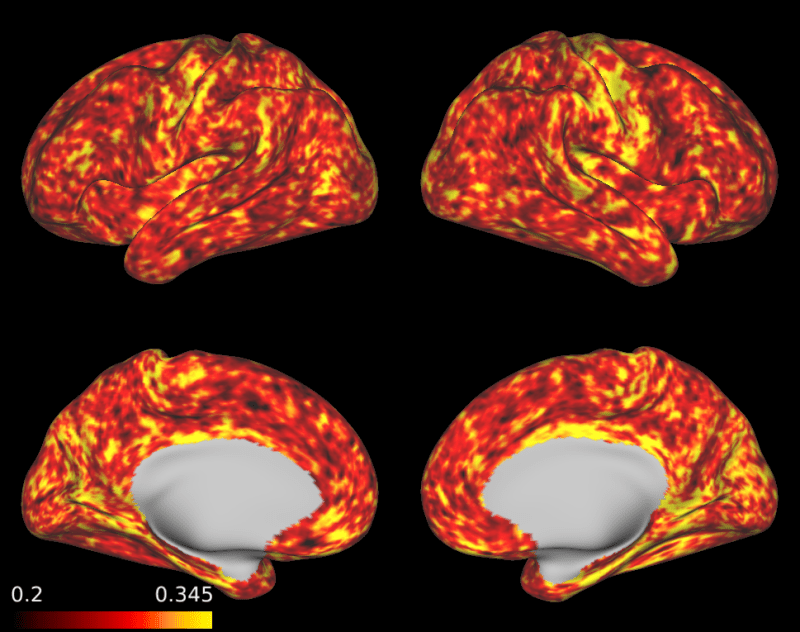} 
 
 \caption{Heritability estimated using the point-wise MLE (top left), point-wise MWLE (top right), PSD-ACE (bottom right), and S-FSEM (bottom left).}\label{fig:h2_psdace}
\end{figure}

The patterns of higher heritability in the central sulci and medial areas near the corpus callosum are similar to \cite{shen2016heritability}, which was based on young adults (22.8 years $\pm$ 2.3) and used 10 mm Laplace-Beltrami pre-smoothing. Their vertex-wise heritability estimates averaged across ROIs ranged from 0.026 to 0.523. Since our data were not pre-smoothed and the GCV-selected bandwidth was small, we suggest our Figure \ref{fig:h2_psdace} has a higher effective resolution than \cite{shen2016heritability} Figure 2.  There are some notable differences. \cite{shen2016heritability} found strong heritability in medial frontal areas in the left hemisphere (labeled as ``right'' using the radiological convention in their paper), whereas we did not find high heritability in these areas in either hemisphere. We found higher heritability in the parahippocampal gyrus and entorhinal cortex (ventral to the medial wall; see annotated Figure S.13 in the Appendix), whereas this pattern was less evident in \cite{shen2016heritability}. The entorhinal cortex is involved in  memory, and interestingly, thinning of the entorhinal cortex may interact with the APOE-$\epsilon$4 gene in Alzheimer's \citep{thompson2011design}. 

\begin{figure}[h!]
 \includegraphics[width=0.5\textwidth]{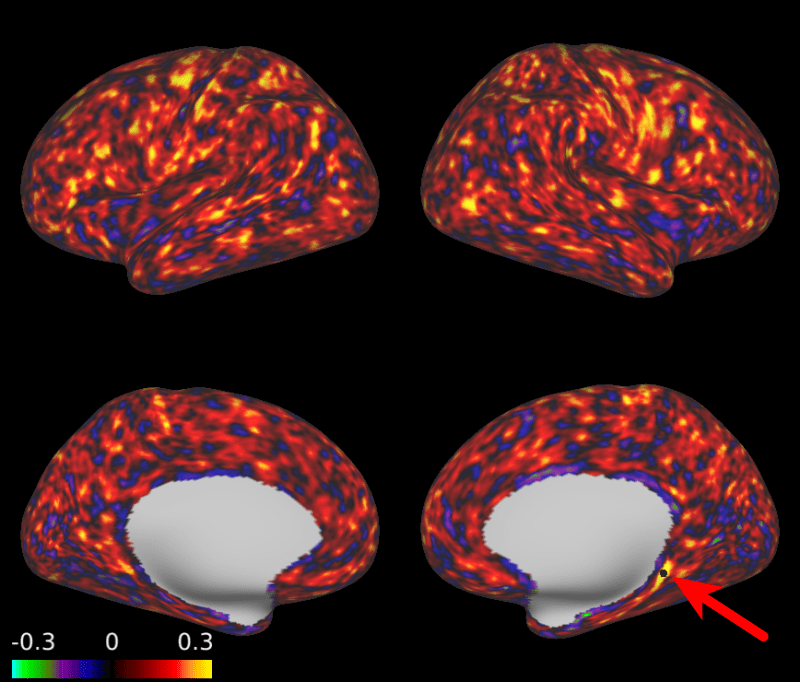}
 \includegraphics[width=0.5\textwidth]{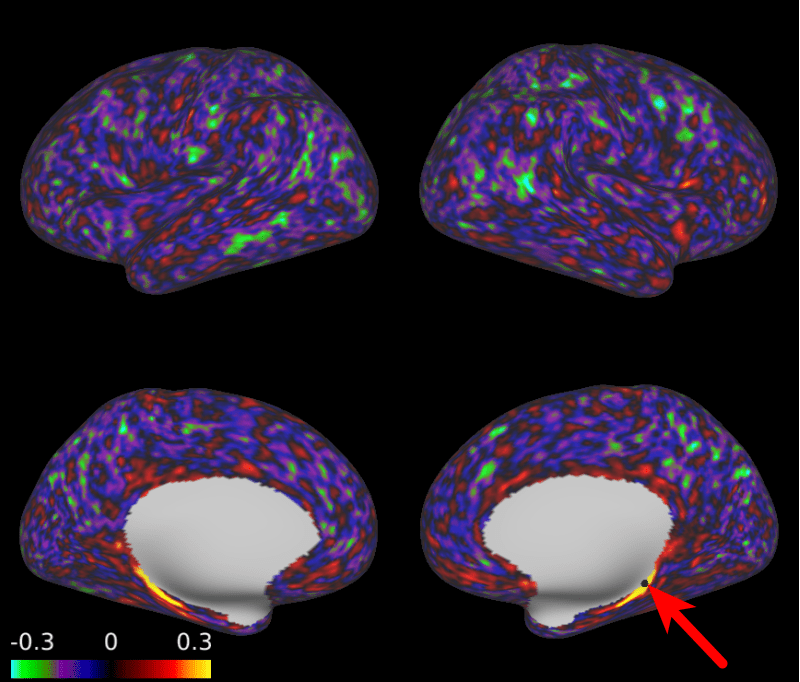}
 
 \includegraphics[width=0.5\textwidth]{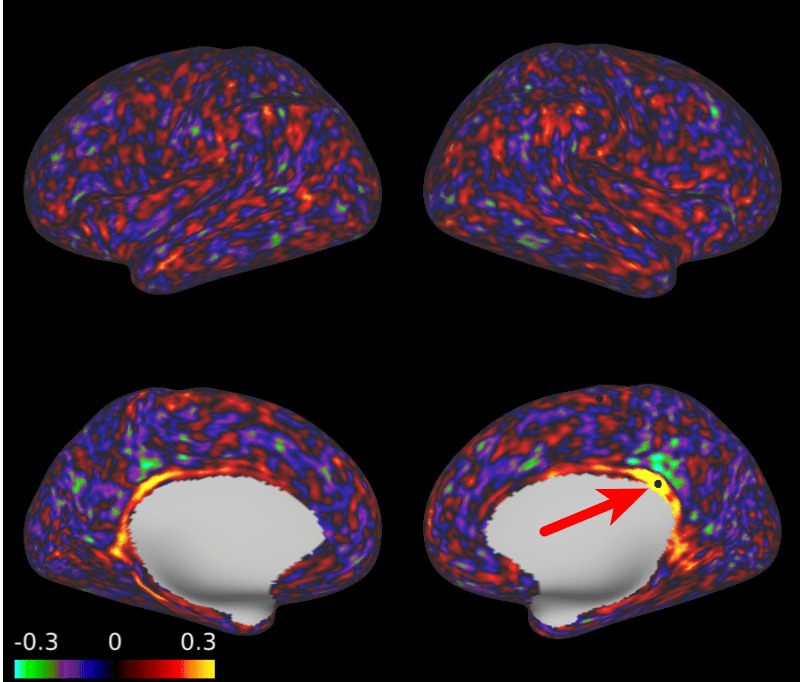}
 \includegraphics[width=0.5\textwidth]{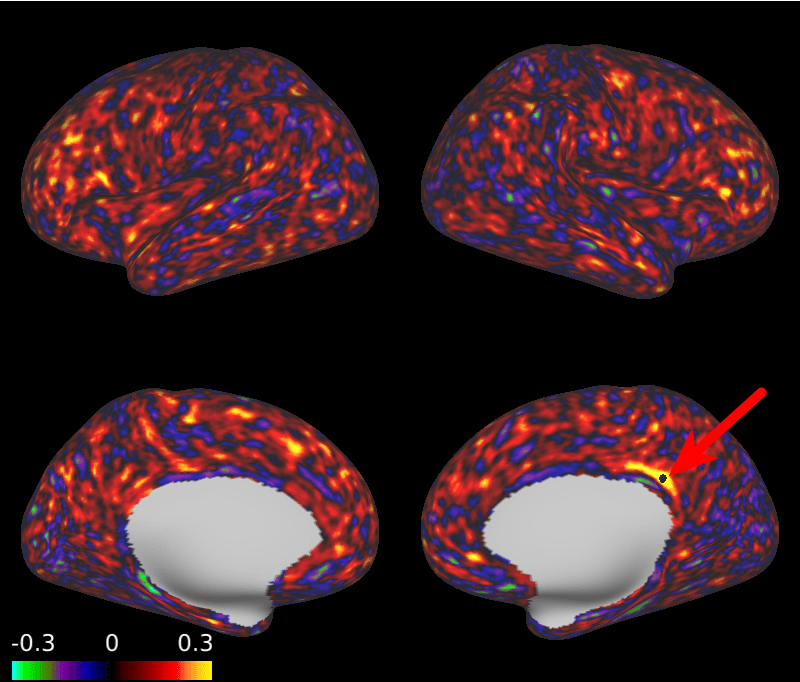}
 
 \caption{Genetic correlation function evaluated at selected seeds, as estimated using PSD-ACE. Surface vertex indices from right cortex of the fs\_LR template, clockwise from top left: 1577, 161, 180, and 239. Animation depicting hundreds of seeds is available in the Supplementary Materials.}\label{fig:exampleseeds}
\end{figure}

The genetic covariance function can be efficiently explored by creating an animation that progresses through different seeds (see \url{https://youtu.be/ew-Pq-Enf1I}).  We have created figures from selected seeds that highlight findings of scientific interest (Figure \ref{fig:exampleseeds}).  The covariances are normalized to define correlations.  First, the seed map for vertex 1577 in the right cortex (top left), located in the parahippocampal gyrus near the boundary with the isthmus cingulate, suggests that this location is a potential hub, as it is relatively highly correlated with many areas of the cerebral cortex. Thus the genetic control of cortical thickness for this location is related to the genetic control over cortical thickness across broad areas of the brain.  
In contrast, the seed map for vertex 161 (top right), also located in the parahippocampal gyrus, exhibits high correlations along a narrow ridge, with substantially lower overall correlations and much more localized genetic control. The parahippocampal gyrus is associated with memory encoding and retrieval, and our analysis indicates heterogeneous genetic patterns in this region. Next, the bottom row of Figure \ref{fig:exampleseeds} illustrates that the local patterns of correlation can differ greatly between nearby locations. Vertex 180 (bottom right) and vertex 239 (bottom left), both located in the isthmus cingulate, have very different local correlation patterns. 

\section{Discussion}
We present a method to estimate the genetic covariance function and heritability of brain traits from neuroimaging data. Our main contribution is the development of an estimation method that can handle a large number of grid points. In simulations, our approach improves estimates of heritability, genetic covariances, and environmental covariances. We apply our method to gain novel insights into the heritability and genetic correlations from the Human Connectome Project. Our approach reveals fine-scale differences in covariance patterns, identifying locations in which genetic control is correlated with large areas of the brain and locations where it is highly localized. This enables insight into the genetic underpinnings of structural networks of cortical thickness. 

Our analysis reveals tradeoffs between bias and variance, which is an important consideration for scientific interpretation. 
Here we discuss three bias-variance tradeoffs in estimates of heritability in neuroimaging: 1) bias from smoothing; 2) bias from constraining the covariance matrix to be PSD; and 3) bias from measurement error. 
With respect to smoothing, we use GCV for a data-based selection of the bias-variance tradeoff, which in general will reduce the mean squared error relative to using an a priori determined degree of smoothing. The impacts of smoothing on bias in twin studies is discussed in \cite{li2012twinmarm}. 
At higher resolutions, there is greater measurement error, which has historically motivated the use of a large amount of smoothing, e.g., a Gaussian kernel with full-width at half-maximum equal to 25-30 mm, or the use of brain traits averaged across a smaller number of regions.  
These approaches decrease the variance of estimators, which can increase power, but can also decrease the spatial precision, which can increase false positives.The small amount of smoothing selected via GCV in our estimates will generally result in more variable estimates than approaches using larger amounts of smoothing, but also preserves fine-scaled changes in correlation. Different approaches have different costs and benefits, and in this respect can complement one another.

A second form of bias arises when imposing PSD constraints, as revealed by large differences between the unconstrained, truncated, and constrained estimators. Truncating the covariance functions to positive eigenvalues/eigenvectors results in lower MISE but large bias, and this large bias motivated our development of the PSD-ACE estimator.  We view PSD-ACE as a compromise that results in dramatically lower variance relative to the symmetric estimators at the cost of some bias and dramatically lower bias relative to truncated estimators at additional computational cost. In practice, one approach is to estimate both PSD-ACE and S-FSEM to compare the estimators with better variance properties versus better bias properties. 

A third form of bias, measurement error, is common in heritability studies.  When repeated scans on the same subject are available,  \cite{ge2017heritability} proposed the use of linear mixed effects models with repeated measures, leading to large improvements in estimates of heritability. A functional ACE model for repeated measurements is an important avenue for future research. In the absence of repeated measurements, we utilize the assumption that the underlying genetic and environmental functions are smooth Gaussian processes, which allows the estimation of measurement error in a manner similar to the nugget effect in spatial statistics. In our simulations, likelihood-based approaches were \emph{more} biased than PSD-ACE in heritability estimates because they conflate measurement error and unique environmental variance.  

In this study, we have not addressed inference. In particular, many of the correlations may not be statistically significant, including the large changes in genetic correlation observed over short distances. For datasets with fewer locations, future research could develop permutation tests to calculate FWER-corrected p-values for genetic correlations. Note that for variance components, \cite{luoetal} proposed a test for the significance using MWLE. Another important avenue for future research is the extension of the PSD-ACE to more general pedigree models, which could allow the estimation of genetic components from large datasets like UK Biobank. This would also be useful in evaluating the replicability of cortical thickness heritability.

We applied our method to tens of thousands of points and produced a detailed atlas of the covariance in cortical thickness related to genetic factors. By determining the degree of smoothing from the data, our approach allows a more detailed spatial resolution. We used the same kernel and bandwidth for all locations across the cortical surface. 
However, the large changes in correlation patterns over small distances, e.g., the bottom row of Figure \ref{fig:exampleseeds}, together with the small bandwidth selected by GCV and 5-fold CV, suggest that future research could explore additional modeling approaches. Locally adaptive procedures have been developed for image smoothing, regression, and maximum weighted likelihood (e.g., \citealt{li2012twinmarm}). A recent method for functional PCA based on the Laplace-Beltrami operator for the cortical surface may better characterize local features than a fixed kernel \citep{lila2017smooth}. The most flexible approach would be to allow jump discontinuities, for example extending \cite{zhu2014spatially}. Developing these approaches to estimate multiple covariance functions in big neuroimaging twin studies is challenging. 


\section*{Supplementary Material}
\label{sec6}
For Matlab code, see \url{https://github.com/BenjaminRisk/SpatialACE}. The genetic correlation video is available at \url{https://youtu.be/ew-Pq-Enf1I}.


\section*{Funding}
This work was supported by the NSF grant DMS-1127914 to the Statistical and Applied Mathematical Science Institute. 
Dr. Zhu was supported by U.S. NIH grants R01 MH086633, R01 MH116527, MH092335, NSF grants SES-1357666 and DMS-1407655, the Cancer Prevention Research Institute of Texas, and the endowed Bao-Shan Jing Professorship in Diagnostic Imaging. This material was based upon work partially supported by the NSF grant DMS-1127914 to the Statistical and Applied Mathematical Science Institute (SAMSI). 

\section*{Acknowledgments}
We thank Sandeep Sarangi at the University of North Carolina Research Computing for his assistance with high-memory, high-performance computing. We thank Dr.\! Richard Smith in the Department of Statistics and Operations Research at UNC for valuable input while Dr.\! Risk conducted his post-doc at SAMSI and his continuing support. Data were provided by the Human Connectome Project, WU-Minn Consortium (Principal Investigators: David Van Essen and Kamil Ugurbil; 1U54MH091657) funded by the 16 NIH Institutes and Centers that support the NIH Blueprint for Neuroscience Research; and by the McDonnell Center for Systems Neuroscience at Washington University.


\appendix

\setcounter{figure}{0}
\renewcommand{\thefigure}{S.\arabic{figure}}  

\setcounter{equation}{0}
\renewcommand{\theequation}{S.\arabic{equation}}

\setcounter{table}{0}
\renewcommand{\thetable}{S.\arabic{table}}

\section{Variance and covariance estimators}

\subsection{Estimating measurement error}\label{sec:SigmaEm}
Estimation of $\sigma^2_{e,L}(v)$ is based on the difference between the total variance estimated from the smoothed MLEs and an estimate of the sum of the additive genetic, common environmental, and unique environmental variances. Let $\bSigma_G = \bSigma_a + \bSigma_c + \bSigma_{e,G}$. For now let us take $h$ as given. We solve the objective function 
\begin{align*}
 \cJ_n(v,v') &= \sum_{i=1}^n \sum_{j=1}^{m_i} \sum_{v_0  \ne v_0' } \left\{ \widehat{R}_{ij} (v_0) \widehat{R}_{ij}(v_0') - \bSigma_G(v,v') \right\}^2 k_{h}(v_0,v) k_{h}(v_0',v')
\end{align*}
where $\widehat{R}_{ij}(v_0)$ are the SMLE residuals. Then the solution can be expressed as 
\begin{align}\label{eq:SigmaG}
\hatbSigma_{G} = \left\{ \bK (\bS_0 - \diag \bS_0) \bK \right\} ./ (\bK \bJ \bK - \bK \odot \bK),
\end{align}
with $\bK_{ij} = k_h(v_i,v_j)$; $\bS_0$ is defined in \eqref{eq:S0S1S2}; ``$./$'' denotes element-wise division; $\bJ$ is the $V \times V$ matrix of ones; and $\odot$ is element-wise multiplication. Let $\hat{\bsigma}_{T}^2 = \hat{\bsigma}_{a \;SMLE}^2 + \hat{\bsigma}_{c \;SMLE}^2 + \hat{\bsigma}_{e\; SMLE}^2$. Then $\hat{\bsigma}^2_{e,L} = \hat{\bsigma}^2_T - \diag \hatbSigma_{G}$.

The bandwidth in \eqref{eq:SigmaG} can not be chosen using GCV because the trace is zero, so we chose $h$ using the matrix $\bS_1$ generated by the monozygotic twin pairs defined in \eqref{eq:S0S1S2}. Note here $\E \bS_1 \approx \bSigma_a + \bSigma_c$, which we assume is a sufficient a proxy for the smoothness of $\bSigma_{e,G}$. The error is calculated as $||\bS_1 - \tilde{\bK} \bS_1 \tilde{\bK} ||_F^2$ with $\tilde{\bK}$ described in Section 2.2. We found that GCV in this manner appeared to produce better results than 5-fold CV on family id with the estimator in \eqref{eq:SigmaG}, where the latter appeared to under-smooth. Additionally, GCV is computationally more tractable.

\subsection{FSEM Covariance Estimators}\label{sec:FSEMCovEstimators}
Here we present the original objective function used in covariance estimation from \cite{luoetal}. Recall the definitions of
$\widehat{U}_{i,j,v_0,v_0'}$ and $\widehat{U}_{i,v_0,v_0'}$ from (2.5) and (2.6) of the main manuscript. Define the objective function,
\begin{align}\label{eq:linearspaceFSEM}
 \cJ_n(v,v') &= \\
 \nonumber &\frac{1}{N} \sum_{i=1}^n \sum_{j=1}^{m_i} \sum_{v_0  \ne v_0' } \left\{ \widehat{U}_{ij} (v_0,v_0') - \bSigma_a(v,v') - \bSigma_c(v,v') - \bSigma_{e,G}(v,v') \right\}^2 k_{h}(v_0 - v) k_{h}(v_0' - v') \\
 \nonumber &+ \frac{1}{n_1} \sum_{i=1}^{n_1} \sum_{v_0 \ne v_0'}  \left\{ \widehat{U}_{i} (v_0,v_0') - \bSigma_a(v,v') - \bSigma_c(v,v') \right\}^2 k_{h}(v_0 - v) k_{h}(v_0' - v') \\
 \nonumber &+ \frac{1}{n_2} \sum_{i=1}^{n_2} \sum_{v_0 \ne v_0'}  \left\{ \widehat{U}_{i} (v_0,v_0') - 0.5 \bSigma_a(v,v') - \bSigma_c(v,v') \right\}^2 k_h(v_0 - v) k_h(v_0' - v').
\end{align}
The objective function is equivalent to local constant regression (e.g., (A.3) in \cite{yao2005functional}), i.e., kernel regression, and we call the solution S-FSEM, where the S denotes symmetric. The solution to this objective function has a closed form.
\begin{align}
 \label{eq:FSEMsigmaA} \widehat{\bSigma}^{S-\!FSEM}_a(v,v') &= 2 \left\{ \widehat{S}_{w1}(v,v') - \widehat{S}_{w2}(v,v') \right\} / \left\{ w^*(v,v') \right\} \\
  \label{eq:FSEMsigmaC} \widehat{\bSigma}^{S-\!FSEM}_c(v,v') &= \left\{ - \widehat{S}_{w1}(v,v') + 2 \widehat{S}_{w2}(v,v')  \right\} / \left\{w^*(v,v') \right\} \\
 \label{eq:FSEMsigmaEg} \widehat{\bSigma}^{S-\!FSEM}_{e,G}(v,v') &= \left\{\widehat{S}_{w0}(v,v') - \widehat{S}_{w1}(v,v') \right\} / \left\{w^*(v,v') \right\} 
 \end{align}
where $\widehat{S}_{w0}(v,v')$ is a function defined using all individuals,
\begin{align*}
 \widehat{S}_{w0}(v,v') = \frac{1}{N} \sum_{i=1}^n \sum_{j=1}^{m_i} \sum_{v_0 \ne v_0'}  \widehat{U}_{i,j,v_0,v_0'} k_h(v - v_0)k_h(v' - {v_0}');
\end{align*}
next, $\widehat{S}_{w1}(v,v')$ is a function defined using MZ twins,
\begin{align*}
 \widehat{S}_{w1}(v,v') = \frac{1}{n_1} \sum_{i=1}^{n_1} \sum_{v_0 \ne v_0'}  \widehat{U}_{i,v_0,v_0'} k_h(v - v_0) k_h( v' - {v_0}');
\end{align*}
and $\widehat{S}_{w2}(v,v')$ is a function defined using the DZ twins,
\begin{align*}
 \widehat{S}_{w2}(v,v') = \frac{1}{n_2} \sum_{i=n_1+1}^{n_2} \sum_{v_0 \ne v_0'}  \widehat{U}_{i,v_0,v_0'} k_h(v - v_0) k_h( v' - {v_0}'),
\end{align*}
and
$$
w^*(v,v') = \sum_{v_0 \ne v_0' } k_h(v - v_0) k_h( v' - {v_0}').
$$
In our application, we use geodesic distance in lieu of $v - v_0$. 

\subsection{Sandwich Estimators}\label{sec:SW}
Here we use kernel regression wherein the objective function differs from \cite{luoetal} in that we include $v_0 = v_0'$, where we subtract the estimates of the measurement error variance from $R_{ij}^2(v_0)$, as in (2.5) of the main manuscript. The solution is similar to \eqref{eq:FSEMsigmaA} and \eqref{eq:FSEMsigmaC} except that the terms include $v_0 = v_0'$ and the normalizing constant is replaced by $w(v,v') = \sum_{v_0,v_0'} k_h(v, v_0) k_h(v', v_{0}')$. 

If we restrict ourselves to  the covariance matrices corresponding to the $V$ locations, then the previous estimators can be written in a simple form that is also useful for computations. Define
\begin{align}
\bS_{0} = {N}^{-1} \left( \bR^T \bR  \right) - \hatbSigma_{e,L}; \label{eq:S0S1S2} \\
\nonumber \bS_{1} = (2 n_1)^{-1} \left( \bR_{11}^T \bR_{12} + \bR_{12}^T \bR_{11} \right);\\
\nonumber \bS_{2}  = (2 n_2)^{-1} \left( \bR_{21}^T \bR_{22} + \bR_{22}^T \bR_{21} \right). 
\end{align}
Then let $\bS_a = 2\bS_1 - 2 \bS_2$, $\bS_c = 2 \bS_2 - \bS_1$, and $\bS_{e,G} = \bS_0 - \bS_1$. The estimators are $\hatbSigma_a^{S-\!SW}\! = \tilde{\bK}_h  \bS_a \tilde{\bK}_h^T$, $\hatbSigma_c^{S-\!SW}\! = \tilde{\bK}_h \bS_c \tilde{\bK}_h^T$, and $\hatbSigma_{e,G}^{S-\!SW}\! = \tilde{\bK}_h \bS_{e,G} \tilde{\bK}_h^T$, where S-SW is an abbreviation for symmetric sandwich. We define the PSD-SW by calculating the EVD and truncating to the positive eigenvalue/eigenvector pairs. 

GCV is based on the observation that 
\begin{align}
\label{eq:GCVSW} \textrm{vec} \{\tilde{\bK}_h \bS_a \tilde{\bK}_h \} = (\tilde{\bK}_h \otimes \tilde{\bK}_h) \textrm{vec} \{\bS_a\}.
 \end{align}
Note that here we apply GCV spatially, which approximates leave-one-location out CV. 

\subsection{Analytic gradients of the PSD-ACE}\label{sec:Gradients}

 Let $\bK$ be the $V \times V$ matrix with entries $\bK_{v,v_0} = k_h(v,v_0)$ (not divided by $w(v)$). Let $\hatbSigma_{e,L}$ be the diagonal matrix of smoothed estimates of the measurement error; let $\bR$ denote the $N \times V$ matrix of residuals for all individuals and all locations; let $\bR_{11}$ denote the $n_1 \times V$ matrix of residuals for the first individual in each MZ pair; let $\bR_{12}$ denote the second individual in each MZ pair;  let $\bR_{21}$ denote the $n_2 \times V$ matrix for the first individual in each DZ pair; and let $\bR_{22}$ denote the second individual in each DZ pair. 
 Then we can derive the partial derivatives of (2.7) of the main manuscript for the elements corresponding to $\bZ_a$:
\begin{align}\label{eq:gradientSigmaA}
 \nabla_a \cJ^{PSD} &= -2\left\{ \bK \bS_a^* \bK - \left( \frac{9}{2} \bZ_a \bZ_a^T + 5 \bZ_c \bZ_c^T  + 2 \bZ_{e,G} \bZ_{e,G}^T\right) \odot \left(\bK \bone_V \bone_V^T \bK\right) \right\} \bZ_a,
\end{align}
where $\odot$ is element-wise multiplication and $\bS_a^* = 2 \bS_0 + 2 \bS_1 + \bS_2$. With respect to $\bZ_c$, we have 
\begin{align}\label{eq:gradientSigmaC}
 \nabla_c \cJ^{PSD} &= -2\left\{ \bK \bS_c^* \bK - \left( 5 \bZ_a \bZ_a^T + 6 \bZ_c \bZ_c^T  + 2 \bZ_{e,G} \bZ_{e,G}^T \right) \odot \left(\bK \bone_V \bone_V^T \bK\right) \right\} \bZ_c,
 \end{align}
 where $\bS_c^* = 2 \bS_0 + 2 \bS_1 + 2 \bS_2$. 
With respect to $\bZ_{e, G}$, we have  
 \begin{align}\label{eq:gradientSigmaEg}
 \nabla_{e,G} \cJ^{PSD} &= -4\left\{ \bK \bS_0 \bK - \left( \bZ_a \bZ_a^T + \bZ_c \bZ_c^T + \bZ_{e,G} \bZ_{e,G}^T \right) \odot \left(\bK \bone_V \bone_V^T \bK\right) \right\} \bZ_{e,G},
 \end{align}
which are used in Algorithm 1 of the main manuscript.

\subsection{Estimating covariance functions from covariance matrices}\label{sec:CovarianceFunctionsAppendix}
In this section, we consider the problem of estimating the covariance functions, which have a continuous domain, from the covariance matrices, defined for the $\approx$60,000 locations. Towards this, we want to estimate an unsmoothed basis for each covariance function based on  $\tilde{\bK}^{-1} \hatbZ_a$, $\tilde{\bK}^{-1} \hatbZ_c$, and $\tilde{\bK}^{-1} \hatbZ_{e,G}$. However, $\tilde{\bK}$ is typically poorly conditioned. Using the inverse for poorly conditioned $\tilde{\bK}$ can lead to ringing artifacts when evaluating the function for some $v \notin \cV_0$.  Instead, we use a robust inverse  as follows. We take the eigenvalue decomposition of $\bK = \bU \bLambda \bU^T$  and define
$\bK^{-} = \bU_Q \bLambda_Q^{-1} \bU_Q^T$,  where   $\bLambda_Q$ denote the diagonal matrix of $Q$ eigenvalues meeting this criteria
 truncated to $\lambda_q > 0.0001$. Then, 
we set $\hatbW_a = \bK^{-} \hatbZ_a$, $\hatbW_c = \bK^{-} \hatbZ_c$, and $\hatbW_{e,G} = \bK^{-} \hatbZ_{e,G}$ 
and estimate the covariance functions for any $v,v' \in \cV$ as follows: 
\begin{align}
\label{eq:conpsda} \hatbSigma_a(v,v')^{PSD-\!ACE} &= \bk_h(v)^T \hatbW_a \hatbW_a^T \bk_h(v'), \\
\label{eq:conpsdc} \hatbSigma_c(v,v')^{PSD-\!ACE} &= \bk_h(v)^T \hatbW_c \hatbW_c^T \bk_h(v'), \\
\label{eq:conpsdeg} \hatbSigma_{e,G}(v,v')^{PSD-\!ACE} &= \bk_h(v)^T \hatbW_{e,G} \hatbW_{e,G}^T \bk_h(v'), 
\end{align}
where 
$\bk_h(v) = [k_h(v,v_1)/w(v),\dots,k_h(v,v_V)/w(v)]^T.
$ 
As shown in  Proposition \ref{prop:1}, these estimators are positive semidefinite functions. Moreover,  this additional step is only necessary for evaluating the covariance function between unobserved locations or for a data partitioning approach. 

\begin{prop}\label{prop:1}
Consider a domain $\cV$ and locations $\{v_1,\dots,v_V\} \in \cV$. Let $k(v,v')$ be a kernel and for any $v \in \cV$, and define $\bk(v) = [k(v,v_1)/w(v),\dots,k(v,v_V)/w(v)]^T$. Let $\bW \in \tR^{V \times d}$ for any positive integer $d$. Define the bivariate function $f(v,v')$ on $\cV \times \cV$:
\begin{align}\label{eq:funpsd}
 f(v,v') = \bk(v)^T \bW \bW^T \bk(v').
\end{align}
For any $[v_1^*,\dots,v_m^*] \in \cV^m$, define the $m \times m$ matrix $\bF$ with elements $F_{ij} = f(v_i^*,v_j^*)$. Let $\ba \in \tR^m$. Then 
\begin{align}\label{eq:definepsd}
 \ba^T \bF \ba \ge 0.
\end{align}
\end{prop}

Note that \eqref{eq:definepsd} is the definition of a positive semidefinite function $f(v,v')$, e.g., p. 46, (2.29) of \cite{hsing2015theoretical}. 

\begin{proof}
Let $\bv^* \in \cV^m$ be given. Let $\bK_*$ be the $V \times m$ matrix with columns $\bk(v_i^*)$. Then for any $\ba \in \tR^m$, we have
 \begin{align*}
  \ba^T \bF \ba &= \ba^T \bK_*^T \bW \bW^T \bK_* \ba \\
  &= \bd^T \bd \\
  &\ge 0
 \end{align*}
 for $\bd = \bW^T \bK_* \ba$. 
\end{proof}

\subsection{MWLE}\label{sec:MWLE}
Throughout this study, we use the biweight kernel:
\begin{align}\label{eq:biweight}
 k_h(v,v_0) = \frac{15}{16h}\left\{ 1 - (d(v,v_0)/h)^2 \right\}^2 \iv_{d(v,v_0)\le h}.
\end{align}
Next, we summarize the maximum weighted likelihood estimate (MWLE) proposed in \cite{luoetal}. Suppose we have obtained $\hat{\bbeta}_{v_0}$ for all $v_0 \in \cV_0$, e.g., point-wise MLE or OLS estimates. Let $\widehat{\bR}(v_0)$ be the fixed effect residuals at location $v_0$ for all subjects. 
Let $d(v, v_0)$ denote the geodesic between two vertices. Let $\bsigma^2(v) = \{\sigma_a^2(v), \sigma_c^2(v), \sigma_e^2(v)\}$. They define a weighted log-likelihood
\begin{align}\label{eq:mwle} 
  \ell_{wtd} ({\bm \sigma}^2(v) | \hatbR) &= \sum_{v_0 \in \cV_0} k_{h}(v, v_0) \ell (\bsigma^2({v}) | \hatbR(v_0)).
\end{align}
Five-fold leave-family-out cross-validation is used to select $h$ in which the residuals are treated as data.

\section{Additional simulation results}

Detailed results containing the additional model estimation methods (S-SW, PSD-SW, PSD-ACE-O) are included in Table \ref{table:msecovfun}. We also present normalized variances of ISE, defined as $\textrm{Normalized \;ISE}^{(t)} =  \sum_{v=1}^V \sum_{v'=1}^V \left\{ \hatbSigma_a^{(t)}(v,v') - \bSigma_a(v,v') \right\}^2/ \sum_{v=1}^V \sum_{v'=1}^V \left\{\bSigma_a(v,v') \right\}^2$ and similarly define normalized MISE. 

\begin{table}[ht]                                            
\centering                                               
\begin{tabular}{|c | r r r| r r r| r r r|}      
\hline      
 Model & \multicolumn{3}{c|}{ $\hatbSigma_a(v,v')$} & \multicolumn{3}{c|}{$\hatbSigma_c(v,v')$}  & \multicolumn{3}{c|}{ $\hatbSigma_{e,G}(v,v')$} \\
 \hline
 & Bias$^2$ & Var & MISE  & Bias$^2$ & Var & MISE & Bias$^2$ & Var & MISE \\
\hline
\hline                                                   
S-FSEM & 1.34 & 1080.05 & 1081.39 & 0.92 & 671.40 & 672.32 & 9.14 & 155.48 & 164.62 \\     
\hline                                                                                     
PSD-FSEM & 160.35 & 403.67 & 564.02 & 102.79 & 238.90 & 341.69 & 10.16 & 149.47 & 159.63 \\
\hline                                                                                     
S-SW & 1.34 & 1080.54 & 1081.89 & 0.92 & 671.71 & 672.62 & 9.13 & 155.52 & 164.65 \\       
\hline                                                                                     
PSD-SW & 160.39 & 404.10 & 564.49 & 102.81 & 239.16 & 341.97 & 10.13 & 149.50 & 159.63 \\  
\hline                                                                                     
PSD-ACE-O & 10.95 & 44.52 & 55.47 & 9.14 & 26.87 & 36.01 & 36.57 & 62.49 & 99.06 \\        
\hline                                                                                     
PSD-ACE & 10.80 & 38.60 & 49.39 & 8.29 & 22.97 & 31.26 & 38.93 & 60.68 & 99.60 \\          
\hline                                                   
\end{tabular}                                            
\caption{Bias, variance, and MISE of the covariance functions multiplied by $V^2$, where $V=1002$.}                                 
\label{table:msecovfun}                               
\end{table}  

\begin{figure}[ht]
 \includegraphics[width=0.3\textwidth]{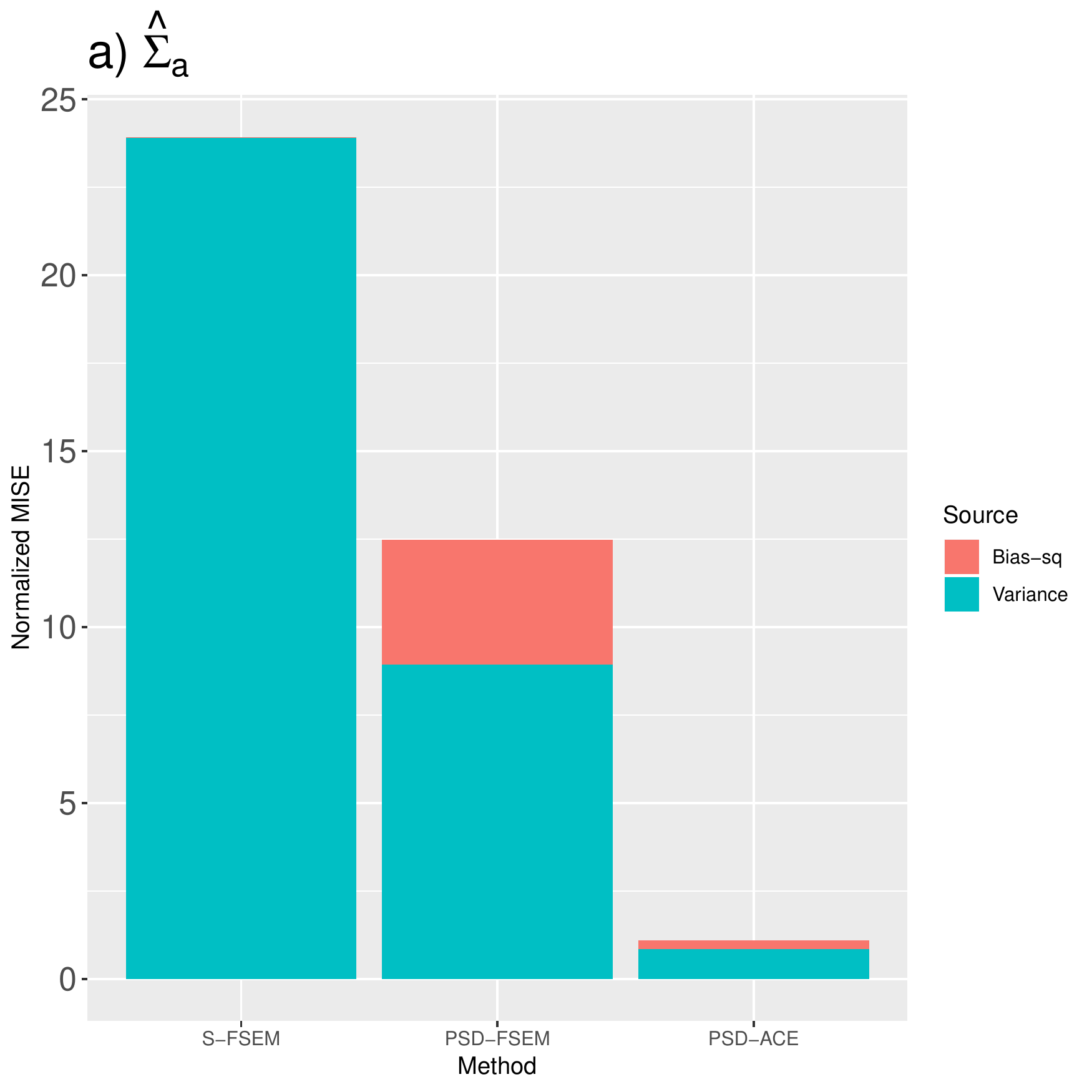}
 \includegraphics[width=0.3\textwidth]{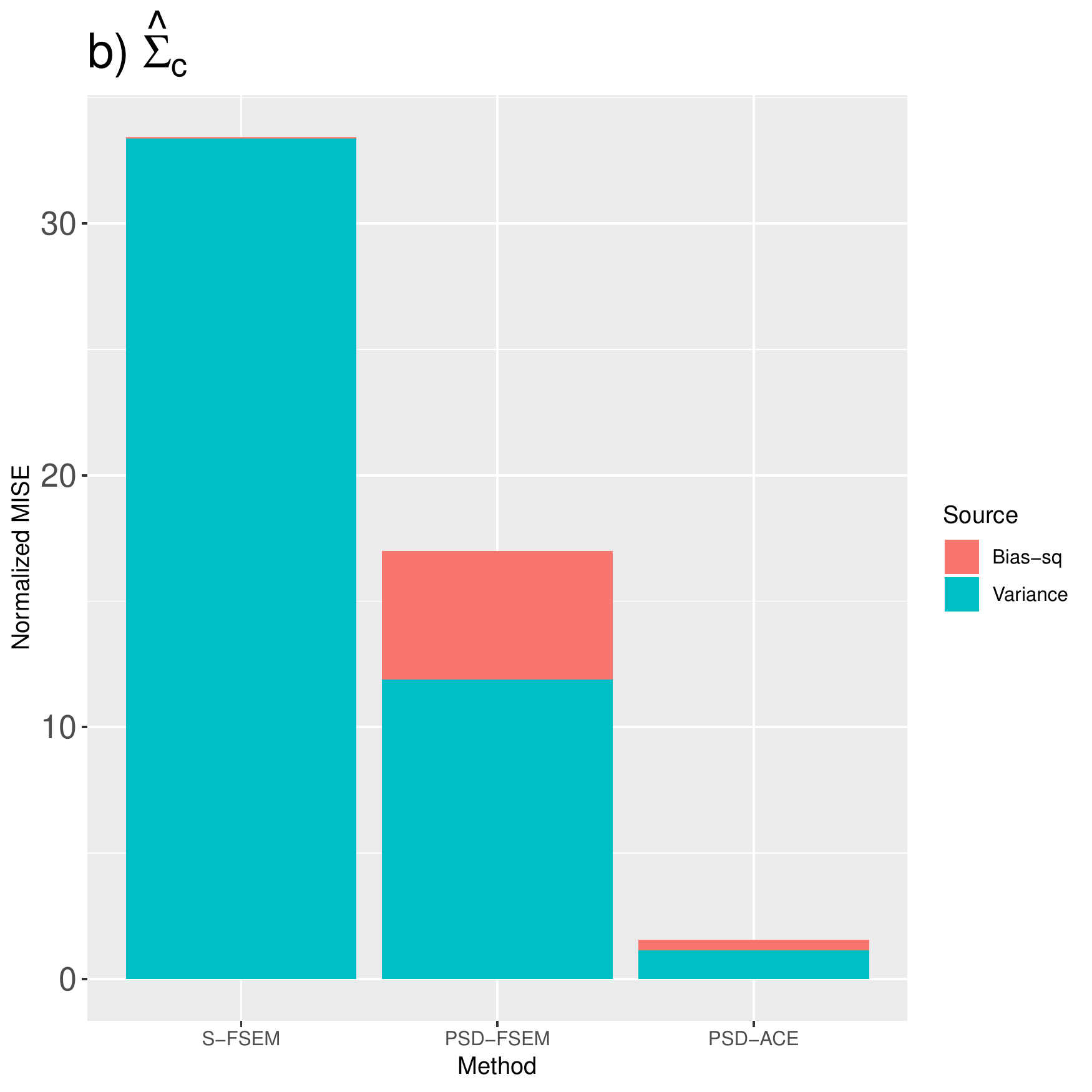} 
 \includegraphics[width=0.3\textwidth]{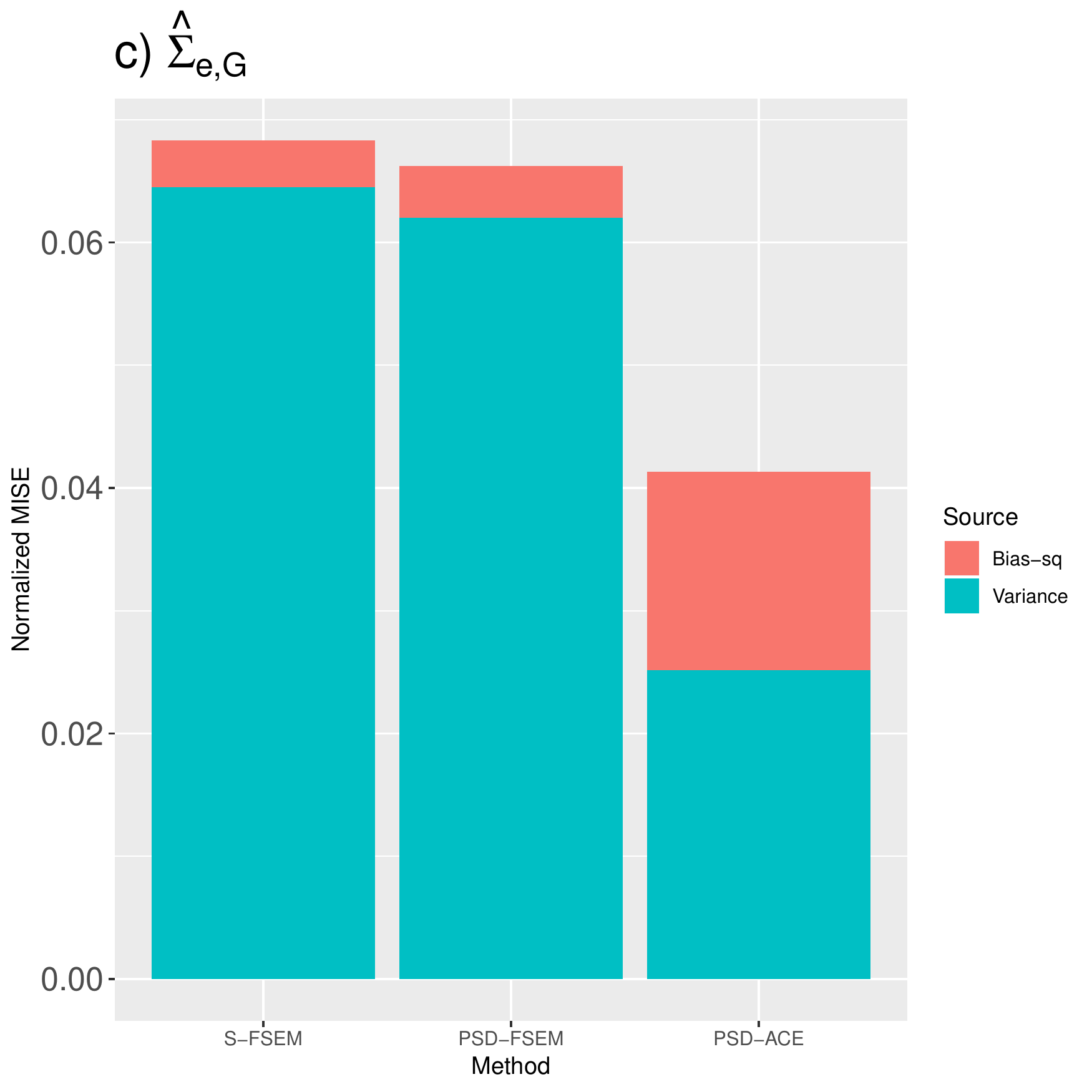} 
 \includegraphics[width=0.3\textwidth]{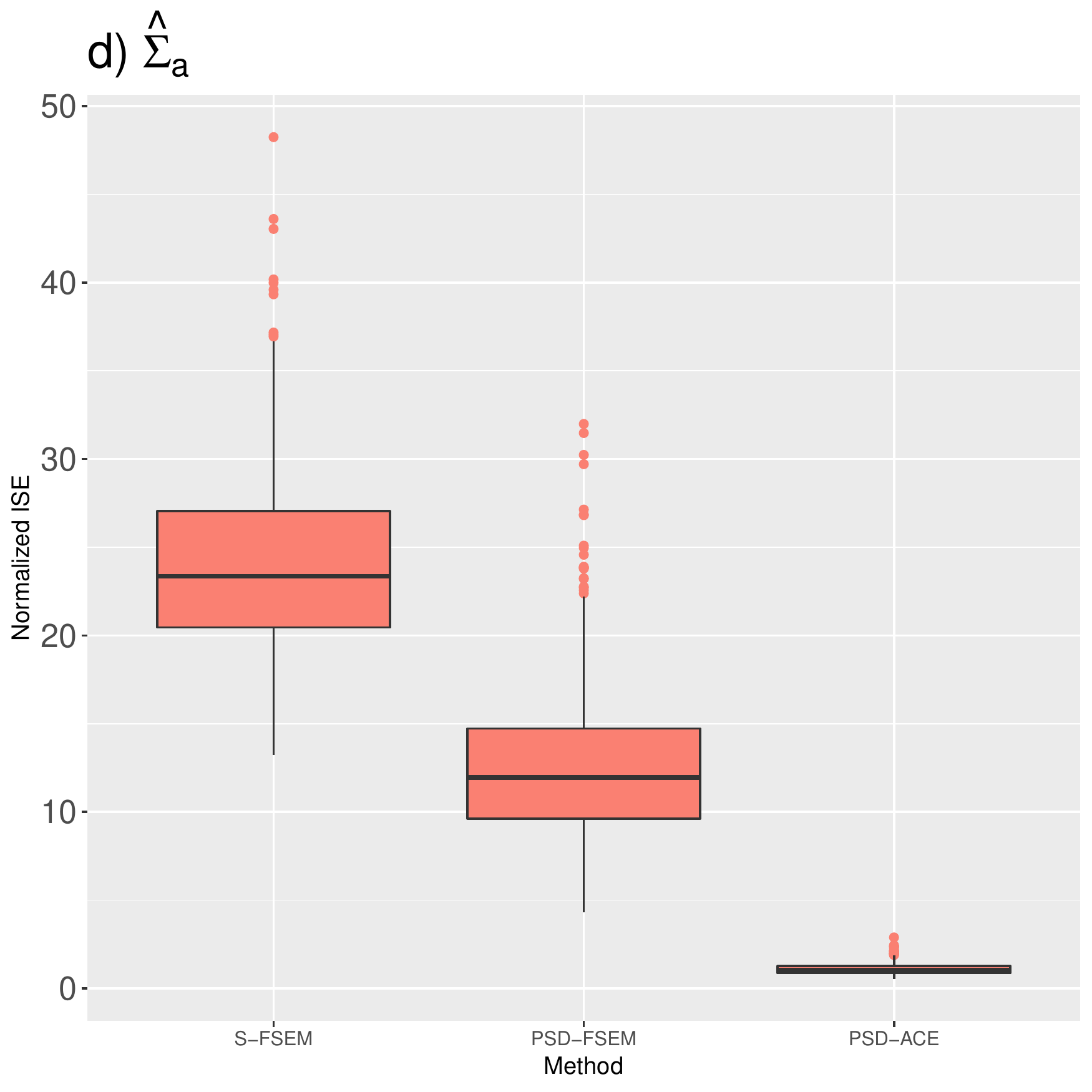}
 \includegraphics[width=0.3\textwidth]{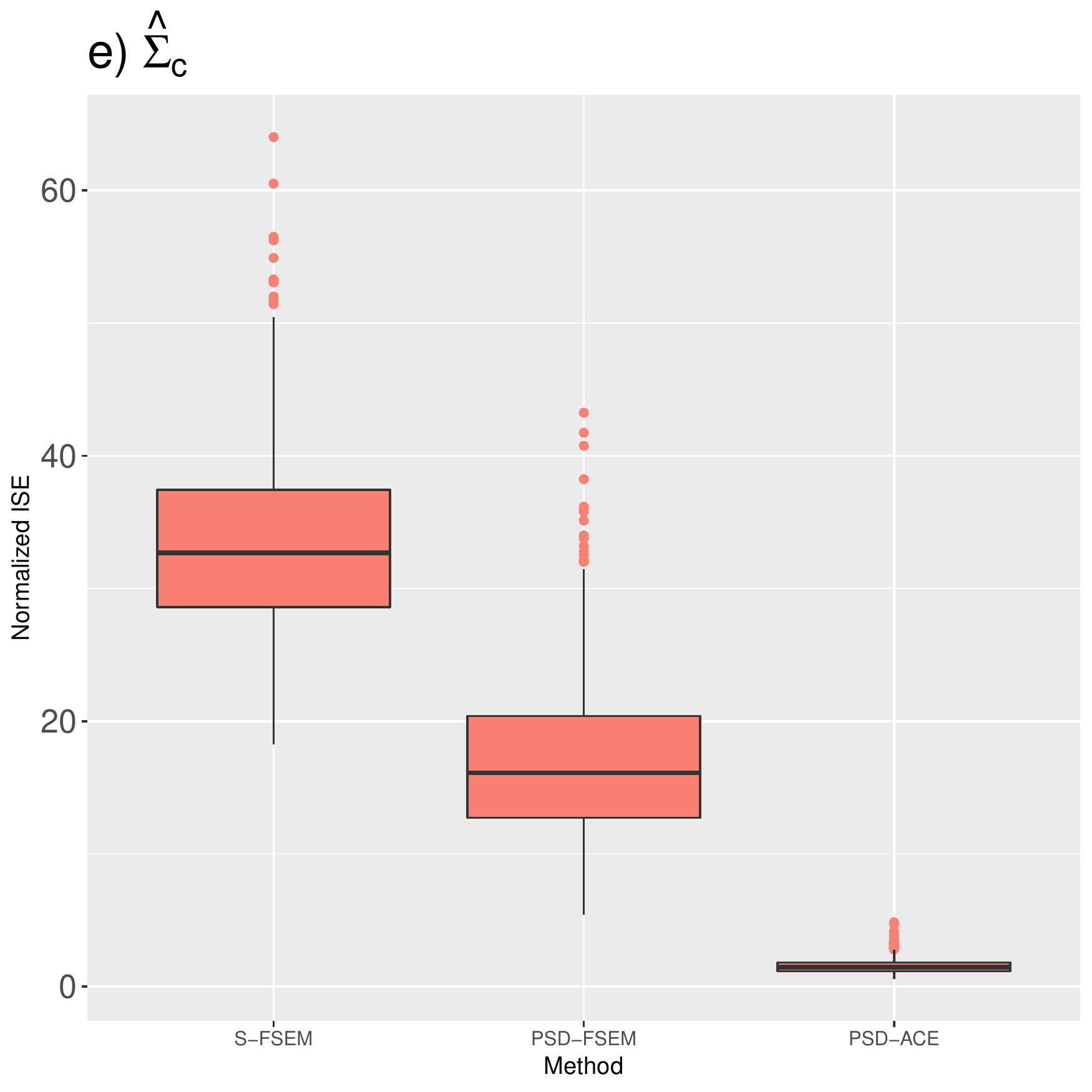}
 \includegraphics[width=0.3\textwidth]{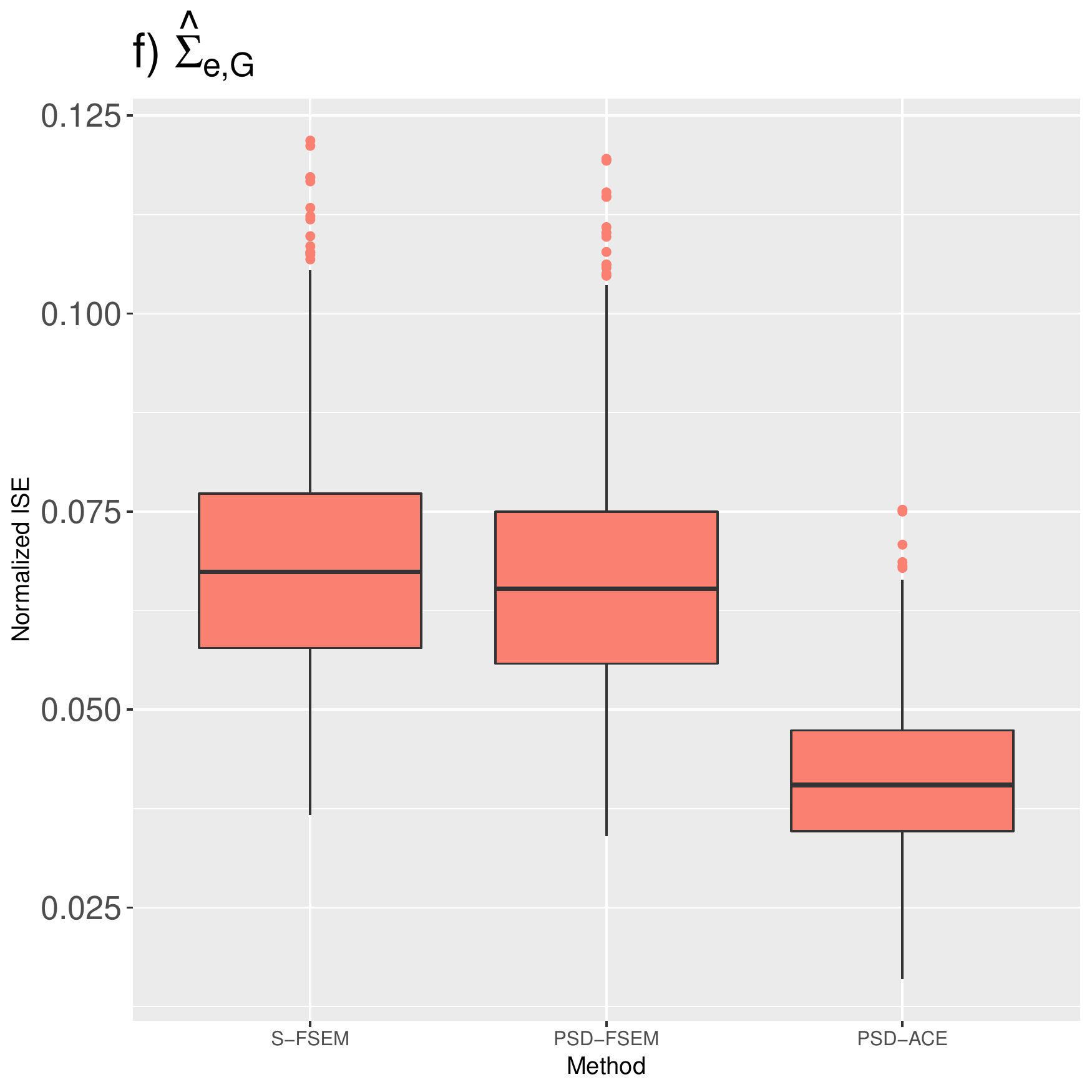}
\caption{Normalized MISE of covariance functions for the S-FSEM, PSD-FSEM (truncated eigenvalues of S-FSEM), and PSD-ACE. Panels d, e, and f depict boxplots of the normalized ISE from 1,000 simulations.}\label{fig:boxplotcovfun}
\end{figure}

\begin{figure}
\includegraphics[width=\textwidth]{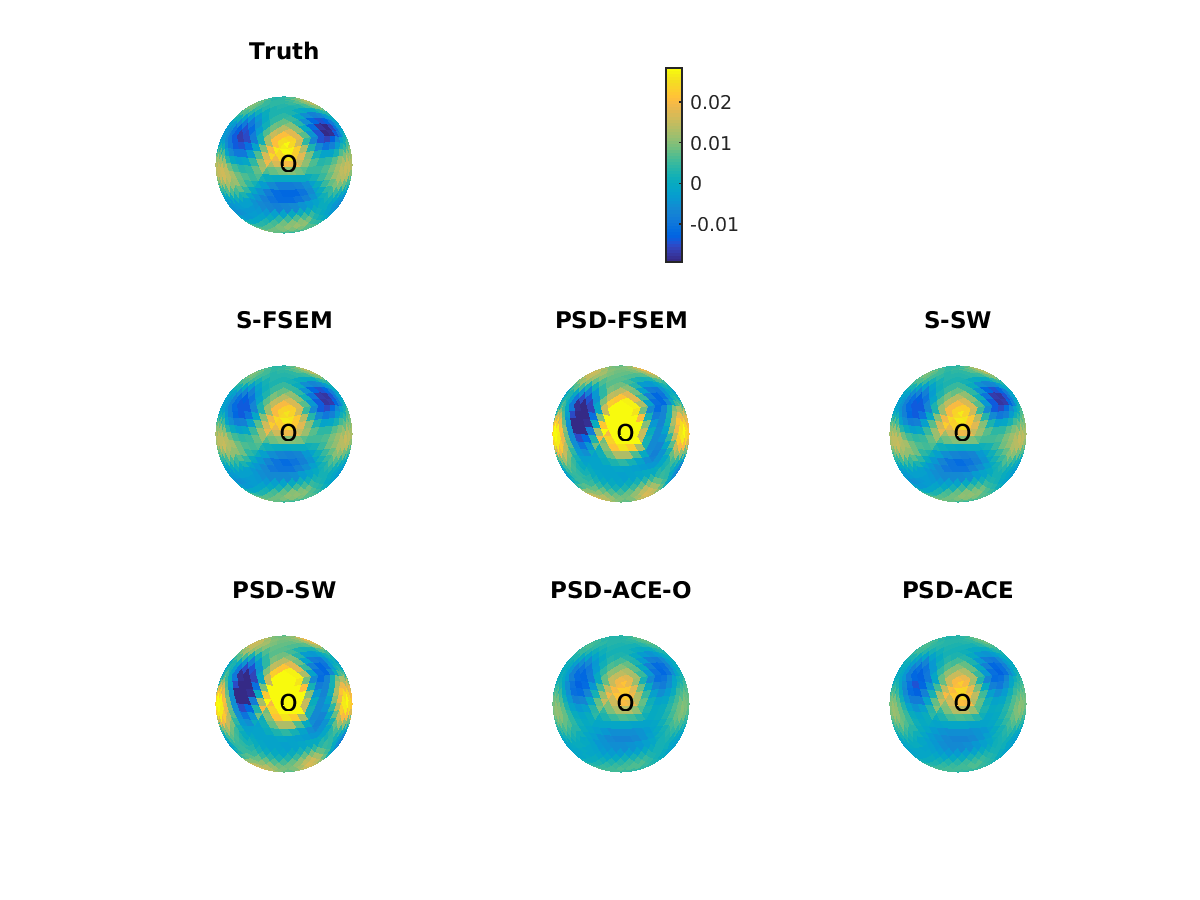}
\caption{Visualizing bias in covariance estimation at a randomly selected seed location. Plotted is the average across 1,000 simulations of $\hatbSigma_a^{(t)}(\cdot,v)$, where $t$ denotes the simulation and the covariance between a fixed location ($v=888$) and the 1,002 locations is evaluated. Note the same color scale is used in all plots.}\label{fig:wsSigmaACov1}
\end{figure}

\begin{figure}[h!]
\includegraphics[width=0.3\textwidth]{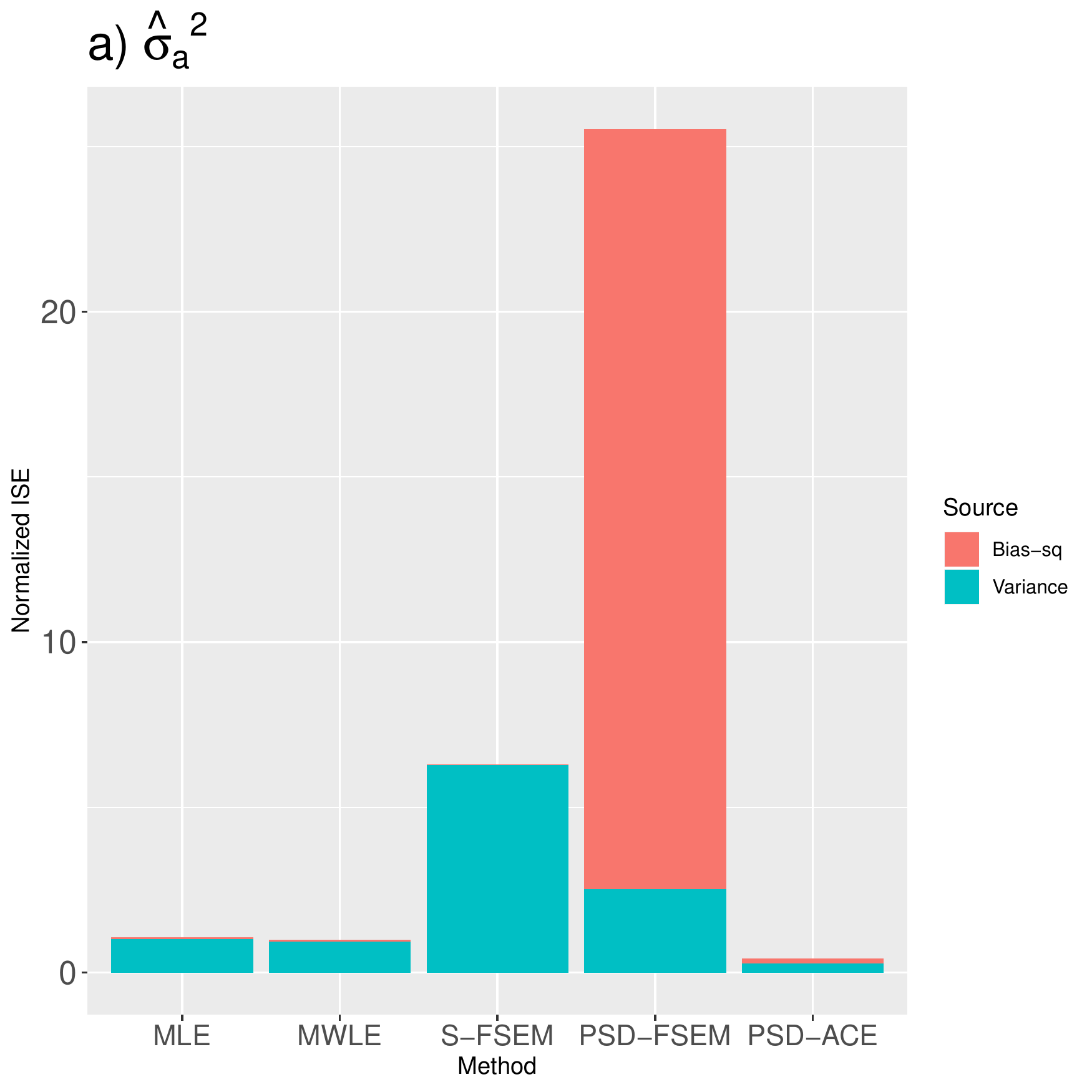}
\includegraphics[width=0.3\textwidth]{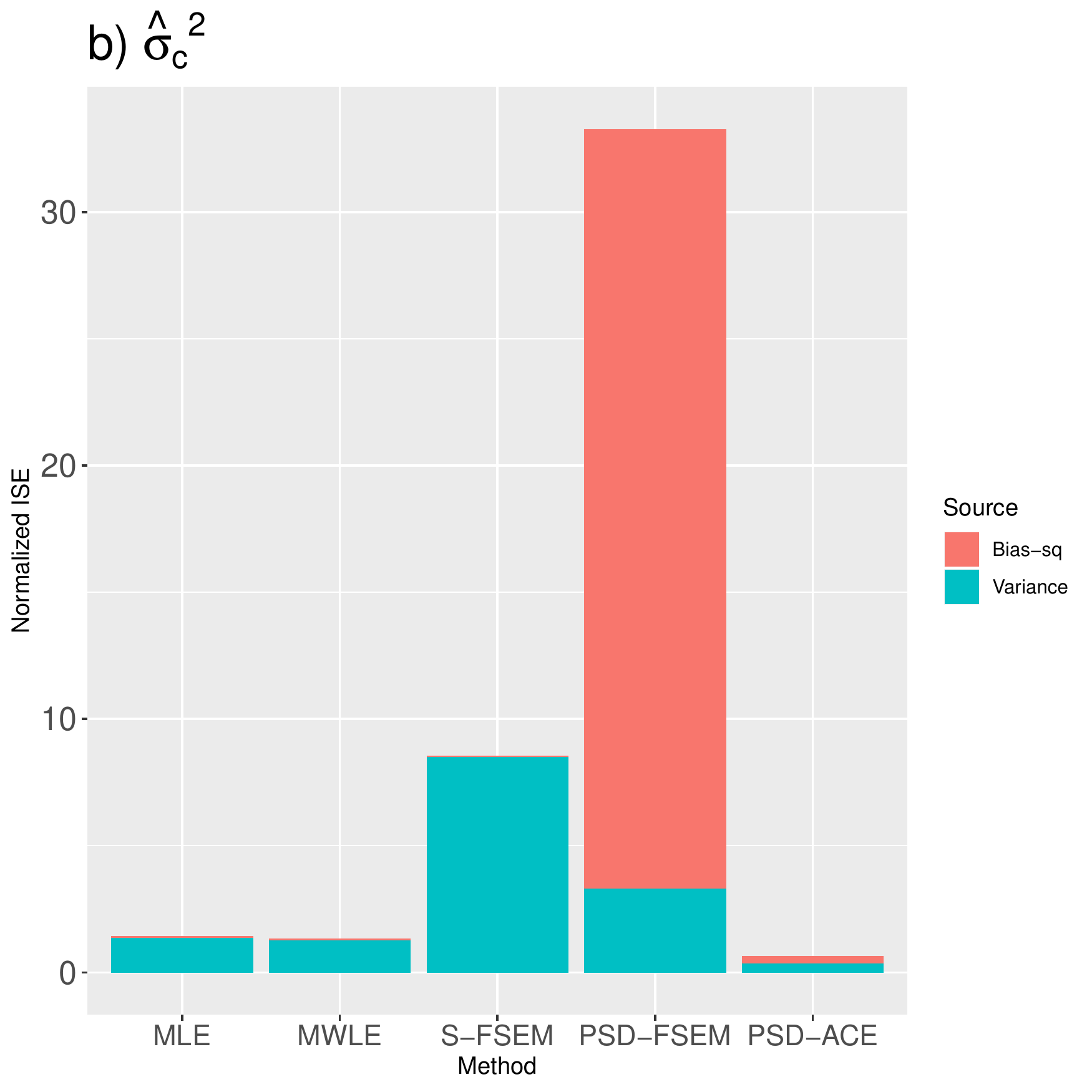}
\includegraphics[width=0.3\textwidth]{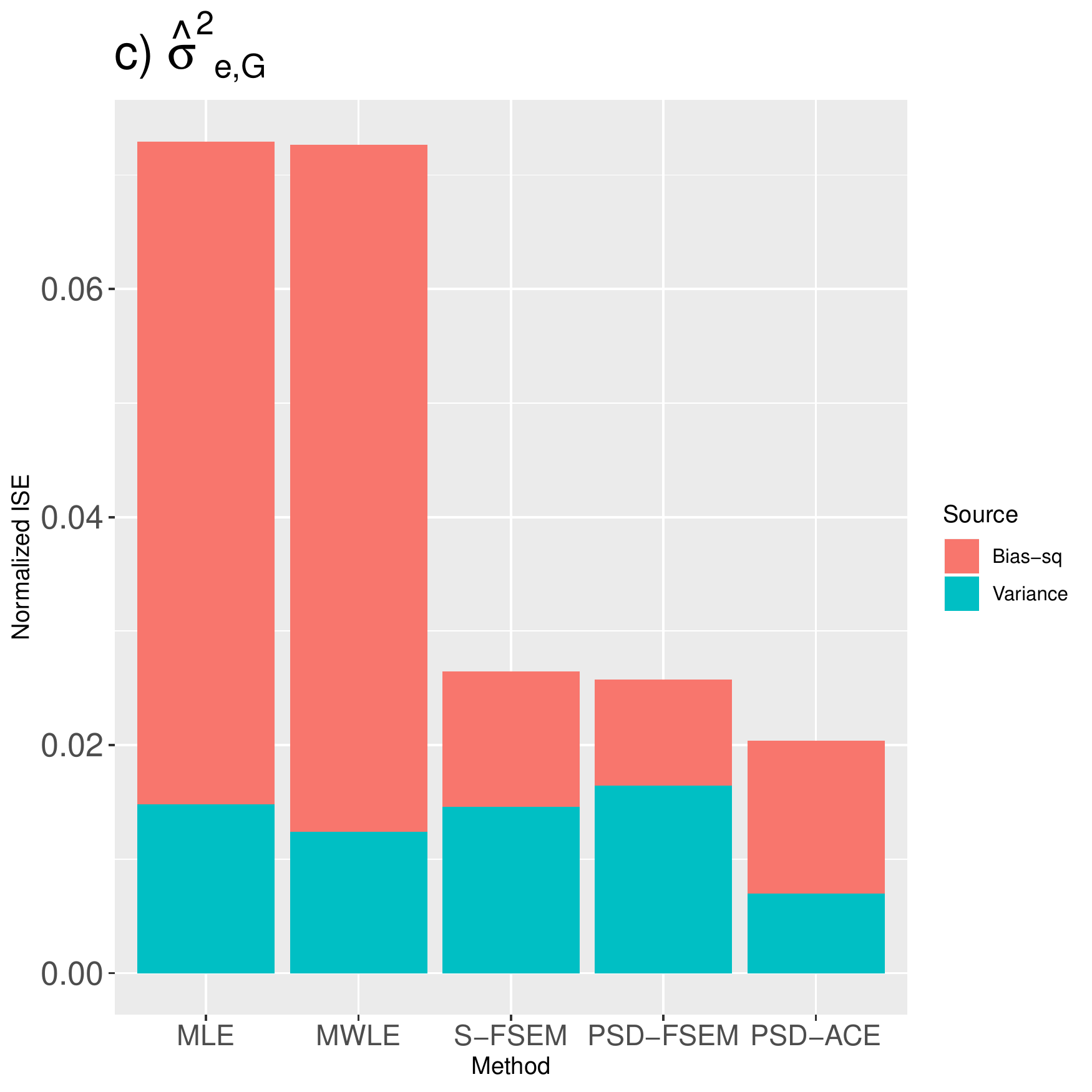}

\includegraphics[width=0.3\textwidth]{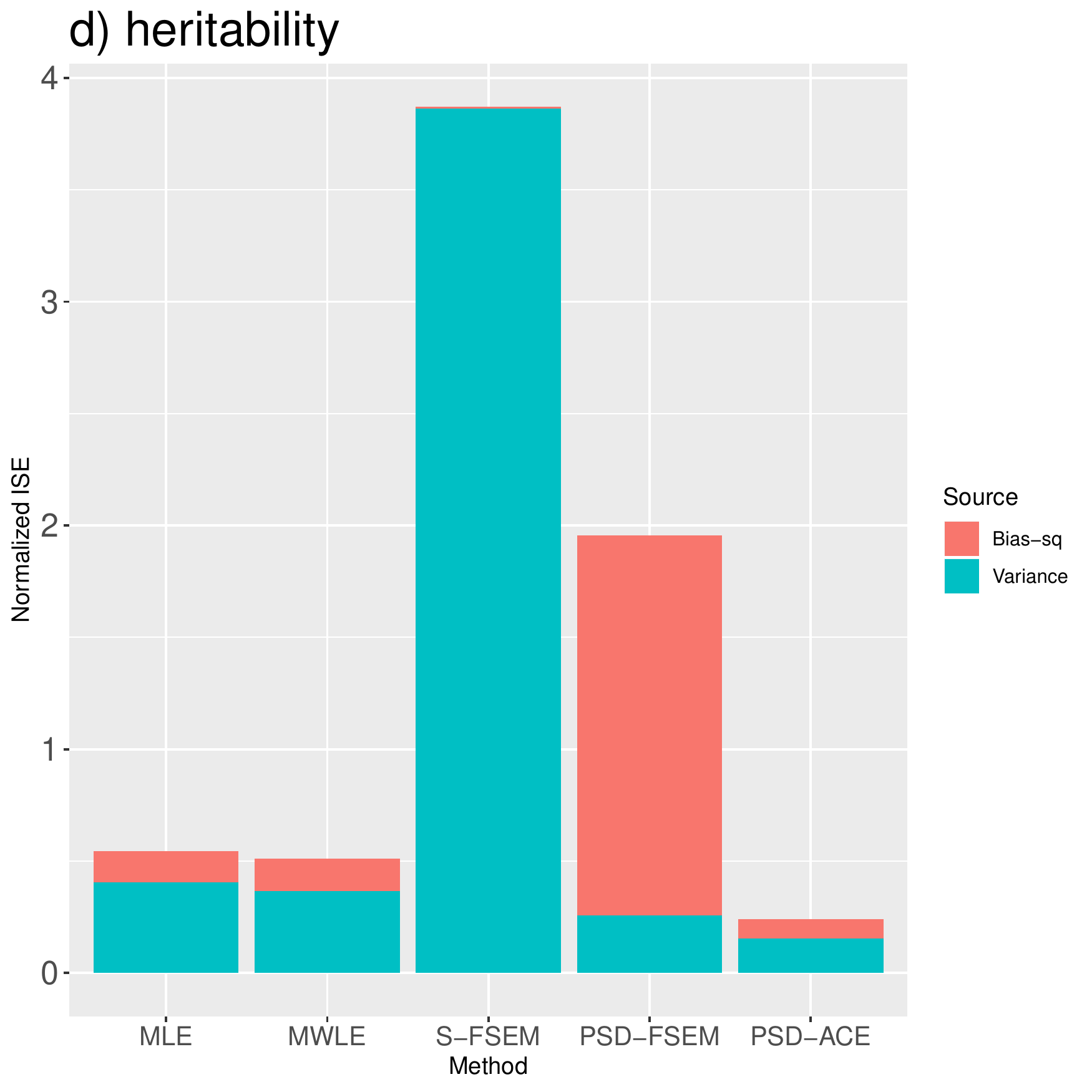}
\includegraphics[width=0.3\textwidth]{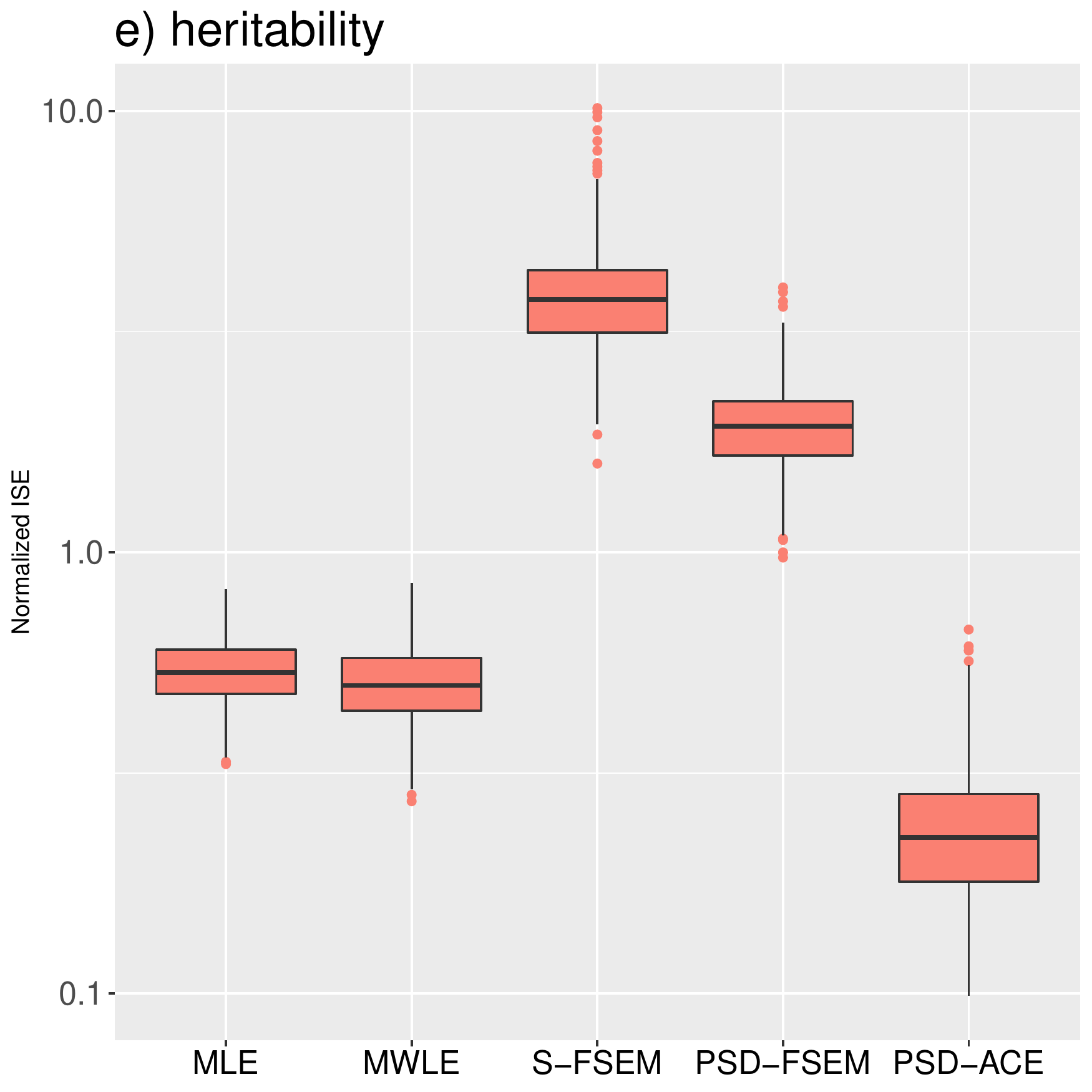}
\caption{Normalized MISE of $\hat{\sigma}_a^2(v)$, $\hat{\sigma}_c^2(v)$, $\hat{\sigma}_{e,G}^2$, and heritability across $V=$1,002 locations. (See Figures S.2-S.5 for a visualization of the bias across space.) Panel e) includes boxplots of the normalized ISE from 1,000 simulations, where the y-axis is on the log10 scale due to large differences between methods.}\label{fig:varianceetc}
\end{figure}

\begin{figure}
\centerline{\includegraphics[width=\textwidth]{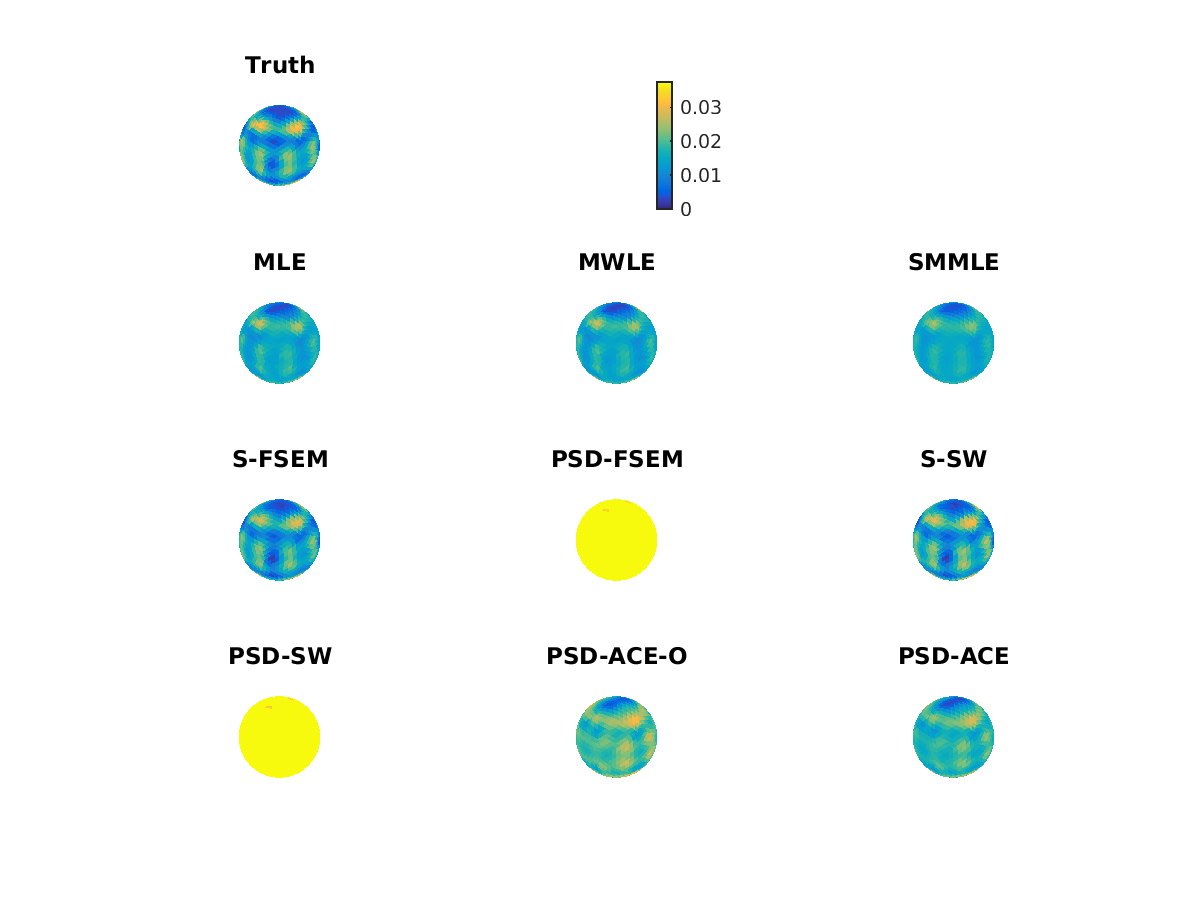}}
\caption{Visualizing bias. Estimates of $\bsigma_a^2(v)$ averaged across 1,000 simulations. Note the same color scale is used in all plots.}\label{fig:imageSigmasqA}
\end{figure}

\begin{figure}
\centerline{\includegraphics[width=\textwidth]{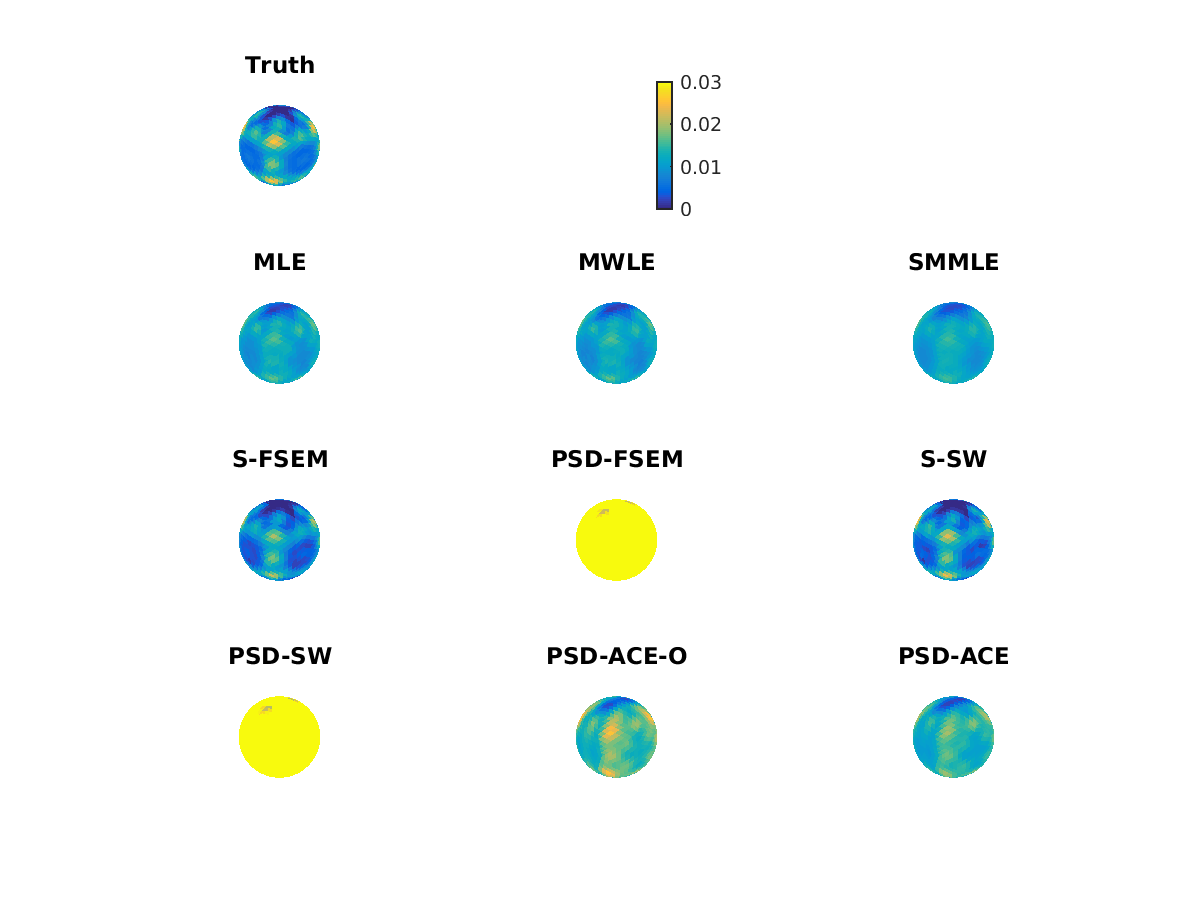}}
\caption{Visualizing bias.  Estimates of $\bsigma_c^2(v)$ averaged across 1,000 simulations. Note the same color scale is used in all plots.}\label{fig:imageSigmasqC}
\end{figure}
       
\begin{figure}
\centerline{\includegraphics[width=\textwidth]{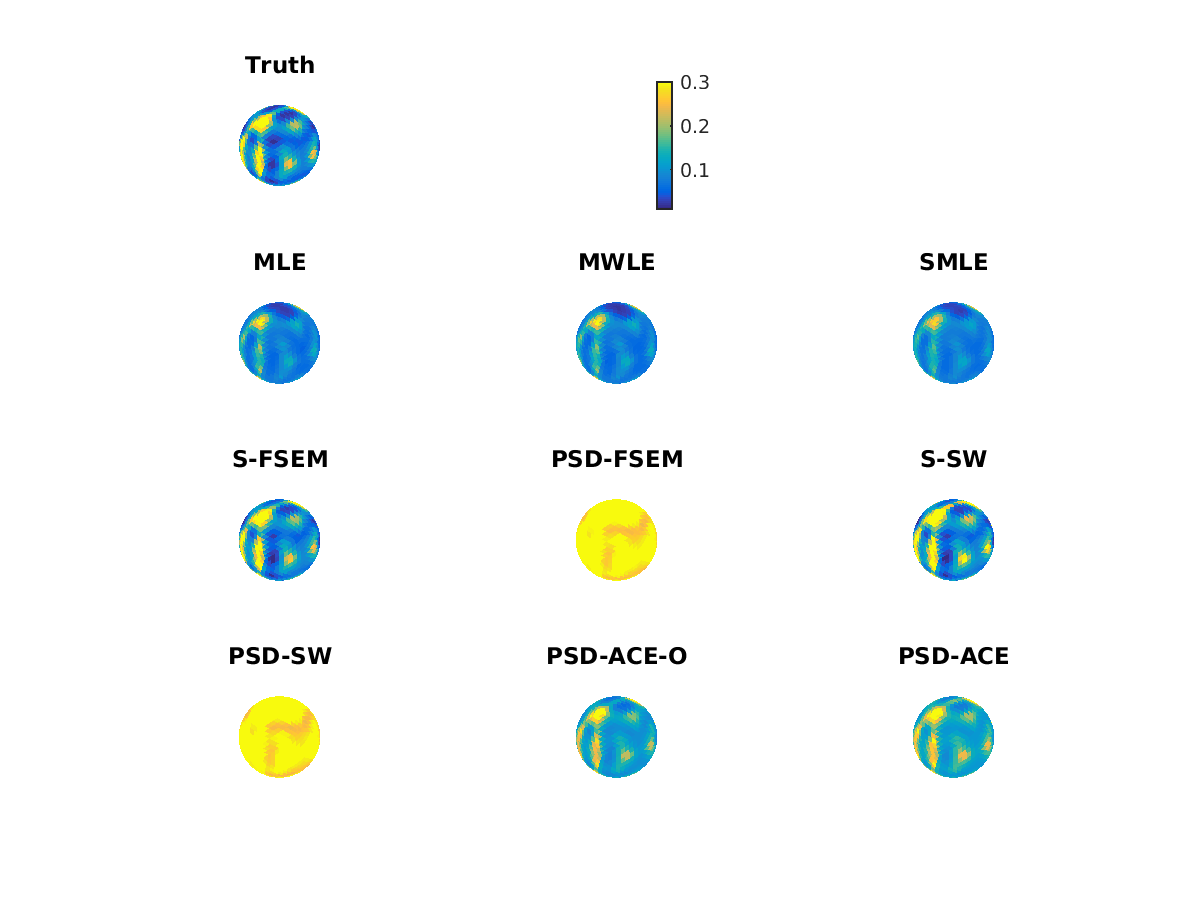}}
 \caption{Visualizing bias in the estimation of heritability. Estimates of $h^2(v)$ averaged across 1,000 simulations. Note the same color scale is used in all plots.}\label{fig:imageh2_fixed}
\end{figure}

\begin{figure}
\centerline{\includegraphics[width=\textwidth]{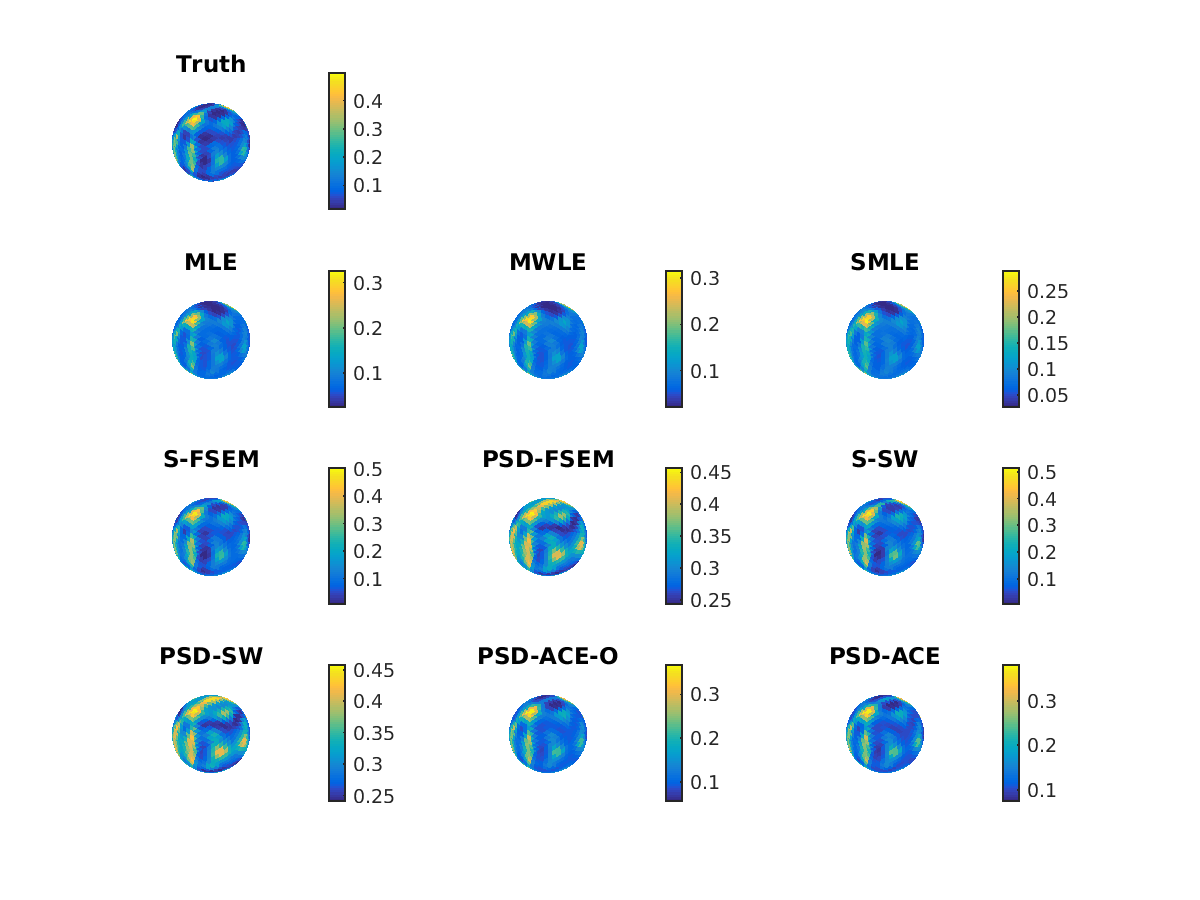}}
 \caption{Visualizing bias in the estimation of heritability. Estimates of $h^2(v)$ averaged across 1,000 simulations. Note a different color scale is used in each plot.}\label{fig:imageh2_varying}
\end{figure}

\begin{table}[ht]                                                                                    
\centering 
\smaller
\begin{tabular}{|c | r r r | r r r | r r r | r r r|}  
\hline      
 Model & \multicolumn{3}{c|}{$ \hat{\sigma}_a^2(v)$} & \multicolumn{3}{c|}{$\hat{\sigma}_c^2(v)$}  & \multicolumn{3}{c|}{$\hat{\sigma}_{e,G}^2(v)$} & \multicolumn{3}{c|}{$\hat{h}^2(v)$} \\
 \hline
 & Bias$^2$ & Var & MISE  & Bias$^2$ & Var & MISE  & Bias$^2$ & Var & MISE  & Bias$^2$ & Var & MISE \\
\hline
\hline                                                                                           
MLE & 0.02 & 0.28 & 0.29 & 0.01 & 0.17 & 0.18 & 0.99 & 0.25 & 1.24 & 3.45 & 10.06 & 13.51 \\     
\hline                                                                                           
MWLE & 0.02 & 0.26 & 0.27 & 0.01 & 0.16 & 0.17 & 1.02 & 0.21 & 1.24 & 3.62 & 9.09 & 12.71 \\     
\hline                                                                                           
SMLE & 0.02 & 0.14 & 0.16 & 0.01 & 0.09 & 0.10 & 1.03 & 0.17 & 1.20 & 4.08 & 4.73 & 8.81 \\      
\hline                                                                                           
S-FSEM & 0.00 & 1.73 & 1.74 & 0.00 & 1.08 & 1.08 & 0.20 & 0.25 & 0.45 & 0.20 & 95.98 & 96.18 \\  
\hline                                                                                           
PSD-FSEM & 6.33 & 0.70 & 7.03 & 3.79 & 0.42 & 4.21 & 0.16 & 0.28 & 0.44 & 42.21 & 6.40 & 48.61 \\
\hline                                                                                           
S-SW & 0.00 & 2.16 & 2.16 & 0.00 & 1.34 & 1.35 & 0.20 & 0.29 & 0.49 & 0.54 & 132.98 & 133.52 \\  
\hline                                                                                           
PSD-SW & 6.11 & 0.80 & 6.91 & 3.65 & 0.48 & 4.13 & 0.14 & 0.29 & 0.44 & 41.59 & 7.62 & 49.21 \\  
\hline                                                                                           
PSD-ACE-O & 0.03 & 0.08 & 0.11 & 0.02 & 0.05 & 0.07 & 0.22 & 0.12 & 0.34 & 1.60 & 4.68 & 6.28 \\ 
\hline                                                                                           
PSD-ACE & 0.04 & 0.07 & 0.12 & 0.04 & 0.04 & 0.08 & 0.23 & 0.12 & 0.35 & 2.11 & 3.84 & 5.95 \\   
\hline     
\end{tabular}                                                                                
\caption{Squared bias, variance, and MISE of the variance and heritability estimates multiplied by $V\!=$1,002. On a side note, SMLE with GCV (approximating leave-one-location out) had lower overall MISE than MWLE with 5-fold CV on family ID (leaving surfaces out). In general, GCV tended to choose a larger bandwidth than 5-fold CV resulting in lower MISE, in addition to computational savings.
}                                                                     
\label{table:mseheritability}                                                                   
\end{table} 

\section{HCP data and model diagnostics}\label{sec:HCPwebsupplement}

\subsection{HCP cortical thickness processing}

\begin{figure}
  \includegraphics[width=\textwidth]{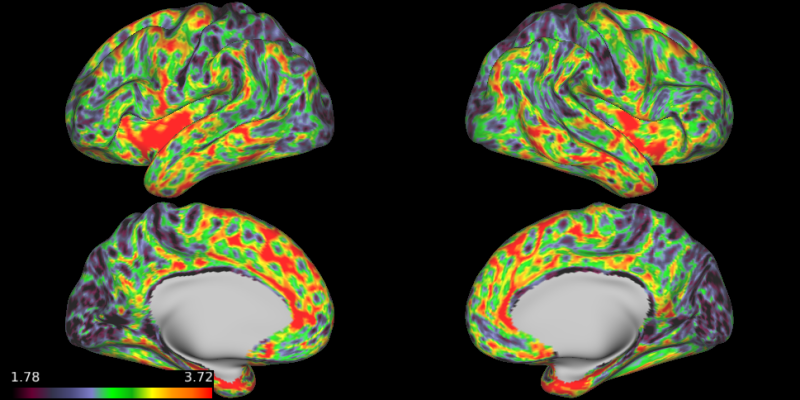}

  \includegraphics[width=\textwidth]{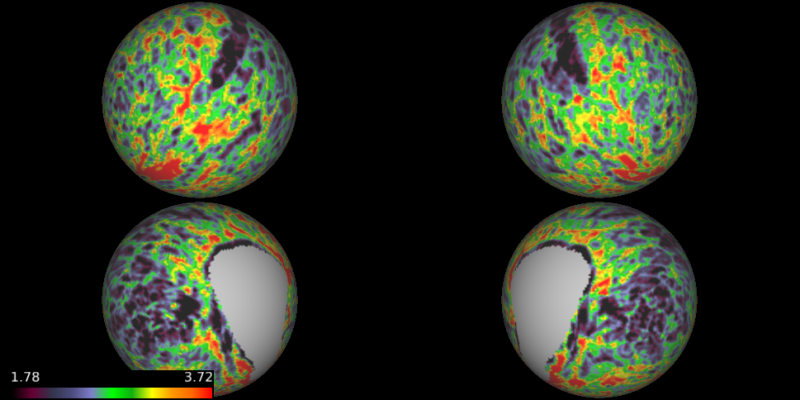}
  
 \caption{Example cortical thickness data from subject 101915. Top four images: data on the subject's inflated surface. The gray regions represent the medial wall. Figures on the left represent the left hemisphere. The top row depicts the lateral view. The second row depicts the medial views. Bottom: data on the 32k\_fs\_LR spherical template.}\label{fig:examplecortthick}
\end{figure}

We used the cortical thickness files $<$SUBJECT ID.thickness\_MSMAll.32k\_fs\_LR.dscalar.nii$>$. We briefly summarize the processing steps performed by the HCP consortium; additional details are in \cite{glasser2013minimal}. The HCP data include high-resolution T1-weighted and T2-weighted structural MRIs with 0.7 mm$^3$ voxels, which are used to construct the cortical surface and estimate cortical thickness. Volume data were processed using Freesurfer's recon-all pipeline, which uses a connected-component algorithm to segment structural images. The T1 image was used to delineate the white matter surface (the inner boundary between the gray matter of the cortical surface and the white matter). A triangular tessellation was overlaid on the white matter surface, inflated, and warped to a sphere, and then diffeomorphic registration performed by aligning the cortical folding patterns with the fsaverage template. We used the MSMAll data, which also utilizes myelination and functional connectivity during registration \citep{robinson2014msm} and alignment to the 164k\_fs\_LR standard surface mesh, in which the left and right hemispheres are in geographic correspondence \citep{van2012parcellations}. The registered surface was then transformed back to each subject's surface space by applying the inverse of the subject-to-atlas mapping. The pial surface was also estimated (the outer boundary of the cortical surface) in which the recon-all pipeline was customized to include the use of both T1w and T2w images.  Cortical thickness is then equal to the distance between the pial and white matter surfaces. 
We utilize the lower resolution version of the cortical thickness data in which each hemisphere of the brain is represented by a sphere with 32,492 vertices at approximately 2 mm spacing, labeled as 32k\_fs\_LR in the HCP data. Data for a single subject is depicted in Figure \ref{fig:examplecortthick}.

\subsection{Selection of model covariates}

\begin{figure}
 \includegraphics[width=\textwidth]{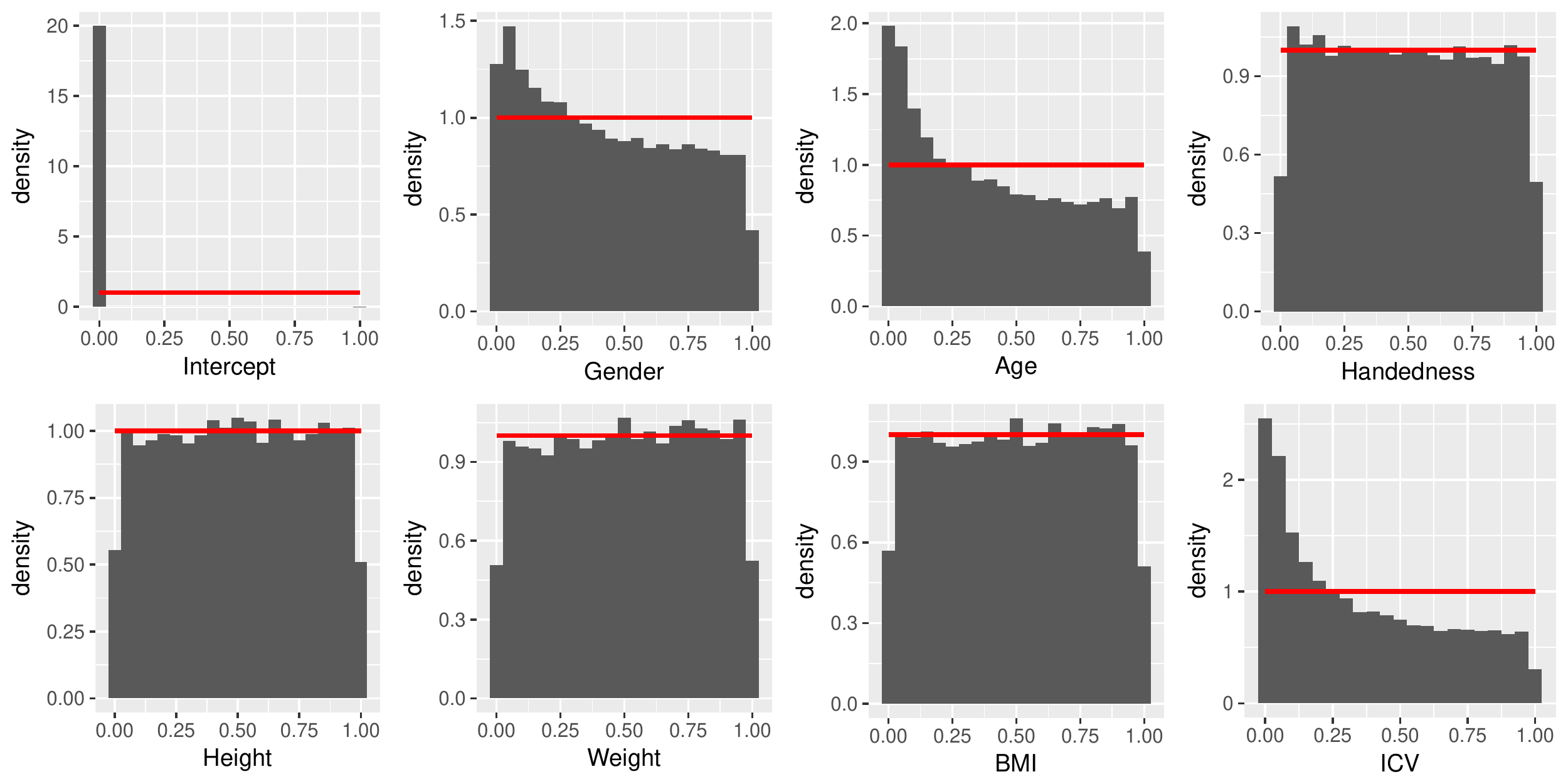}
 \caption{Variable selection in the ACE model. Covariates handedness, height, weight, and BMI were excluded from subsequent analysis because the distribution of their p-values from all vertices in the right hemisphere (V=29,716) was approximately uniform, whereas gender, age, and intracranial volume (ICV) were controlled for in all estimates of heritability.}\label{SuppFigure:PvalueCovariates}
\end{figure}

We considered controlling for the following covariates: age, gender, handedness, height, weight, body mass index, and total intracranial volume (TIV). We inspected the p-values from the Wald statistics of point-wise MLE estimates from the 29,716 vertices (excluding the medial wall) in the right hemisphere and found that handedness, height, weight, and body mass index were roughly uniformly distributed, whereas the p-values for age, gender, and TIV were right skewed (Figure \ref{SuppFigure:PvalueCovariates}). Consequently, we included age, gender, and TIV in all subsequent models. 

\subsection{Geodesic distance}

We calculate geodesic distance along the 32k\_fs\_LR spherical template in MSMAll, which represents a common space in which cortical thickness at a given vertex represents the same location across subjects relative to the aligned cortical folding patterns of the sulci and gyri (as used in FreeSurfer) and the patterns of myelination and functional connectivity (the additional information used in MSMAll). These distances have been used in previous spatial studies \citep{bernal2013spatiotemporal,risk2016spatiotemporal}. There are two alternative approaches: 1) use the geodesic distance along the group-averaged cortex rather than the 32k\_fs\_LR spheres, or 2) use the registered individual surfaces wherein vertices are in correspondence but distances between vertices vary by subject. Regarding the first alternative, using a group template is not advised because folding patterns are averaged resulting in a smoothed surface with distances affected in undesirable ways. Regarding the second alternative, it is unclear whether using the individual-specific distances would improve or degrade performance, and it creates additional computational and mathematical challenges. The subject-specific distances would require a separate kernel matrix for every subject, causing scalability issues that prevent the application of Algorithm 1. Additionally, we define our covariance function for a common domain $\cV$. In a subject-specific approach, the nodes $v_1,\dots,v_V$ are common across subjects but the map to $\tR^3$ is subject-specific. Then conceptually we would be attempting to define a common covariance function for different manifolds. 

\subsection{Bandwidth selection, rank selection, and sensitivity to rank selection}

\begin{figure}
 \includegraphics[width=0.75\textwidth]{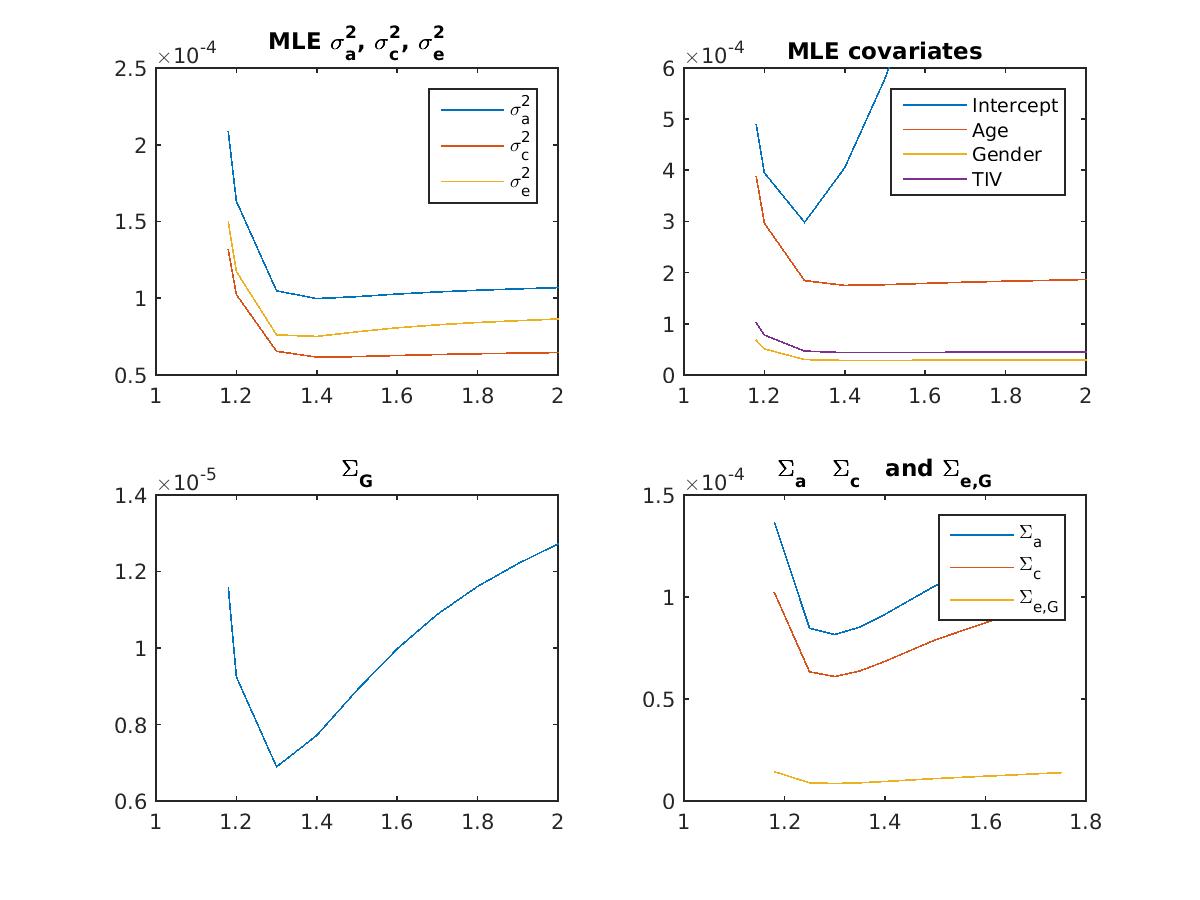}
 \caption{Bandwidth selection using GCV to determine the smoothing of the MLE estimates in the HCP data (Step 2) (top panels), the smooth Gaussian process variance $\Sigma_G$ (Step 3) (bottom left), and the covariance functions, $\Sigma_a$, $\Sigma_c$, and $\Sigma_{e,G}$, in Step 4 (bottom right).}\label{fig:GCV_HCP}
\end{figure}

In the MWLE analysis, 5-fold CV selected a bandwidth = 1.8 arclengths (in degrees). The bandwidth chosen in Step 2 of PSD-ACE was 1.4 for each of the three variance parameters; the bandwidth chosen in Step 3 for $\hatbSigma_G$ equaled 1.3; and the bandwidths chosen for $\hatbSigma_a$, $\hatbSigma_c$, and $\hatbSigma_{e,G}$ were 1.3 (Figure \ref{fig:GCV_HCP}). Note the triangles on the 32k\_fs\_LR tessellation of the sphere are not equal in area. A typical vertex has two closest neighbors at a distance of approximately 1.15, then two neighbors at 1.2, then two more neighbors near 1.3, which compose the set of vertices of the triangles containing the focal vertex; then the next closest distance is approximately 2, which represents a vertex in a set of triangles that do not contain the focal vertex. For bw=1.3, average weights are 0.878, 0.044, 0.044, 0.015, 0.015, 0.002, and 0.002 for the focal and neighboring vertices. 

\begin{figure}
  \includegraphics[width=0.75\textwidth]{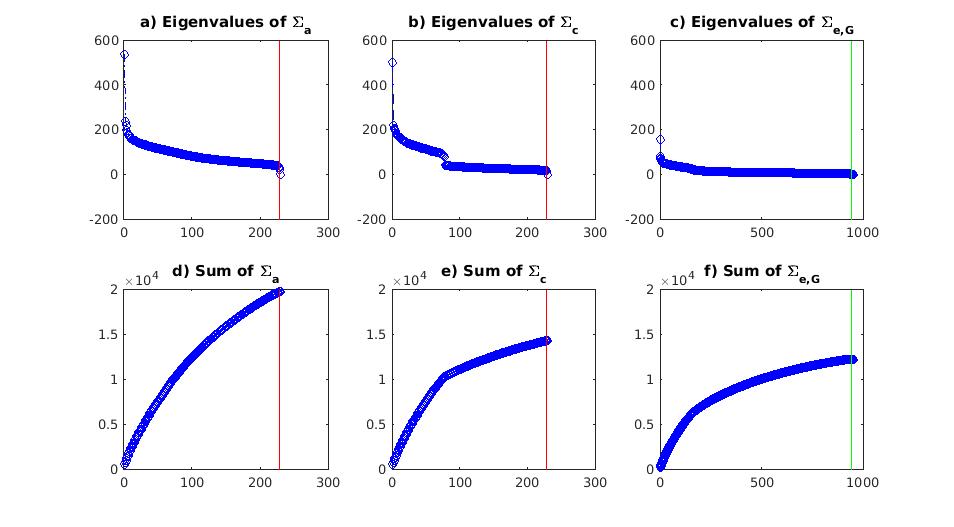}
  \caption{Eigenvalues of $\hatbSigma_a^{SW}$, $\hatbSigma_c^{SW}$, and $\hatbSigma_{e,G}^{SW}$ for the HCP data. The red vertical line in a), b), d), and e) corresponds to the number of twin pairs.  The green line in c) and f) corresponds to the total number of individuals minus the number of monozygotic twin families. Since we used the power algorithm to iteratively estimate the eigenvalues, we stopped the algorithm when the eigenvalue was numerically zero.}\label{SuppFigure:HCP_Eigenvalues}
\end{figure}

The eigenvalues from the decomposition of $\hatbSigma_a^{SW}$, $\hatbSigma_c^{SW}$, and $\hatbSigma_{e,G}^{SW}$ are depicted in Figure \ref{SuppFigure:HCP_Eigenvalues}. Note the eigenvalues/vectors were estimated using the Matlab function $\texttt{eigs}$ to calculate a reduced number of eigenvalues; naive implementation of $\texttt{eig}$ for large datasets may be impracticable. 

We also assessed the sensitivity of PSD-ACE to the selected ranks. We re-ran Step 6 with the ranks reduced by 10 for each covariance (i.e., the ranks were set to 219, 219, and 933 for the additive genetic, common environmental, and unique environmental covariances). There were negligible changes in heritability (max difference = 0.0006), additive genetic correlations (max diff = 0.0017), common environmental correlations (max diff = 0.0009), and unique environmental correlations (max diff = 0.0035). The results appear to be relatively robust to the rank selection. This is expected because the PSD-ACE is initiated from the PSD-FSEM, in which the eigenvectors are ordered by their variance, such that excluding the smallest eigenvalue/eigenvector pairs should have negligible impacts.

\subsection{Assessing convergence}
\begin{figure}
\includegraphics[width=0.45\textwidth]{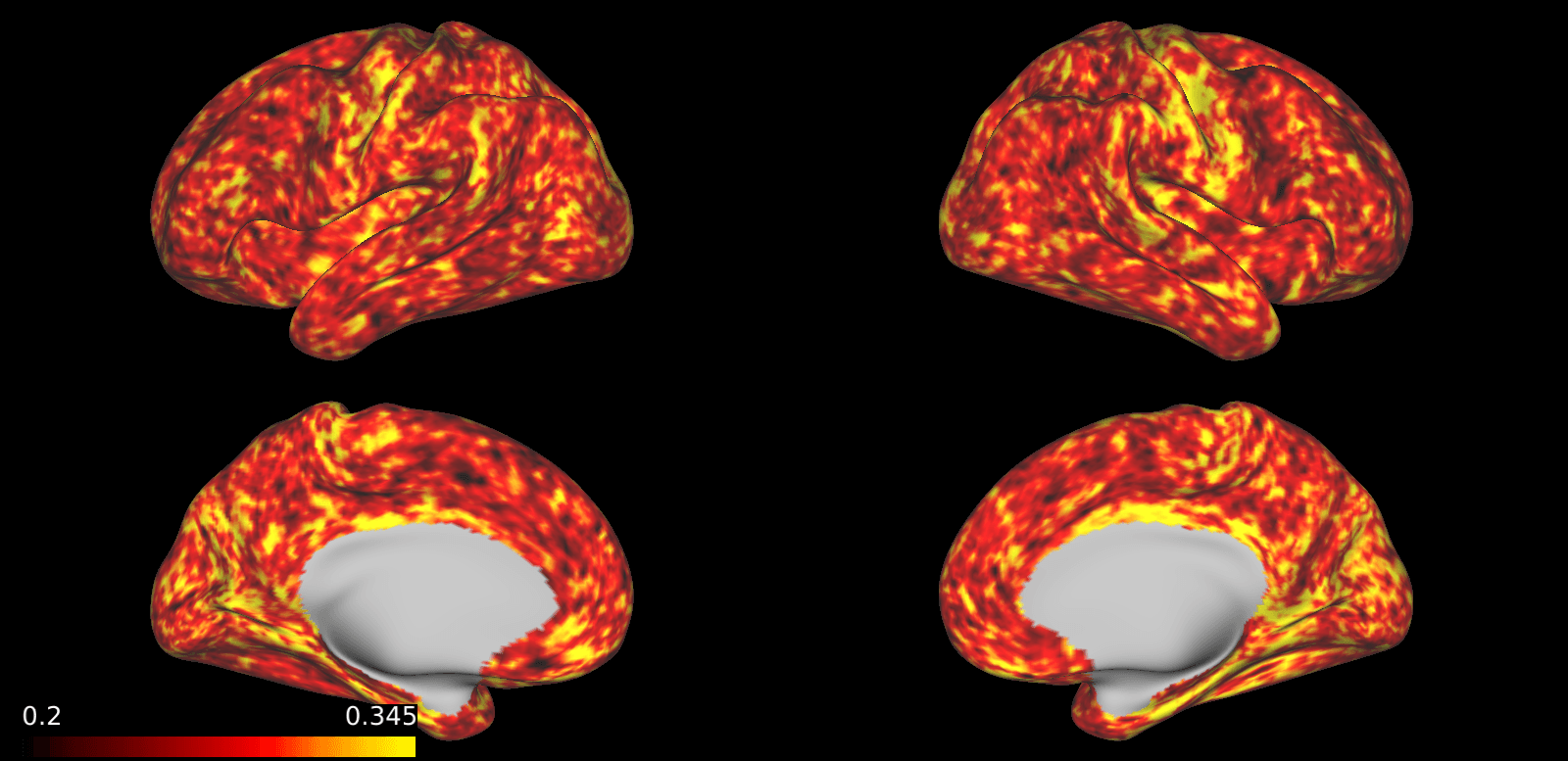}
\includegraphics[width=0.45\textwidth]{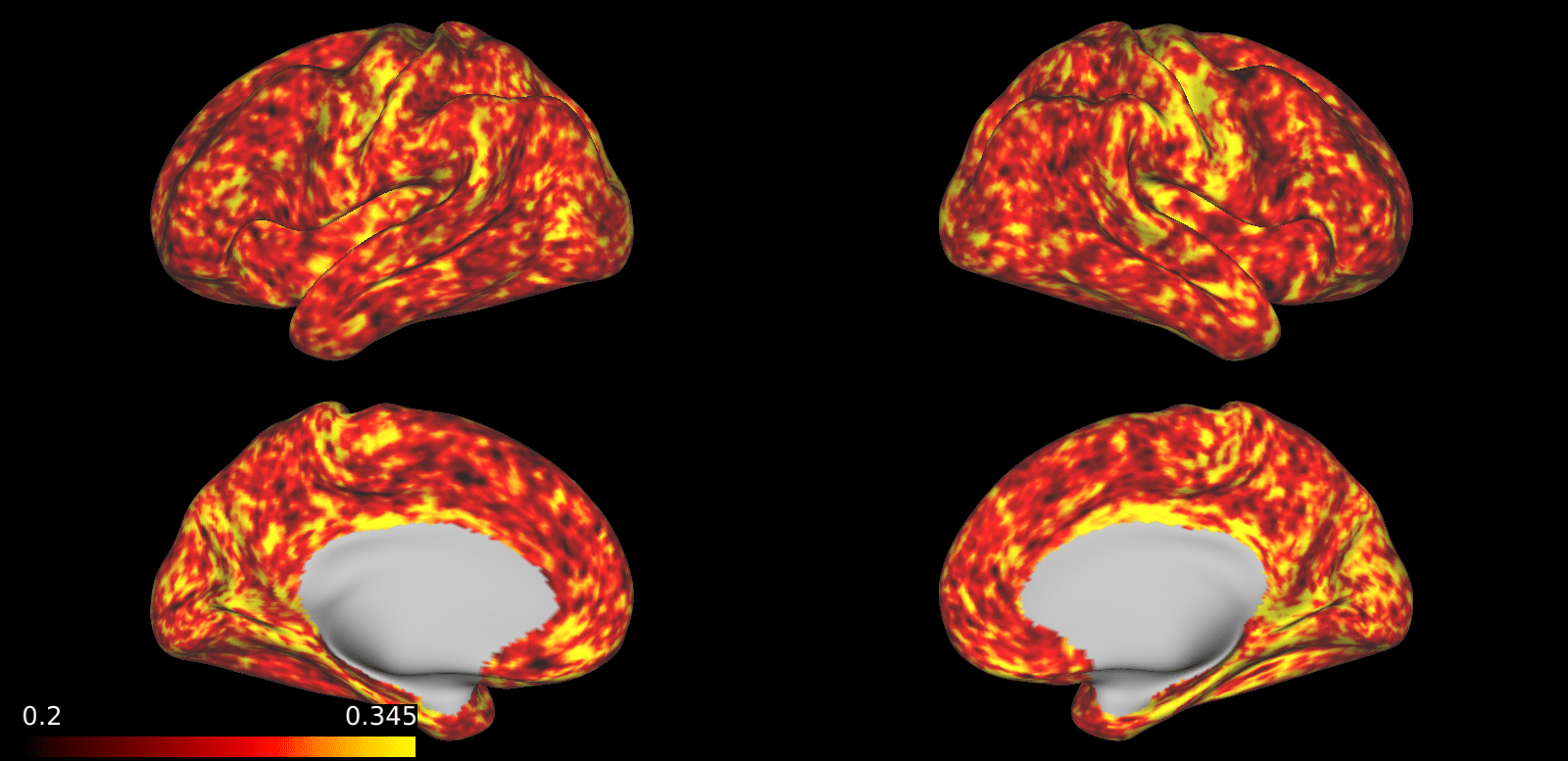}
\caption{Heritabilities estimated from the HCP data using 1000 iterations (left, in Figure 4 of the main manuscript) and an additional 600 iterations (right), demonstrating convergence.}\label{SuppFigure:HCP_convergence}
\end{figure}

We ran 1000 iterations of Algorithm 1, which took 25 hours. The magnitude of the gradient after 1000 iterations was 0.02\% of the magnitude in the first iteration; for the additional iterations, this decreased to 0.01\%. With the additional iterations, there were negligible changes in the heritability estimates (max difference across all vertices = 0.0048), additive genetic correlations (max diff in correlation = 0.0007), common environmental correlations (max diff = 0.0008), and unique environmental correlations (max diff = 0.0082). A visual examination of the heritability results suggest they are nearly indistinguishable (Figure \ref{SuppFigure:HCP_convergence}). Thus we used the results from 1000 iterations.

\section{Additional HCP results}

\begin{figure}
\includegraphics[width=0.8\textwidth]{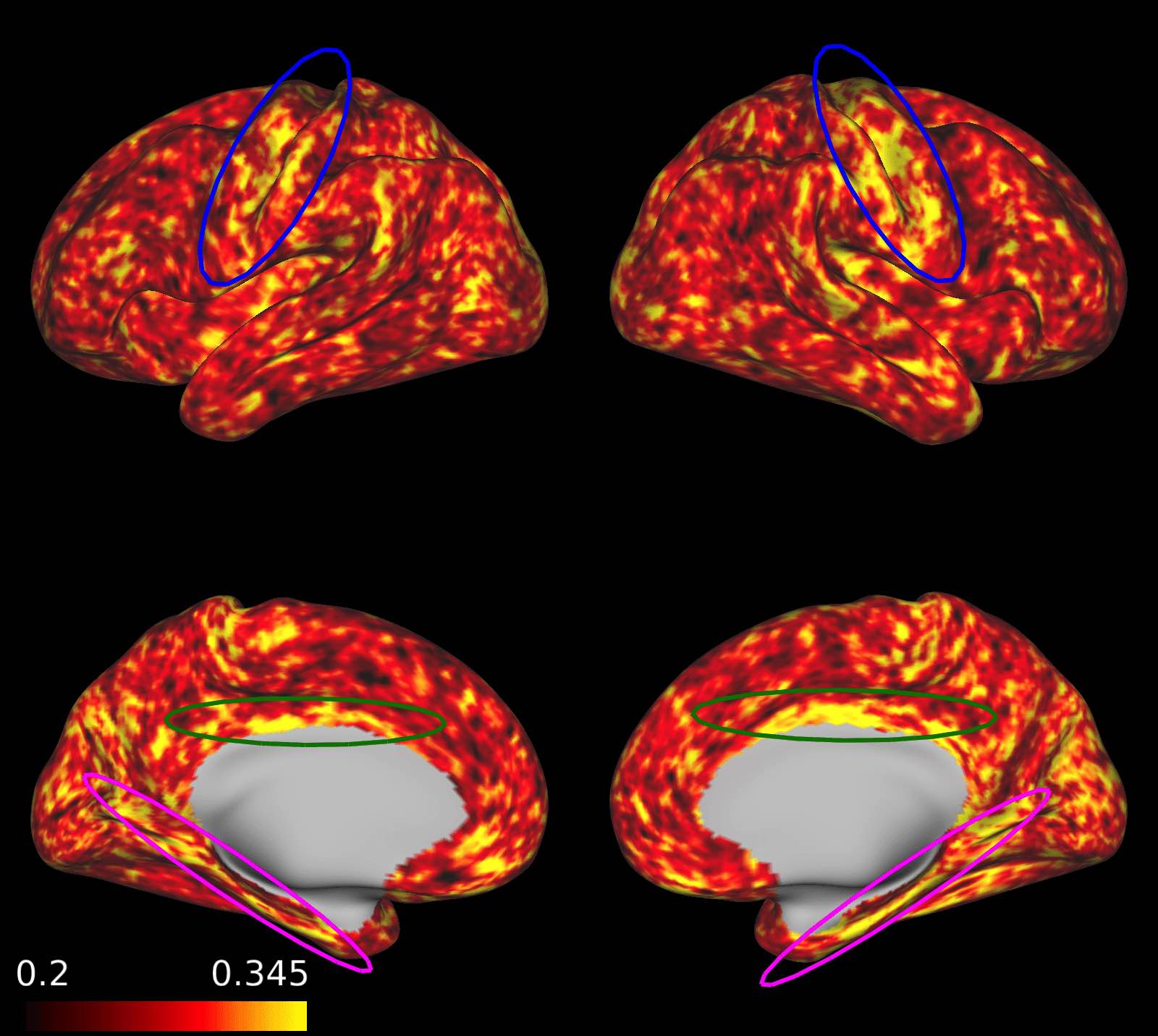}
\caption{Heritability from the PSD-ACE. The blue ovals correspond to the central sulci, the green ovals correspond to medial areas dorsal to the corpus callosum, and the fuchsia ovals correspond to the parahippocampal gyrus and entorhinal cortex.}\label{fig:h2_psd_annotated}
\end{figure}

In Figure \ref{fig:h2_psd_annotated}, brain regions described in the text are highlighted. The blue and green ovals are areas of higher heritability in \cite{shen2016heritability}, while the fuchsia ovals correspond to areas that had more pronounced heritabilities than in \cite{shen2016heritability}. 

\begin{figure}
 \includegraphics[width=0.5\textwidth]{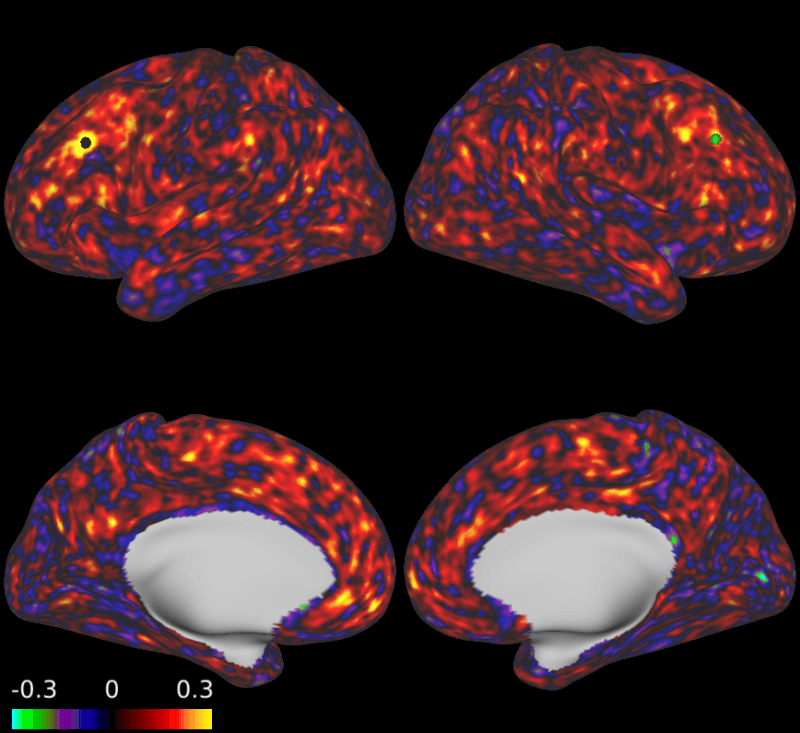}
 \includegraphics[width=0.5\textwidth]{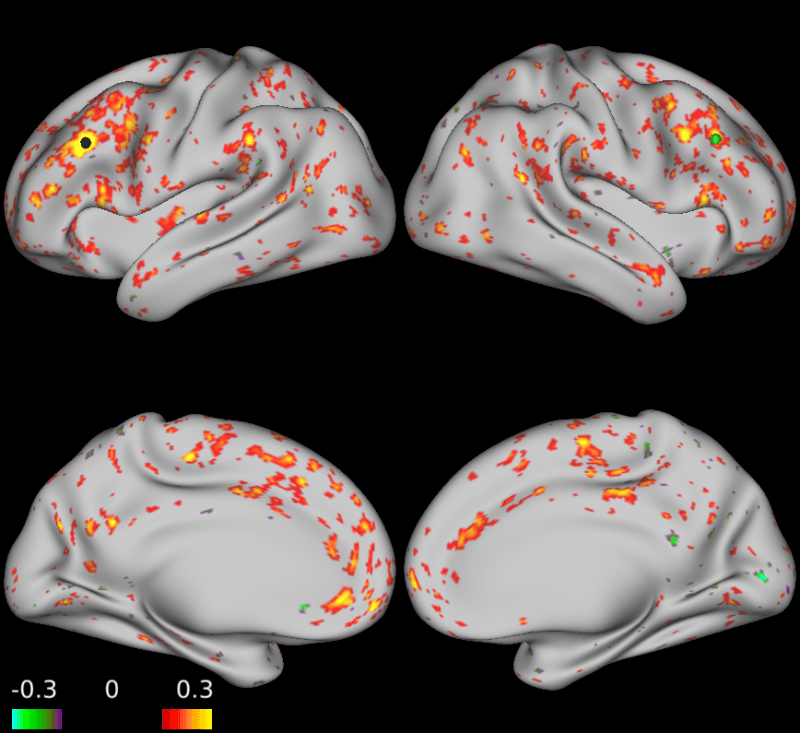}
 \caption{Correlation between a middle frontal seed vertex in the left hemisphere (black dot, surface vertex for the left cortex of the fs\_LR template: 30,456) and all other locations. The contralateral location is highlighted in green. The right images are thresholded at correlations whose absolute values are greater than 0.15.}\label{fig:middlefrontalseed}
\end{figure}

We evaluated the covariance function at a location in the middle frontal gyrus and  (Figure \ref{fig:middlefrontalseed}), similar to Figure 3 in \cite{rimol2010cortical}. We detected some contralateral correlation, but to a lesser extent than noted in \cite{rimol2010cortical}. Additionally, our correlation is much more localized. Overall, our heritability results, as well as those from \cite{shen2016heritability}, bear little resemblance to \cite{rimol2010cortical}, which used the additive and dominant genetic and unique environmental (ADE) model with broad-sense heritability, a cohort of middle-aged men, and higher smoothing, and consequently had considerably higher overall heritability.

\bibliography{SpatialACE.bbl}

\end{document}